\documentclass[twocolumn]{aastex62}
\usepackage{bm}
\usepackage{multirow}
\usepackage{gensymb}
\usepackage{hyperref}
\hypersetup{
    colorlinks=true,
    linkcolor=blue,
    filecolor=magenta,      
    urlcolor=cyan,
    pdftitle={Overleaf Example},
    pdfpagemode=FullScreen,
    }
\usepackage[]{color}

\usepackage{natbib}
\newcommand{\xstar}{\textsc{\scriptsize XSTAR}}
\newcommand{\cloudy}{\textsc{\scriptsize CLOUDY }}
\newcommand{\athena}{\textsc{Athena\scriptsize ++ }}

\def\CIVdbl{{\rm C~}\kern 0.1em{\sc iv}~$\lambda\lambda 1548, 1550$} 
\def\OVIIIi{\hbox{{\rm O}\kern 0.1em{\sc viii}}}
\def\SiXIVi{\hbox{{\rm Si}\kern 0.1em{\sc viii}}}
\def\OVIII{\hbox{{\rm O}\kern 0.1em{\sc viii}~{\rm Ly}$\alpha$}}
\def\SiXIV{\hbox{{\rm Si}\kern 0.1em{\sc viii}~{\rm Ly}$\beta$}}
\def\FeXXV{\hbox{{\rm Fe}\kern 0.1em{\sc xxv}}}
\def\FeXXVI{\hbox{{\rm Fe}\kern 0.1em{\sc xxvi}}}
\def\FeXXVK{\hbox{{\rm Fe}\kern 0.1em{\sc xxv}~{\rm K}$\alpha$}}
\def\FeXXVIK{\hbox{{\rm Fe}\kern 0.1em{\sc xxvi}~{\rm K}$\alpha$}}

\newcommand{\beq}{\begin{equation}}
\newcommand{\seq}{\end{equation}}

\newcommand{\gv}[1]{\ensuremath{\mbox{\boldmath$ #1 $}}} % for vectors of Greek letters
\renewcommand{\d}[2]{\frac{d #1}{d #2}} % for derivatives
\newcommand{\pd}[2]{\frac{\partial #1}{\partial #2}}
\newcommand{\f}{\frac}  %shorter fraction
\usepackage{letltxmacro,amsmath}
\LetLtxMacro{\originaleqref}{\eqref}
\renewcommand{\eqref}{Eq.~\originaleqref}
\defcitealias{Begelman83}{BMS83}
\defcitealias{Dannen20}{D20}

\received{January 2020}
\revised{April 2020}
\accepted{April 2020}
\shorttitle{Multiphase AGN disk winds}
\shortauthors{Waters, Proga, \& Dannen}
\begin{document}
\title{Multiphase AGN winds from X-ray irradiated disk atmospheres}

\correspondingauthor{Tim Waters}
\email{tim.waters@unlv.edu}
\author[0000-0002-5205-9472]{Tim Waters}
\affiliation{Department of Physics \& Astronomy \\
University of Nevada, Las Vegas \\
4505 S. Maryland Pkwy \\
Las Vegas, NV, 89154-4002, USA}
\affiliation{Theoretical Division, Los Alamos National Laboratory}
\author[0000-0002-6336-5125]{Daniel Proga}
\affiliation{Department of Physics \& Astronomy \\
University of Nevada, Las Vegas \\
4505 S. Maryland Pkwy \\
Las Vegas, NV, 89154-4002, USA}
\author[0000-0002-5160-8716]{Randall Dannen}
\affiliation{Department of Physics \& Astronomy \\
University of Nevada, Las Vegas \\
4505 S. Maryland Pkwy \\
Las Vegas, NV, 89154-4002, USA}

%%%%%%%%%%%%%%%%%%%%%%%%%%%%%%%%%%%%%%%%%%
\begin{abstract}
The mechanism of thermal driving for launching mass outflows is interconnected with classical thermal instability (TI).  In a recent paper, we demonstrated that as a result of this interconnectedness, radial wind solutions of X-ray heated flows are prone to becoming clumpy.    
In this paper, we first show that the Bernoulli function determines whether or not the entropy mode can grow due to TI in dynamical flows.  Based on this finding, we identify a critical `unbound' radius beyond which TI should accompany thermal driving.  Our numerical disk wind simulations support this result and reveal that clumpiness is a consequence of buoyancy disrupting the stratified structure of steady state solutions.  Namely, instead of a smooth transition layer separating the highly ionized disk wind from the cold phase atmosphere below, hot bubbles formed from TI rise up and fragment the atmosphere.  These bubbles first appear within large scale vortices that form below the transition layer, and they result in the episodic production of distinctive cold phase structures referred to as irradiated atmospheric fragments (IAFs).  Upon interacting with the wind, IAFs advect outward and develop extended crests.  The subsequent disintegration of the IAFs takes place within a turbulent wake that reaches high elevations above the disk. We show that this dynamics has the following observational implications: dips in the absorption measure distribution are no longer expected within TI zones and there can be a less sudden desaturation of X-ray absorption lines such as \OVIII\ as well as multiple absorption troughs in \FeXXVK.
\end{abstract}

%%%%%%%%%%%%%%%%%%%%%%%%%%%%%%%%%%%%%%%%%%%%%%%%%%%%%%%%%%%%%%%%%%%%%
\section{Introduction} \label{sec:intro}

The radiation associated with disk-mode accretion in active galactic nuclei (AGNs) may power a wide variety of the ionized outflows that have been observed \citep[e.g.,][]{Giustini19}.  In order of decreasing kinetic power, these include extremely high-velocity outflows, broad absorption line quasar outflows, ultrafast outflows, several other classes of broad and narrow absorption line winds, X-ray obscurers, and warm absorbers
\citep[][and references therein]{Hidalgo20,Laha20}.
If the radiation primarily transfers its momentum to the gas, the resulting wind is referred to as being line-driven, dust-driven, or radiation pressure driven, depending on if the dominant source of opacity is, respectively, opacity due to spectral lines, dust opacity, or electron scattering opacity. It is common to employ a simple treatment of the gas thermodynamics (e.g., an isothermal approximation) when modeling such winds, which precludes the possibility that the gas can become multiphase due to thermal instability (TI; \citealt{Field65}).  To explain X-ray obscurers and warm absorbers, which seem to be multiphase components within a more highly ionized outflow \cite[e.g.,][]{Kaastra2002,Kaastra2014}, a more compelling wind launching mechanism results from radiation primarily transferring its energy rather than its momentum to the gas, i.e. thermal driving.  The appeal of this explanation is due to the fundamental connection between thermal driving and TI, our focus in this paper. 

This connection was first utilized in a pioneering study by \citeauthor{Begelman83} (\citeyear{Begelman83}; hereafter \citetalias{Begelman83}) to reveal how the irradiation of optically thin gas results in a thermally driven wind beyond the Compton radius,
\beq
R_{\rm IC} = \f{GM_{\rm bh} \bar{m}}{k\,T_C} = 5.24\times 10^{-2}\, \f{\mu M_7}{T_C/10^8{\rm\,K}}\,{\,\rm pc},
\label{eq:RIC}
\seq
the distance where the sound speed of gas heated to the Compton temperature $T_C$ exceeds the escape speed from the disk.  (Here $M_{\rm bh}$ is the black hole mass, $M_7 = M_{\rm bh}/10^7{M_\odot}$, and $\bar{m} = \mu m_p$ is the mean particle mass of the plasma.) 
Compared to $T_C$, the thermalized gas in the disk itself is much colder at $T\sim 10^4\,\rm{K}$.  
Prior to the work of \citetalias{Begelman83}, it had been shown by 
\cite{Krolik81} that the process of heating gas from $\sim 10^4\,\rm{K}$ to $T_C$ is highly prone to TI. 
In particular, \cite{Krolik81}, \cite{Kallman82} and \cite{Lepp85} calculated a variety of so-called S-curves in the AGN environment, revealing that they commonly have thermally unstable regions (`TI zones') at temperatures where recombination cooling, bremsstrahlung, and Compton heating are the dominant heating and cooling processes. 

As \citetalias{Begelman83} commented, the effective scale height of an irradiated accretion disk is set by the highest stable temperature on the cold branch of the S-curve (of order $10^5\,{\rm K}$) rather than by the conditions inside the disk.  Above this temperature, the presence of a TI zone leads to runaway heating and an accompanying steep drop in density, thereby facilitating the heating of atmospheric gas to Compton temperatures.  The magnetohydrodynamical (MHD) turbulent nature of the disk is not expected to alter this picture because the Compton radius is at distances where the local dissipation energy (and hence the magnetic energy) is far less than that due to radiative heating.  While self-gravity and dust physics may be important within the accretion disks at these radii, they should be dynamically unimportant within the irradiated disk atmosphere and disk wind, which have comparatively low density and are hotter than dust sublimation temperatures.

The spectral energy distribution (SED) determines several properties of the irradiated gas, in particular the Compton temperature and the net radiative cooling rates relevant for assessing the number and temperature ranges of TI zones.  How the existence and properties of TI zones depend on the underlying SED was systematically explored in series of papers by \cite{Chakravorty08,Chakravorty09,Chakravorty12}
with further dependencies assessed by \cite{Ferland13}, \cite{Lee13} and \cite{Rozanska14}.    
SEDs for individual systems are inferred by fitting the observed spectra to theoretical models for the intrinsic components of an AGN spectrum \citep[e.g.,][]{Done12,Jin12,Ferland20}.  In this work, we present calculations for an S-curve obtained in this way for the well studied system NGC~5548 \citep[see][]{Mehdipour15}.  While the velocity, density, and temperature structure of thermal disk winds can be sensitive to the SED, our main results regarding TI are expected to hold so long as the S-curve features at least one TI zone.

SED fitting calculations for NGC~5548 performed by different groups using photoionization modeling are in agreement that the corresponding S-curves have prominent TI zones 
\citep[e.g.,][]{Arav15,Mehdipour15}.  
The SED found by \cite{Mehdipour15}, which has been implemented into \cloudy \citep{Ferland17}, yields a Compton temperature of $T_{\rm C} \approx 10^8\,\rm{K}$.   The Compton radius for NGC~5548 is therefore about $R_{\rm IC} = 0.2\,\rm{pc}$, which is within the inferred distance of the warm absorbers observed in this system 
\citep{Arav15,Ebrero16}
and other similar systems like NGC~7469 \citep[e.g.,][]{Grafton-Waters20}.  It is this correspondence that is suggestive of warm absorbers, which are found in more than half of all nearby AGN \citep{Crenshaw03}, being probes of the multiphase gas dynamics taking place within the thermally driven winds expected at these distances.   

In a recent study applying a thermally driven disk wind model to explain the origin of warm absorbers, \cite{Mizumoto19} find that steady state solutions (discussed here in \S{2}) generally have velocities and column densities in the right range ($300-3000\,{\rm km/s}$ and $10^{20} - 10^{22}\,{\rm cm^{-2}}$) to account for the sample of warm absorbers collected by \cite{Laha14}.  However, the ionization parameters were found to be systematically too large, so they concluded that there must be an additional process such as condensation due to TI that can explain the presence of lower ionization gas at high velocity.  Note that most previous studies considering the role of TI in explaining the warm absorber phenomenon have done so in the context of a thermal wind being driven from a geometrically thick disk (i.e., a torus)  \citep[e.g.,][]{Krolik01,DK17,Kallman19}, not from a geometrically thin disk.

To further understand the interconnectedness between TI and heating by irradiation, in recent papers we took a step back from a disk wind geometry by adopting an even simpler radial wind setup. This allowed us to first perform controlled studies of global 1D solutions that result from using heating and cooling processes obtained from photoionization calculations.  Specifically, in \cite{Dyda17}, we investigated the luminosity and SED-dependence of isotropically irradiated radial winds.  We verified that the critical luminosity identified by \citetalias{Begelman83} does indeed determine whether or not a flow will occupy the S-curve (i.e. reach thermal equilibrium) or evolve to a lower temperature below the S-curve due to adiabatic cooling 
\citep[see also][]{Higginbottom18}.  While this deviation from thermal equilibrium was expected based on the flow timescales of warm absorbers \citep{Kallman10}, it was important to show that it can 
be quantified using time-dependent calculations because many photoionization studies assume that gas only occupies the S-curve
\citep[e.g.,][]{Chakravorty08,Chakravorty09,Chakravorty12,Ponti12,Ebrero16}. In \citeauthor{Dannen20} (\citeyear{Dannen20}, \citetalias{Dannen20} hereafter), we followed up this work with both 1D and 2D radial wind simulations that reveal under what conditions TI can be triggered in an outflow.  We showed that for a small range of the ratio of the local escape speed to the sound speed, corresponding to parsec-scale distances for typical AGN parameters, unsteady clumpy outflow solutions are obtained instead of smooth, steady-state solutions.  

In this paper, we return to disk wind simulations to assess if clumpy disk wind solutions can be obtained 
as simple extensions to the clumpy radial wind solutions of \citetalias{Dannen20}.  
The full answer is complicated but the short answer is `yes' --- we show that irradiated disk atmospheres 
are unstable to linear, isobaric perturbations, the outcome being the formation of what we will refer to 
as irradiated atmospheric fragments (IAFs).  Our simulations initially reach a smooth state that is itself multiphase when
viewed from lines of sight intersecting both the disk atmosphere and disk wind. 
A transition from a smooth to clumpy solution occurs as IAFs form and evolve into larger scale filaments.
The wind becomes multiphase over a large solid angle once these filaments begin to break into clumps and disintegrate upon mixing with the fast outflow, with the final gas distribution spanning a broad range of ionization parameters.

This paper is organized as follows.
In \S{2}, we describe the dynamics of steady state thermally driven wind models after presenting the framework that is now widely used to obtain these solutions.   
In \S{3}, we present our results.  We first connect our study of TI in radial winds to disk winds and derive a characteristic radius beyond which thermally driven outflows are expected to be clumpy.  
Then we present our numerical simulations of clumpy disk winds followed by an analysis of the the absorption measure distribution and several X-ray absorption lines along two illustrative lines of sight.
We focus our discussion in \S{4} on the possibility of TI occurring at smaller radii.
Finally, in \S{5} we summarize our understanding of these new disk wind solutions.

\section{Overview of steady state solutions} 
In this section, we summarize the flow structure of steady state thermal 
wind solutions obtained using a hydrodynamical modeling framework 
in which the effects of radiation are treated in the optically thin limit.    
We first review this framework because the overall takeaway from this work 
is that these solutions serve as initial conditions for multiphase disk 
wind solutions. 
Additionally, these thermal wind models are complementary to the latest 
efforts to model MHD winds together with the internal dynamics of the disk in full 3D, 
using either MRI turbulent disks \citep[]{Zhu18,JLF20} or strongly magnetized disks, 
where a magnetic diffusivity is included to permit accretion \citep[e.g.,][]{SF18}.   
While the thermodynamics used in the above works is highly simplified compared to 
that employed here, it may be feasible to combine both approaches \citep[see e.g.,][]{Waters18}.

\subsection{Modeling framework}
The `classical disk wind solution' approach taken here is to not explicitly treat 
the disk physics.  Rather, a boundary condition (BC) along the midplane is applied, 
providing a reservoir of mass and angular momentum.  
A self-consistent choice of parameters and domain size will prevent the atmosphere 
from becoming Compton thick, so that it becomes possible to model how irradiation 
launches a thermal wind using an optically thin heating and cooling function.
As discussed below, given an SED, these solutions are governed by just three dimensionless parameters.
Formulating the problem in terms of these parameters enables a comparison with our prior results on radial winds, 
as well as with an already large body of published results.

In X-ray irradiated environments, the density BC cannot be chosen arbitrarily; 
it is constrained by the value of the pressure ionization parameter, 
$\Xi \equiv (4\pi J_{\rm ion}/c)/p$, 
where $J_{\rm ion}$ is the mean intensity of the ionizing radiation (defined as energies between $1$ and $10^3\,{\rm Ry}$; see \cite{Krolik81}) local to a given parcel of gas with pressure $p$.  For parsec-scale gas irradiated by an effective point source, $4\pi J_{\rm ion}$ is the ionizing flux $F_X$, making $\Xi$ 
the ratio of the radiation pressure $p_{\rm rad} = F_X/c$ and gas pressure. 
Taking the central source to have a total ionizing luminosity $L_X$, 
we have $F_X = L_X\, e^{-\tau(r)}/4\pi r^2$, 
where $\tau(r) = \int^r_0 n(r') \sigma(r') \,dr'$ is the optical depth along 
the line of sight from the source at a distance $r$.
Hence, without invoking an optically thin approximation, the density is poorly 
constrained due to the highly nonlinear dependence on optical depth.  
AGN warm absorbers are located at parsec scale distances where this approximation 
may apply, thereby reducing the ionization parameter to
$\Xi = L_X/(4\pi r^2 c\,n\,k\,T)$.

The density in these models is therefore determined relative to the midplane BC 
on the ionization parameter, $\Xi_0 = L_X/(4\pi r_0^2 c\,n_0\,k\,T_0)$, where $r_0$ 
is the radius at which $n = n_0$ and $T_0$ is the equilibrium temperature 
on the S-curve corresponding to $\Xi_0$; $n_0$ thus evaluates to
\beq
n_0 = 2.0\times 10^6 \f{(L_X/10^{44}\,{\rm erg\,s^{-1}})}{\Xi_0\,(T_0/10^5\,{\rm K})\,(r_0/{\rm pc})^2}\:{\rm cm^{-3}}.
\label{eq:n0}
\seq 
This normalization density is considered to be typical of the density at heights in the 
irradiated layers of the accretion disk where the plasma first becomes 
Compton thin.  As indicated above, in the classic disk wind solution approach, 
this value is assigned to the midplane of the computational domain at 
$(r,\theta) = (r_0, 90^\circ)$.\footnote{In \cite{Waters18}, we describe 
a modified approach that introduces a shielded adiabatic accretion disk 
into the setup, which in 3D would allow combining these irradiated disk solutions 
with MRI turbulent disks.}
The inner and outer boundaries of the computational domain are set relative to $r_0$, 
and the midplane density profile is given by 
$n(r) = n_0(r_0/r)^2$ to keep $\Xi(r) = \Xi_0$ along the midplane.

The fiducial radius $r_0$ can be expressed in terms of the hydrodynamic escape 
parameter (HEP), defined as
\beq
{\rm HEP} = \f{v_{\rm esc}^2}{c_s^2} = \f{G M_{\rm bh} (1-\Gamma)}{r c_s^2}.
\seq   
Here, $c_s = \sqrt{\gamma kT/\bar{m}}$ is the adiabatic sound speed, 
$v_{\rm esc}$ the effective escape speed from the disk, 
and $\Gamma \equiv L/L_{\rm Edd}$ is the Eddington parameter for a system with total luminosity $L$.
The $\sqrt{1-\Gamma}$ reduction to the escape speed $v_{\rm esc} = \sqrt{GM_{\rm bh} (1-\Gamma)/r}$ is due to radiation pressure 
from the central engine.
Using \eqref{eq:RIC}, we have 
\beq
r_0 = \f{1}{\gamma \rm{HEP_0}}\,\f{T_C}{T_0}\,R_{\rm IC}(1-\Gamma),
\label{eq:r0}
\seq  
where ${\rm HEP_0}$ is the HEP assigned to $T_0$; this is the main parameter 
controlling the strength of the thermal wind.  If we insist on having a strong wind 
beyond $r_0$, we require $r_0 \geq R_{\rm IC}(1-\Gamma)$ [\citet{Proga02}; 
this criterion is just that of \citetalias{Begelman83} for $\Gamma = 0$], giving the upper bound 
\beq
\rm{HEP_0} \lesssim \f{1}{\gamma} \f{T_C}{T_0} \:\:\:\text{(requirement for a strong wind)}.
\label{eq:HEPbound}
\seq

To complete the specification of a thermal wind model, we must determine 
the heating and cooling rates corresponding to the radiation field.  
In recent years, we have developed methods to compute these from the observationally inferred intrinsic SED as self-consistently as is currently possible \citep{Dyda17, Dannen19}
using the photoionization code \xstar~\citep{Bautista01,KB01}.  These calculations require introducing the density ionization parameter,
$\xi = L_X/n_H\, r^2$,
where $n_H = (\mu_H/\mu) n$ is the hydrogen number density 
(with $\mu m_p \equiv \rho/n$ and $\mu_H m_p \equiv \rho/n_H$ for mass density $\rho$).
In this work, we use rates derived from the unobscured SED for NGC~5548 \citep[see][]{Mehdipour15}, 
which has $f_X \equiv L_X/L = 0.36$.  The S-curve corresponding to this SED has a Compton temperature $T_C = 1.01\times10^8\,{\rm K}$.  
Notice that for a fixed SED and elemental abundances (needed to determine the heating and cooling rates in photoionization calculations), the three dimensionless parameters governing thermal disk wind solutions are $\Gamma$, $\Xi_0$, and ${\rm HEP_0}$.  

An important property of thermal wind solutions is that they apply to any mass black hole system if the SED is assumed to be unchanged.\footnote{The implicit dependence of the cooling rate ($C_{23}$ in \eqref{eq:tcool}) on density and temperature can make these solutions sensitive to the S-curve and hence to the SED, but because AGN SEDs are overall similar yet uncertain for any given system, a good first approximation can be found by simply adopting a representative S-curve and applying it to different systems.}
The only term that can potentially break the scale-free nature of the hydrodynamic equations is the non-adiabatic source term. When non-dimensionalized, it enters the equations multiplied by $t_0/t_{\rm cool}(r_0)$, where $t_0$ is a characteristic dynamical time such as the orbital period or $R_g/c$ ($R_g = GM_{\rm bh}/c^2$) and $t_{\rm cool}$ is the cooling time defined by  
\beq
t_{\rm cool} \equiv \f{\mathcal{E}}{\Lambda} = 6.57 \f{T_5}{n_4\,C_{23}}\:{\rm yrs}.
\label{eq:tcool}
\seq
Here $\mathcal{E} = c_{\rm v} T$ is the gas internal energy density, 
$\Lambda =  10^{-23}\, (n/\bar{m})\,C_{23}$ is the total cooling rate 
in units of ${\rm erg\,s^{-1}g^{-1}}$ with $C_{23}$ the rate 
in units of $10^{-23}\, {\rm erg\,cm^{3}\,s^{-1}}$, and $n_4 = n/10^4\,{\rm cm^{-3}}$. 
The minimum value of $t_{\rm cool}$ is reached along the midplane at $r_0$ when this is taken as the location of the inner boundary.
Using \eqref{eq:n0} and \eqref{eq:r0}, we find
\beq
t_{\rm cool}(r_0) = 2.22 \, \f{\mu\,M_7 (1-\Gamma)^2}{f_X\,\Gamma ({\rm HEP_0}/100)^2} \f{\Xi_0}{\,C_{23}} \, \:{\rm hrs}, %\left(\f{{\rm HEP_0}}{100}\right)^{-2}
\label{eq:tcool_r0}
\seq
where we have eliminated $L_X$ in favor of $f_X$ using $L_X = f_X \Gamma L_{\rm Edd}$. 
Because both $t_0$ and $t_{\rm cool}(r_0)$ are proportional to $M_{\rm bh}$, the equations have no mass dependence.  This is ultimately due the density scaling as $n_0 \propto M_{\rm bh}^{-1}$ in the above ionization parameter framework.  

%%%%%%%%%%%%%%%%%%%%%%%%%%%%%%%%%%%%%%%%%%%%%%%%%%%%%%%%%%%%%%%%
\begin{figure}[htb!]
 \centering
 \includegraphics[width=\columnwidth]{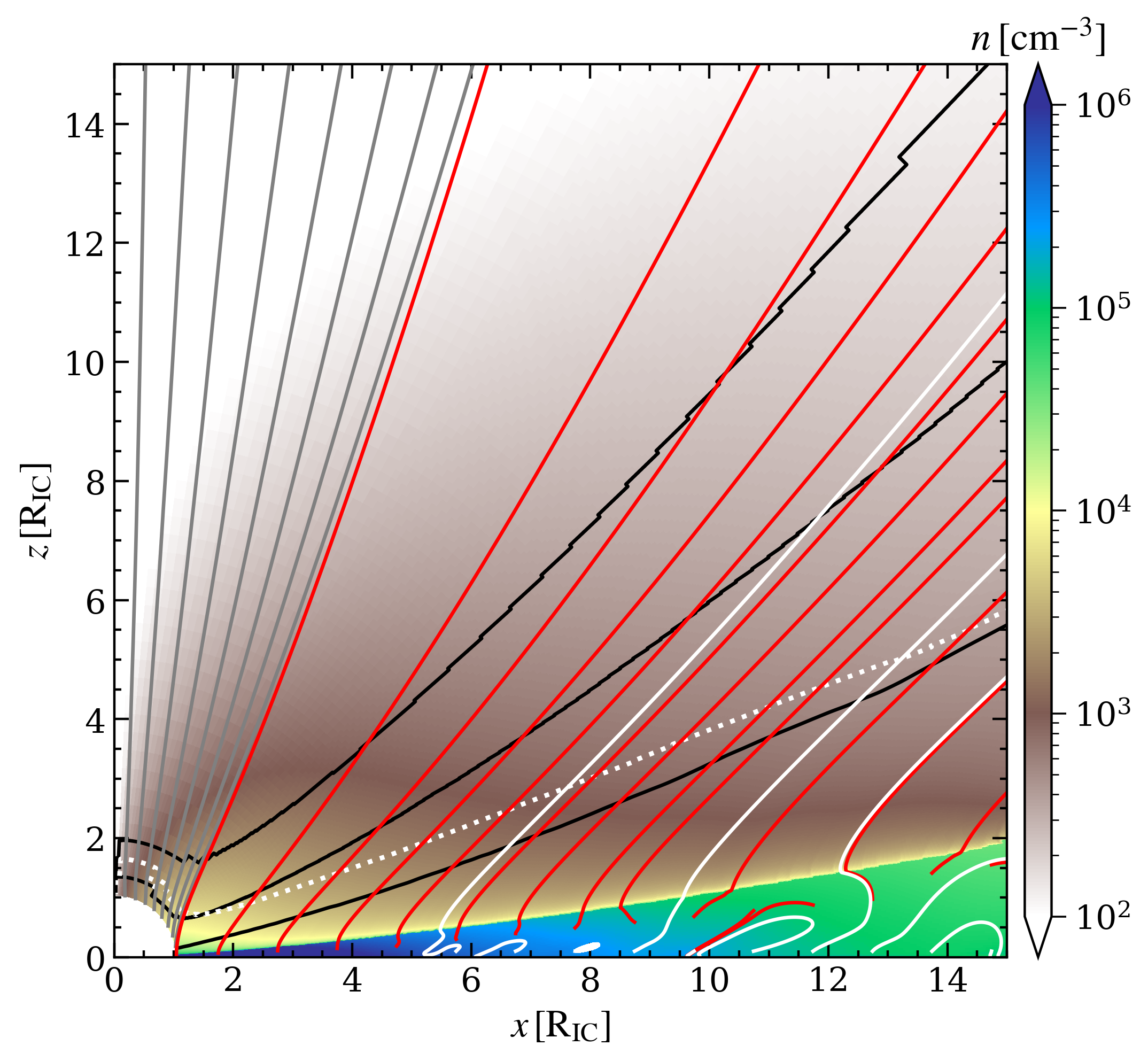}
 \caption{An example of the steady state density and velocity structure of thermally driven disk wind solutions near the Compton radius $R_{\rm IC}$.  The solution consists of a tenuous outflow (brownish gas) above a somewhat denser corona (yellowish gas) that lies above a dense and nearly hydrostatic cold phase atmosphere (blue and green gas). 
A set of streamlines are overlaid and color-coded according to where they originate: 
along the inner spherical boundary (gray), 
parallel to atmosphere/wind interface (red), 
and parallel to the midplane 
at $z=0.1\,R_{\rm IC}$ (white).  
As the red streamlines show, the base of the disk wind is in the upper layers of the atmosphere where poloidal velocities are less than $10\,{\rm km/s}$.  
The lowest black contour denotes $500\,{\rm km/s}$ and the next two 
$1000\,{\rm km/s}$ and $1500\,{\rm km/s}$.
The white dotted line marks the sonic surface.
The circulation apparent in the cold phase atmosphere increases with radius and permits the formation of a vortex for $r > 15\, R_{\rm IC}$ (see Fig.~\ref{fig:vortex}), which in turn triggers a multiphase outflow at even larger radii.  
}
\label{fig:density_map1}
\end{figure}
%%%%%%%Fig.~\ref{fig:density_map1}

%%%%%%%%%%%%%%%%%%%%%%%%%%%%%%%%%%%%%%%%%%%%%%%%%%%%%%%%%%%%%%%%
\begin{figure}[htb!]
 \centering
 \includegraphics[width=0.99\columnwidth]{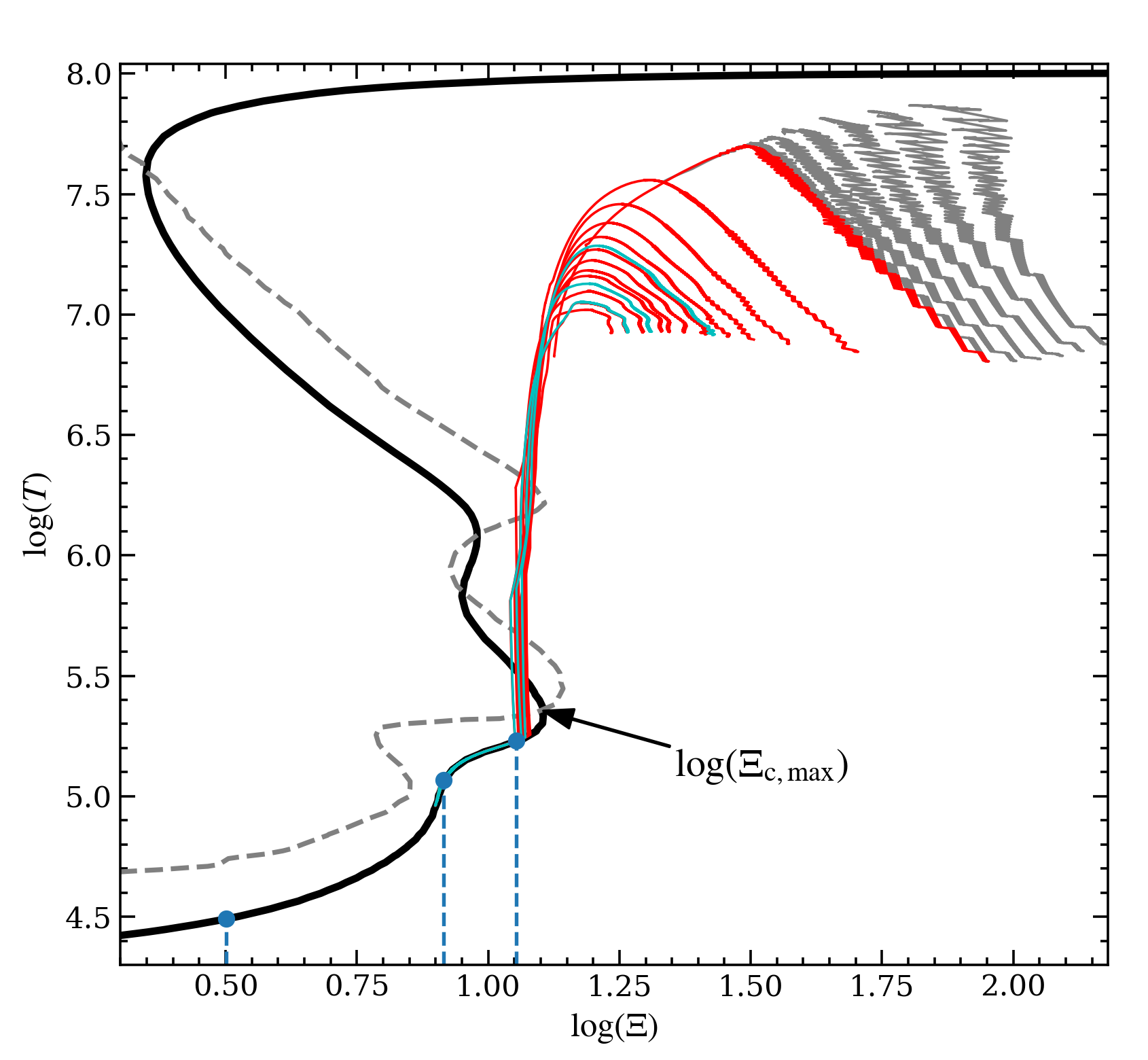}
 \caption{Phase diagram showing the S-curve corresponding to the SED of NGC~5548 
 (solid black curve).  Blue dots and vertical lines mark $\Xi_0$ (the midplane BC on $\Xi$) 
 for each of the solutions in Fig.~\ref{fig:Xi0runs}, 
 $\log(\Xi_0) = 0.5$, $\log(\Xi_0) = 0.92$, and $\log(\Xi_0) = 1.05$. 
 The Balbus contour is plotted as a dashed gray curve; all points to the left of it 
 are thermally unstable and to the right thermally stable according to Balbus' criterion 
 for TI \citep{Balbus86}. The two unstable regions encountered while tracing the S-curve 
 starting from low temperatures are referred to as the upper and lower TI zones.  
 The base of the lower TI zone is conventionally denoted as $\Xi_{\rm c,max}$ and is marked.
Overplotted on this $(T,\Xi)$-plane are the phase-`tracks' for each 
of the streamlines in Fig.~\ref{fig:density_map1}. (The white streamlines are plotted in cyan here.)  
Notice that the disk wind streamlines pass through both TI zones.
}
\label{fig:Scurve}
\end{figure}
%%%%%%%Fig.~\ref{fig:Scurve}

\subsection{Basic flow structure of irradiated disks}
\label{sec:background}
As shown in Fig.~\ref{fig:density_map1}, the density structure that results from this modeling framework consists of three distinct regions: a dense disk atmosphere (blue to green regions), a less dense corona (yellow regions), and a tenuous disk wind (brown to whitish regions).  That the solution consists of these basic components was already apparent from the early study by 
\citeauthor{Woods96} (\citeyear{Woods96}; see also \citealt{Proga02,J-G02}), but the focus subsequently shifted to simulating only the corona and disk wind regions to determine if the velocities of these models can reach those observed in low-mass X-ray binaries 
\citep[e.g.,][]{Luketic10,Higginbottom15,Higginbottom17}.  As discussed in \S{2.3}, this is accomplished by placing $\Xi_0$ just below $\Xi_{\rm c,max}$ on the S-curve.  In Fig.~\ref{fig:Scurve}, we mark the location of $\Xi_{\rm c,max}$ on the S-curve used in this work.  We also show three values of $\Xi_0$; the midplane BC of the solution shown in Fig.~\ref{fig:density_map1} corresponds to the middle blue line at $\log(\Xi_0) = 0.92$ ($\xi_0 = 50$).   

The velocity structure can also be grouped into three sets of streamlines (although these groupings do not correspond to that of the density field): 
those originating along the inner boundary of the computational domain, those with footpoints nearly parallel to the atmosphere/wind interface, and those along the midplane.    
We show this streamline geometry in the Fig.~\ref{fig:density_map1}, 
and we plot the temperature along these streamlines on the $(T,\Xi)$-plane in Fig.~\ref{fig:Scurve}.  
As the gas becomes less bound, the velocity field transitions from streamlines with appreciable poloidal divergence for $1\,R_{\rm IC} \lesssim r \lesssim 5\,R_{\rm IC}$, to nearly parallel streamlines for $r > 5\,R_{\rm IC}$.  Also, contours of constant velocity (black lines denote $500\,{\rm km/s}$, $1000\,{\rm km/s}$, and $1500\,{\rm km/s}$) become more widely spaced. The disk atmosphere is characterized by streamlines that turn back into the midplane or by segments of streamlines near the footpoints that curve before transitioning into the wind, indicative of bound gas with some amount of circulation.
The caption of Fig.~\ref{fig:density_map1} provides further details on the velocity dynamics.

The highest temperatures are reached near the inner boundary, 
corresponding to the gray streamlines closest to the Compton branch of the S-curve in Fig.~\ref{fig:Scurve}. 
This hot gas is responsible for launching the fastest outflow at high latitudes.  
We will refer to this as the radial wind region because the streamlines 
do not originate from the disk.   
We note that the midplane BC is the source of matter for gas along the inner boundary, as the pressure gradients above the atmosphere establish the density profile along $r_{\rm in}$, but this gas then becomes the base of a disconnected wind.
Temperature decreases with distance along these wind streamlines because of 
the near spherical expansion, which permits the timescale for adiabatic cooling 
to be shorter than the timescale for radiative heating.  The radiative heating 
rate scales similarly to $t_{\rm cool}^{-1}$.
Outside the atmosphere, $t_{\rm cool}$ is shortest in the disk corona, 
explaining why gas becomes nearly Comptonized only in this region.  
Radiative heating timescales are shortest here due to the nearly $1/r^2$ 
scaling of density in the radial wind region.  

The streamlines originating near the atmosphere/wind interface show similar behavior 
on the $(T,\Xi)$-plane to the radial wind region streamlines only 
at large distances.  
The footpoints of these streamlines are located on the S-curve slightly below $\Xi_{\rm c,max}$,
and the temperature sharply rises with distance upon entering the region of runaway heating.
All streamlines pass through both the lower and upper TI zones 
(the regions falling to the left of the Balbus contour, the dashed line
in Fig.~\ref{fig:Scurve}). 
The temperature peaks occuring between $\sim10^7\,{\rm K}$ 
and $5\times 10^7\,{\rm K}$ are at points beyond the sonic surface (dotted white line in Fig.~\ref{fig:Scurve}), 
and the decreasing temperature thereafter is again due to adiabatic cooling, 
which is less efficient for non-spherical expansion.  
There is a shear layer associated with the steep rise in temperature at the atmosphere/wind interface,
but we checked that the velocity gradient in this layer is too small to make it unstable to the
Kelvin-Helmholtz instability (KHI).  
Specifically, we introduced constant pressure perturbations with a range of wavenumbers throughout
the atmosphere to test stability to both TI and KHI.  Stability to TI is discussed further in \S{\ref{sec:TIstability}} and \S{\ref{sec:Be}}, but it is likely that even a higher velocity gradient shear layer will be stable to KHI because sound waves are highly damped in the presence of radiative heating/cooling.

%%%%%%%%%%%%%%%%%%%%%%%%%%%%%%%%%%%%%%%%%%%%%%%%%%%%%%%%%%%%%%%%
\begin{figure*}[htb!]
 \centering
 \includegraphics[width=2.\columnwidth]{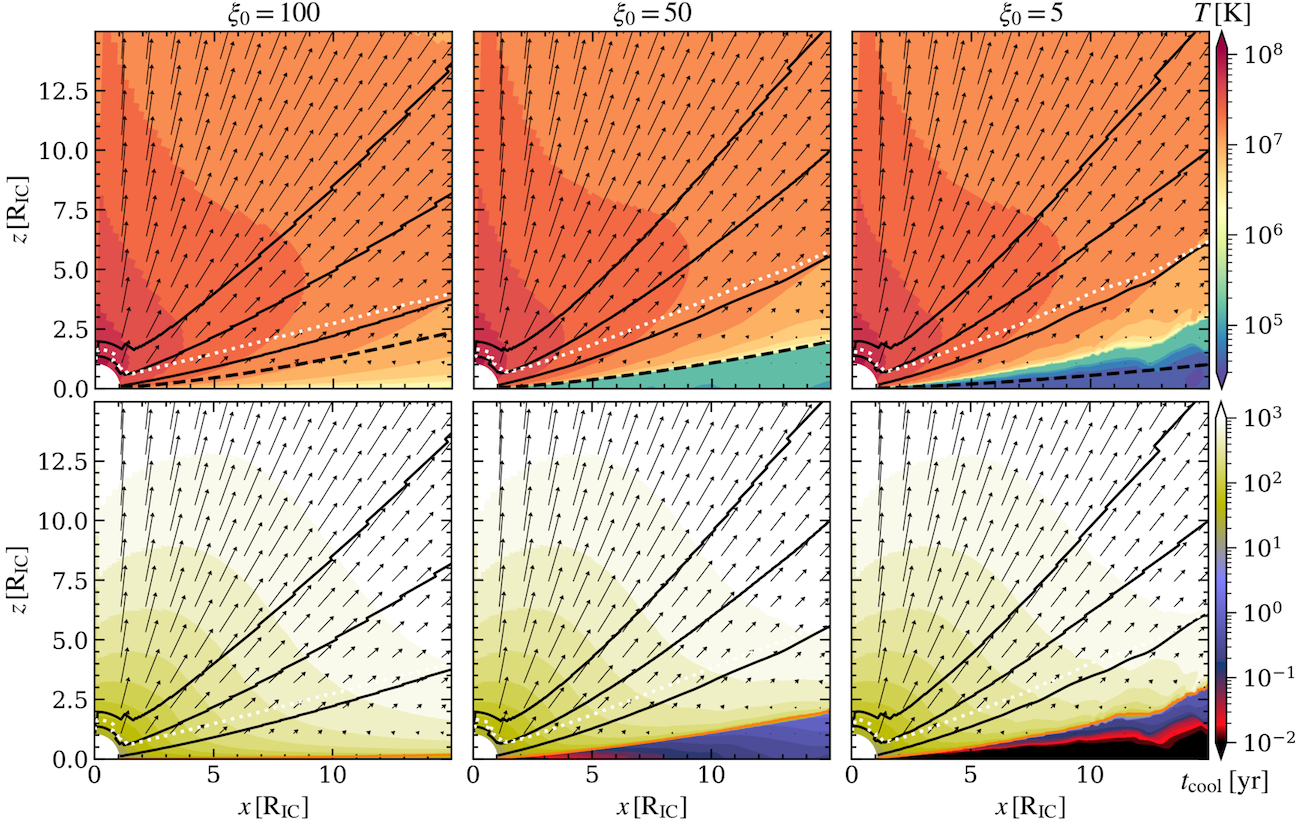}
 \caption{
{\it Top panels}: Temperature maps showing how the height of the disk atmosphere 
varies  with the value of the ionization parameter assigned to the midplane 
of the compuational domain ($\xi_0 = 100.0$, 50.0, and 5.0 correspond 
to $\Xi_0 = 11.324$, 8.235, and 3.171, respectively).  
It has become standard practice to exclude an atmosphere entirely 
by placing $\Xi_0$ right below $\Xi_{\rm c,max}$, which results in a solution 
like that shown in the left panel.  However, the presence of a cold phase atmosphere 
is necessary to capture TI in the simulations.  In this work, we present results 
for $\xi_0 = 50$ solutions.  Even smaller values of $\xi_0$ results in a lower temperature 
atmosphere that extends to greater heights above the midplane 
(compare middle and right panels). This $T$-dependence of the height of the disk atmosphere 
is opposite that of an underlying disk, as seen by comparing the classical disk scale height 
$h = \epsilon r$, marked with the black dashed lines, with the height of the atmosphere.  
The temperature and velocity structure of the disk wind, meanwhile, are comparatively 
insensitive to $\Xi_0$.  To show this clearly, we again plot the sonic surface as well as 
the velocity contours from Fig.~\ref{fig:density_map1}.  
{\it Bottom panels}: 
Maps of the cooling time (in years), calculated for NGC~5548 
($M_{\rm bh} = 5.2\times 10^7M_\odot$).  The velocity structure is again overlaid 
and the orange contour marks where $t_{\rm cool}/t_{\rm ff} = 0.01$.
}
 \label{fig:Xi0runs}
\end{figure*}
%%%%%%%Fig.~\ref{fig:Xi0runs}

\subsection{Scale height dependence of irradiated disk atmospheres on $\Xi_0$}
Recent work applying these solutions to explain the observed changes in wind velocities accompanying state changes in low-mass X-ray binaries has pointed out how the extent of the atmosphere is sensitive to $\Xi_0$ (see the appendix of \citealt{Tomaru19}).  This dependence is illustrated in Fig.~\ref{fig:Xi0runs}, where we compare temperature maps for solutions 
with $\xi_0 = 100$, $\xi_0 = 50$, and $\xi_0 = 5$ ($\Xi_0 = 11.32$, $\Xi_0 = 8.24$, and $\Xi_0 = 3.17$), corresponding to different cold phase locations along the S-curve 
(marked with vertical blue lines in Fig.~\ref{fig:Scurve}). 
Most previous studies have focused on solutions similar to those shown in the left panel, where there is no disk atmosphere at all
\citep[e.g.,][]{Luketic10,Higginbottom15,Higginbottom17,Higginbottom18,Tomaru18,Waters18}.  This corresponds to taking $\Xi_0$ somewhat below $\Xi_{\rm c,max}$.
Including the cold phase atmosphere incurs much greater computational expense because the cooling time becomes orders of magnitude smaller, as shown in the bottom panels.  
Once the atmosphere is present, taking a lower $\Xi_0$ results 
in it extending to larger heights above the disk midplane before the transition into an outflowing corona and disk wind occurs at $T > T(\Xi_{\rm c,max})$.     

A comparison of these solutions reveals that the vertical heights 
of accretion disks and their irradiated atmospheres scale oppositely with temperature. 
The black dashed lines on the temperature maps in Fig.~\ref{fig:Xi0runs} mark, 
for the same midplane temperature, the classical scale height of a disk, 
$h = \epsilon\, r$, where 
$\epsilon = v_{\rm esc}/c_{\rm iso} = 1/\sqrt{\gamma \rm{HEP}}$ 
(with $c_{\rm iso} = \sqrt{k T/\bar{m}}$ the isothermal sound speed).
A cooler disk implies a thinner, less pressurized disk, but a cooler irradiated 
atmosphere requires it to extend to higher latitudes and be of higher pressure.  
The latter property follows immediately from the definition of the pressure ionization
parameter, $\Xi \equiv p_{\rm rad}/p$.  For a fixed radiation flux, 
i.e. at a given radius in the atmosphere, gas in thermal equilibrium at a lower $\Xi$ has a higher gas pressure.  The greater extent of the cold phase 
follows because the atmosphere is nearly in hydrostatic equilibrium, 
so a fixed pressure gradient is required to balance the gravitational 
and centrifugal forces.  When starting from a higher pressure, therefore, 
cold phase gas must extend to larger heights to `climb' the S-curve 
at this particular pressure gradient. 

It is the runaway heating process that occurs when gas reaches $\Xi_{\rm c,max}$ 
on the S-curve that is responsible for the atmosphere transitioning into 
a disk wind (\citetalias{Begelman83}). Thus, while the height of the atmosphere 
is sensitive to the value of $\Xi_0$, the temperature at the base of the disk wind 
is the same in each of the panels of Fig.~\ref{fig:Xi0runs}.  
It is therefore expected that the run of temperature with distance in the wind is not sensitive to $\Xi_0$.  Because the velocity structure of the disk wind 
is set by the temperature at $\Xi_{\rm c,max}$, it also is nearly the same 
from panel to panel.  To see this clearly, we overplot the sonic surface 
(white contour) and three constant velocity surfaces.

\subsection{Stability of the disk wind vs. atmosphere to TI}
\label{sec:TIstability}
To emphasize how TI can operate differently in the cold phase atmosphere versus in the disk wind,
it is useful to draw a comparison with studies 
of the circumgalactic medium (CGM), where TI is invoked 
as a condensation process only.  Focus has been given to the importance 
of the ratio $t_{\rm cool}/t_{\rm ff}$ 
(with $t_{\rm ff} = r\sqrt{2}/\sqrt{GM_{\rm bh}/r}$ the free-fall timescale) 
in assessing the stability of CGM atmospheres to TI
\citep[e.g.,][]{Sharma12,Gasparietal15,Voit17}.  We computed this ratio 
for the solutions here, because the relevant dynamical time is the orbital period, 
which is just $t_{\rm dyn} = \pi\sqrt{2}\,t_{\rm ff}$.  We find that $t_{\rm cool}/t_{\rm ff} < 1$ 
in the entire computational domain, falling to less than $0.01$  below the orange contour 
plotted in the bottom panels of Fig.~\ref{fig:Xi0runs}.  This ratio is 
a qualitative diagnostic of whether cooling times are short enough to permit condensation 
out of the hot phase, which they are.  
The fundamental criterion for triggering TI, however, is that gas 
occupy the TI zone long enough for perturbations to grow.
As explained in \S{3}, dynamical effects prohibit this in the disk wind.  

Gas in the disk atmosphere already occupies the cold phase, so condensation 
clearly cannot take place.  It is still relevant that 
$t_{\rm cool}/t_{\rm ff} \ll 1$ in this gas, however, because it implies that 
$t_{\rm heat}/t_{\rm ff} \ll 1$.  That is, the only way for TI to operate 
in the atmosphere (at least initially) is for perturbations to undergo runaway heating 
to form heated layers or pockets of warmer gas.  We identified this as the dominant
process triggering clumpiness in the radial wind solutions presented 
in \citetalias{Dannen20}.  
Because the atmosphere is cooler than the temperature of the lower TI zone, 
perturbations can only grow near the disk wind interface where $T$ reaches 
$T(\Xi_{\rm c,max})$.  

\section{Results}
The solutions shown in Fig.~\ref{fig:Xi0runs} only reach a steady state
if the outer boundary is located within about $15\,R_{\rm IC}$.
We now proceed to show that with larger domain runs, 
the innermost flow region can still reach a steady state
but the outer regions will not.
This is because the inner flow region is just within 
the radius where TI can first be triggered due to dynamical changes in the atmosphere.  
We note that $15\,R_{\rm IC}$ corresponds to about $1\,\rm{pc}$ for $M_{\rm bh} \approx 10^7\,M_\odot$ (see \eqref{eq:RIC}) and is close to the inner radius 
of our clumpy disk wind simulations presented in \S{\ref{sec:results}}. 
This solution regime therefore corresponds to scales associated with the
torus in the AGN literature \citep[e.g.,][]{RAR17,Combes19,Honig19}.
We choose not to speculate here on this correspondence other than to
point out the following possibility.
The inverse dependence between the atmospheric height and the value of $\Xi_0$ pointed
out in \S{\ref{sec:background}}, which is an overlooked property of thermal wind solutions,
can in principle accommodate a smooth transition from ionized to atomic and molecular gas due to the small ram pressure of the atmospheric gas residing on the cold phase branch of our S-curve.

We emphasize again that these solutions do not depend on the black hole mass, so for AGNs with $M_{\rm bh} \approx 10^6\,M_\odot$, these solutions are within the inferred distance to the torus and the dynamics can therefore be considered independently of outflow models developed for the dusty torus region \citep[e.g.,][]{DKB12,Wada2012,CK16,CK17,WHV19}.
We briefly comment on the effects of adding additional physical processes to these solutions in \S{\ref{sec:Limitations}}.

\subsection{From clumpy radial winds to clumpy disk winds}
Here we examine the dynamics of steady-state disk wind solutions beyond $15\, R_{\rm IC}$ after connecting them
to our previous work on radial winds, where we discussed how the flow 
can be stable to TI despite passing through a TI zone.
It is beyond the scope of this work to develop a comprehensive theory of `dynamical TI'.  
It is clear, however, that the list of qualitative requirements identified 
by \citetalias{Dannen20} for triggering TI in radial winds holds also for disk winds.  
Therefore, we first recall our list from \citetalias{Dannen20}.

\subsubsection{Necessary conditions for triggering TI in outflows}
\label{sec:3bullets}
\begin{itemize}
\setlength\itemsep{-0.25em}
\item For radiation pressure due to a central source, the gas pressure along a streamline 
must decrease as or slower than $r^{-2}$ upon entering a TI zone to allow 
$\Xi \propto 1/(r^2 p)$ to remain constant or decrease.  Otherwise gas will quickly exit 
the TI zone, as is clear from Fig.~\ref{fig:Scurve}.  In more complicated radiation fields, 
the gas pressure gradient along streamlines will still likely need to permit $\Xi = 4\pi J/p$ 
to vary similarly with distance in the TI zone.  
\item The flow must not be so fast that perturbations do not have time to become nonlinear 
during their passage through a TI zone.  
\item The stretching of perturbations due to acceleration terms (from the nonlinear term $\gv{v}\cdot\nabla\gv{v}$) must be less efficient than the amplification of perturbations 
from TI.  
\end{itemize}
Bullets (i) and (ii) relate to pressure gradients in the flow field and (iii) 
to velocity gradients, none of which are accounted for in the local theory of TI.\footnote{While 
the local dispersion relation for TI is unchanged in a uniform velocity field 
(due to Galiliean invariance), bullet (ii) arises because TI zones may span only 
a small range of pressures on the S-curve, and the flow pressure typically 
drops off rapidly.}
(For prior applications of dynamical TI in global flows see \citealt{Balbus89} and \citealt{MP13};
see also the 3D simulations of \citealt{KP09}.)  

\subsubsection{Connection between TI and the Bernoulli function}
\label{sec:Be}
\begin{figure}[htb!]
 \centering
 \includegraphics[width=0.95\columnwidth]{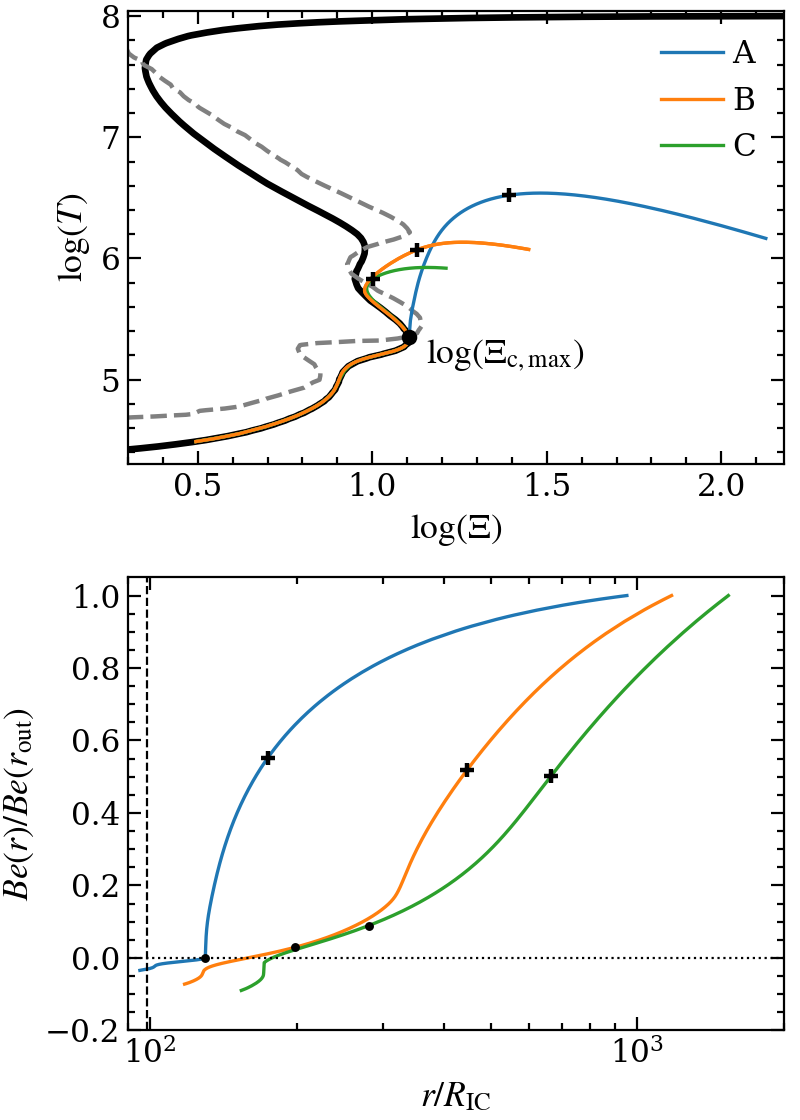}
 \caption{
{\it Top panel:} 
 Steady state versions of our 1D radial wind solutions
 from \citetalias{Dannen20} plotted on the $(T,\Xi)$-plane.  
 Plus signs denote sonic points and a black dot marks $\Xi_{\rm c,max}$.
 Model A is stable to TI and exits the TI zone, whereas models B and C 
 are unstable and settle onto the S-curve inside the TI zone.  Only models B and C will turn 
 into clumpy wind solutions if they are perturbed (by applying linear perturbations 
 at the base, for example). 
{\it Bottom panel:} The Bernoulli function (normalized to its value at $r_{\rm out}$) 
versus radius.  Plus signs mark sonic points. Black dots mark the locations where 
$\Xi = \Xi_{\rm c,max}$, the entry to the lower TI zone.  As discussed in \S{\ref{sec:Be}}, 
models B and C are unstable solutions because they enter the TI zone with $Be > 0$.  
From the dynamical criterion that a sign change in $Be$ at the location where 
$\Xi = \Xi_{\rm c,max}$ implies stability, we identified the characteristic 
`unbound' radius $R_{\rm u}$ beyond which a transition to instability should occur 
(see \S{\ref{sec:ru}}).  Like $R_{\rm IC}$, $R_{\rm u}$ is a property of the S-curve; 
the vertical line marks $R_{\rm u} = 98.3\,R_{\rm IC}$.
}
\label{fig:Be_analysis}
\end{figure}

In \citetalias{Dannen20}, we presented four qualitatively different types of flow solutions 
(Models A, B, C, \& D) in 1D. The model A solution had the highest $\rm{HEP_0}$ 
and is susceptible to TI only during a transient phase of evolution, 
while models B, C, \& D are clumpy wind solutions that never reached a steady state. 
After improving our numerical methods (see Appendix~A), all of these solutions reach 
a steady state, which allowed us to perform a followup analysis based on the Bernoulli theorem.
This analysis is presented in Fig.~\ref{fig:Be_analysis} and amounts to 
the following empirical result: 
\begin{itemize}
\item The above conditions for an outflow to be unstable to TI cannot be met 
if the flow enters a TI zone with a negative value of the Bernoulli function,
\beq 
Be = \f{v^2}{2} + \f{c_s^2}{\gamma - 1} - \f{GM_{\rm bh}}{r}(1-\Gamma).
\label{eq:Be}
\seq
\end{itemize}

%%%%%%%%%%%%%%%%%%%%%%%%%%%%%%%%%%%%%%%%%%%%%%%%%%%%%%%%%%%%%%%%
%%%%%%%Fig.~\ref{fig:Scurve}

To understand this result, we consider what must happen to trigger TI starting 
from a steady state flow field.  First recall that isobaric TI only amplifies 
the entropy mode [unless the S-curve has a negative slope on the $(\xi,T)$-plane; 
see \citealt{Waters19a}], and that entropy modes, unlike sound waves, 
are advected with the flow.  Therefore, only those streamlines that already pass through 
a TI zone can amplify perturbations.  Next recall the close connection between 
Bernoulli's theorem and the steady state entropy equation.  
For a net radiative cooling function $\mathcal{L}$ having units ${\rm erg~g^{-1}s^{-1}}$,
the heat equation can be expressed as $T\,Ds/Dt = -\mathcal{L}$ (for entropy per unit mass $s$ having units ${\rm erg~g^{-1}K^{-1}}$ and with $D/Dt$ the Lagrangian derivative).  In a steady state,
\beq
\gv{v}\cdot \nabla s = -\f{\mathcal{L}}{T}.
\seq
Meanwhile, Bernoulli's theorem is
\beq
\gv{v}\cdot \nabla Be = -\mathcal{L}.
\seq
By eliminating $\mathcal{L}$, we arrive at Crocco's theorem 
($\nabla Be = T\nabla s$ along streamlines) for steady state solutions, 
but it is clear from the derivation that the advected value of the Bernoulli function 
closely tracks that of entropy.  An abrupt jump in the value of $Be$ therefore implies 
a similarly large change in the steady state entropy profile. 

The key insight revealed by this Bernoulli analysis is the effect this profile 
has on an advected entropy mode that can become unstable to TI.  Only for an entropy profile 
that does not suffer a dramatic jump at the entry to a TI zone can the 3 bullet points from
\S{\ref{sec:3bullets}} be satisfied.
That is, TI can amplify an entropy mode into the nonlinear regime and thus seed 
the formation of an IAF only if the mode is not disrupted, which requires the entropy 
and $Be$ profiles to be smooth at the TI zone. % , and this inhibits the mode's growth.  

\subsubsection{Characteristic radius at which thermally driven outflows become clumpy}
\label{sec:ru}
The sign of $Be$ is negative in the disk atmosphere because this cold phase gas is virialized.  
It becomes positive in the disk wind because runaway heating makes the enthalpy large 
and accelerates the flow, i.e., the first two terms in \eqref{eq:Be} dominate the third.  
For an entropy mode to avoid an abrupt change as it enters a TI zone, $Be$ must not change signs, 
hence our previous bullet point.  The radius at which $Be$ first becomes positive just as $T$ reaches 
$T_{\rm c,max} \equiv T(\Xi_{\rm c,max})$ from below defines the characteristic `unbound radius'
beyond which thermally driven winds are expected to be clumpy. The exact radius is only implicitly 
defined by the Bernoulli function, but as shown in Appendix~B, a lower bound is
\beq
R_{\rm u} = \f{2}{\gamma} \f{\gamma-1}{\gamma + 1} \f{T_C}{T_{\rm c,max}} \,R_{\rm IC}\,(1-\Gamma).
\label{eq:Ru}
\seq
Eliminating $R_{\rm IC}$ using \eqref{eq:RIC} shows $R_{\rm u}$ to be proportional to `the other' 
characteristic distance set by the radiation field for a given black hole mass,  
$G M_{\rm bh} \bar{m}/k T_{\rm c,max}$.
Like $R_{\rm IC}$, therefore, it is a property of the S-curve 
for any particular system; the ratio $T_C/T_{\rm c,max} = 4.68\times10^2$ for our S-curve, 
giving $R_{\rm u} = 140\, R_{\rm IC}\,(1-\Gamma)$ for $\gamma = 5/3$.  The vertical line 
in Fig.~\ref{fig:Be_analysis} marks this radius for our radial wind solutions 
(which have $\Gamma = 0.3$), showing that it is indeed a good estimate of the parameter space that 
we uncovered by trial and error in \citetalias{Dannen20}.  
Because $R_{\rm u}$ is simply the approximate radius beyond which gas entering the TI zone 
has enough thermal energy to escape the system, 
it characterizes both radial wind and disk wind solutions.

%%%%%%%%%%%%%%%%%%%%%%%%%%%%%%%%%%%%%%%%%%%%%%%%%%%%%%%%%%%%%%%%
\begin{figure}[htb!]
 \centering
 \includegraphics[width=\columnwidth]{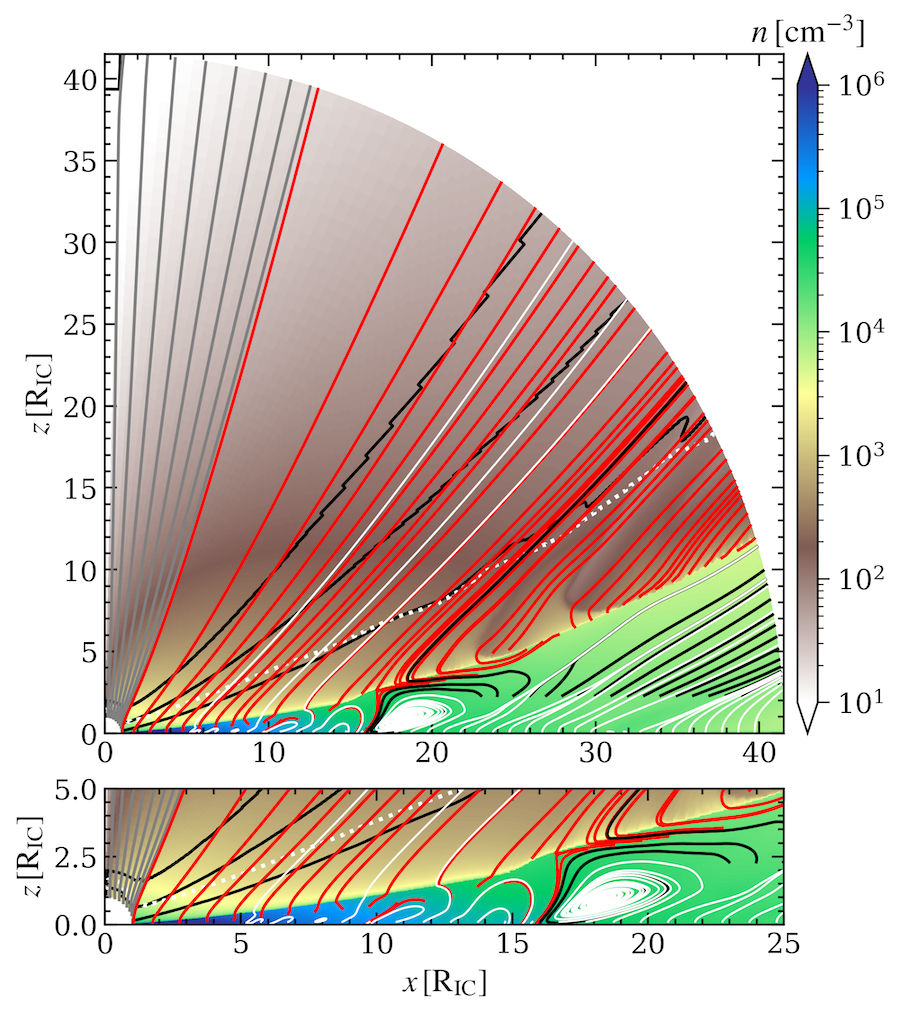}
 \caption{The same $\xi_0 = 50$ solution from Figs.~\ref{fig:density_map1} and \ref{fig:Xi0runs} 
 but showing the full domain, which extends to $40\,R_{\rm IC}$. In addition to the streamlines overlaid 
 in Figs.~\ref{fig:density_map1}, we add a set parallel to the midplane with footpoints 
 $z=2.3\,R_{\rm IC}$ spaced $1\,R_{\rm IC}$ apart along $x$ (black).  Beyond $25\,{\rm R_{IC}}$, 
 these streamlines are all outflowing and do not enter the disk wind, showing that the disk wind becomes 
 dynamically disconnected from the interior of the outflowing atmosphere.  
 Within $25\,{\rm R_{IC}}$, some black streamlines flow 
 into a vortex centered at $r\approx 18.5\,R_{\rm IC}$.  This vortex spans about $9\,R_{\rm IC}$, 
 corresponding to $\approx 1.5\,{\rm pc}$ for NGC~5548 ($M_{\rm bh} = 5.2\times 10^7\,M_\odot$) and $0.03\,{\rm pc}$ for an AGN with $M_{\rm bh} = 10^6\,M_\odot$. 
 In the lower panel, we zoom-in somewhat to better show the vortex, 
 which turns out to be critical for the onset of TI.
}
\label{fig:vortex}
\end{figure}
%%%%%%%Fig.~\ref{fig:vortex}

\subsubsection{The role of a cold phase vortex}
Well within the distance $R_{\rm u}$, the gas dynamics in the cold phase atmosphere changes significantly, 
as we show in Fig.~\ref{fig:vortex}. For $r \gtrsim 25\,R_{\rm IC}$, the atmosphere transitions into a slow outflow.  
Prior to that, a large scale scale vortex appears in the flow at $16\,R_{\rm IC} \lesssim r \lesssim 25\,R_{\rm IC}$, 
and this dynamics is associated with a steepening of the slope of the atmosphere/wind interface
at $15\,R_{\rm IC} \lesssim r \lesssim 16\,R_{\rm IC}$. 
We are not aware of any previous authors reporting on the presence of such a counter-clockwise vortex 
in the atmosphere.  This circulation is driven by pressure gradients accompanying the steepest bend 
in the S-curve at about $\log(T) = 5.1$.  

The vortex turns out to be the critical element that connects the geometry of radial winds 
to that of disk winds and allows TI to operate.  In radial winds, the streamlines in the cold phase atmosphere 
are necessarily normal to the atmosphere's surface, whereas in disk winds they are nearly parallel to it 
beyond $25\,R_{\rm IC}$. Within $15\,R_{\rm IC}$, atmosphere streamlines can connect to the disk wind, 
as Fig.~\ref{fig:density_map1} shows.  However, as discussed above, these streamlines are stable to TI.  
Beyond $25\,R_{\rm IC}$, the base of the disk wind is disconnected from the atmosphere's interior.  The temperature at the base is 
that of the lower TI zone, $T_{\rm c,max}$, but pressure falls off with distance, meaning $\Xi$ increases outward 
and gas does not enter the TI zone.  Thus, starting from a steady state at least, the only path for matter 
to enter the wind from the atmosphere at radii $r > 15\,R_{\rm IC}$ is through the vortex.  
We sum this up with a final bullet point:
\begin{itemize}
\item In the case of thermally driven disk winds, pressure gradients along the S-curve must drive a vortex 
in the atmosphere, as this allows perturbations to enter the TI zone and become nonlinear.
\end{itemize}

%%%%%%%%%%%%%%%%%%%%%%%%%%%%%%%%%%%%%%%%%%%%%%%%%%%%%%%%%%%%%%%%
\begin{figure*}[htb!]
 \centering
 \includegraphics[width=2.0\columnwidth]{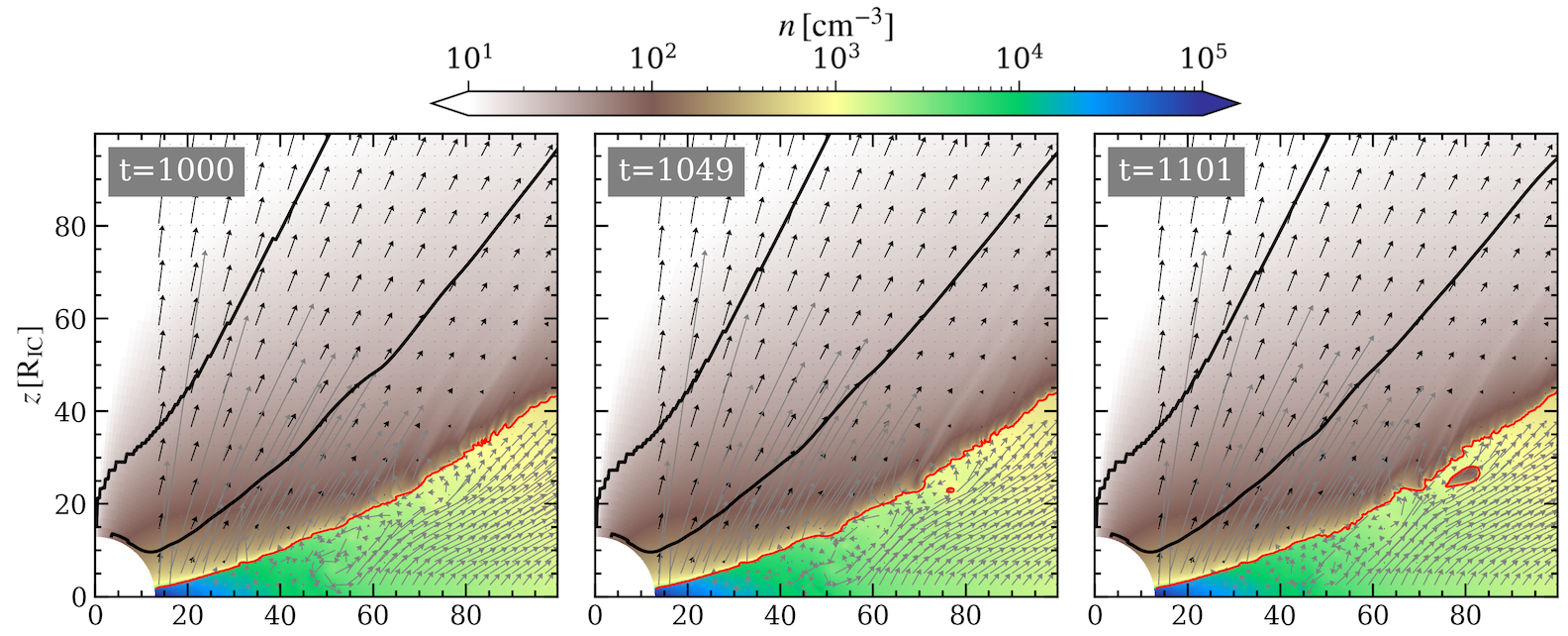}
 \includegraphics[width=2.0\columnwidth]{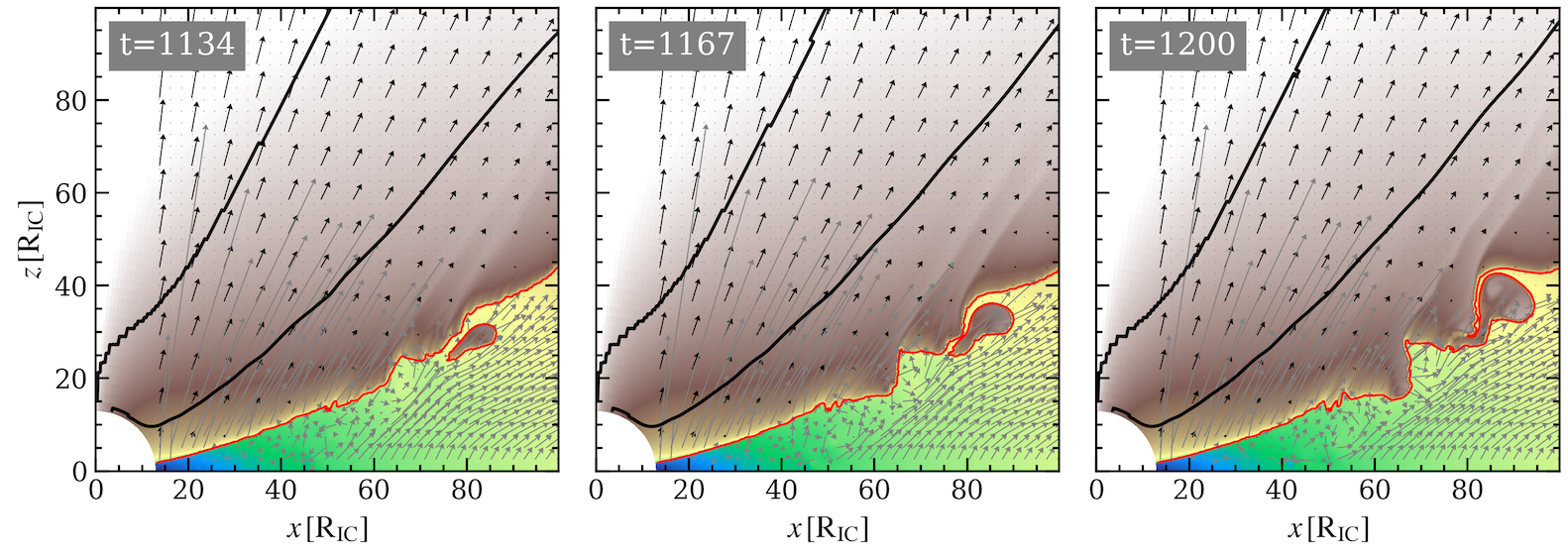}
 \caption{Density maps for our mid-res run capturing the onset of TI and the subsequent formation of an IAF.
 Velocity arrows and contours at $500\,{\rm km\,s^{-1}}$ and $10^3\,{\rm km\,s^{-1}}$ are overlaid as in previous plots.  
 The $Be=0$ contour is shown in red and it coincides with the contour $T=T_{\rm c,max}$ marking the position of the lower TI zone.
 To visualize the slow velocity field within the disk atmosphere, we overplot a second set of velocity arrows (gray) that have $10\times$ smaller magnitude.  
 This reveals a vortex at $(x,z) \approx (75,25)\,R_{\rm IC}$, the formation site of a hot spot (first visible at $t=1049$).  This spot forms due to TI (see Fig.~\ref{fig:hotspot_analysis}).
}
 \label{fig:IAF_midres}
\end{figure*}
%%%%%%%Fig.~\ref{fig:IAF_midres}

%%%%%%%%%%%%%%%%%%%%%%%%%%%%%%%%%%%%%%%%%%%%%%%%%%%%%%%%%%%%%%%%
\begin{figure*}[htb!]
 \centering
 \includegraphics[width=1.85\columnwidth]{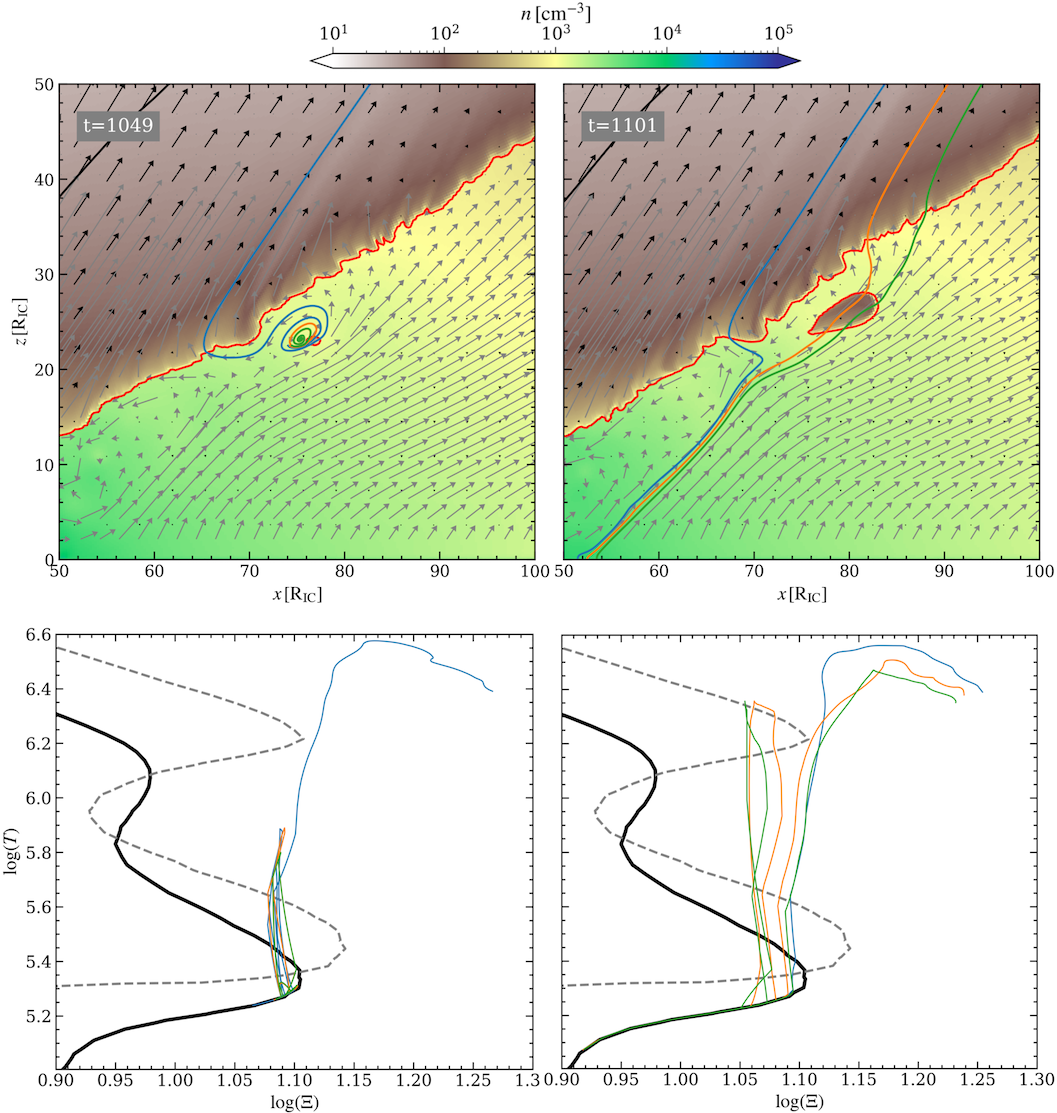} 
 \caption{Analysis of hot spot/bubble dynamics.
 {\it Top Panels:} 
 Colormaps of the number density as in Fig.~\ref{fig:IAF_midres} but zooming in on a $50\,R_{\rm IC} \times 50\, R_{\rm IC}$ region 
 centered on the hot spot.  Three streamlines are overplotted (blue, orange, and green).  In the left panel, the footpoints are located within the vortex and the orange and green streamlines are wrapped around the vortex.  In the right panel, they are located at smaller radii to illustrate that as the hot spot becomes a hot bubble, it separates from the vortex. 
 {\it Bottom Panels:}
 Phase diagrams similar to Fig.~\ref{fig:Scurve} showing the tracks for the streamlines in the panels above.  For the hot spot, the streamlines enter the lower TI zone only (left diagram).  For the hot bubble, the streamlines enter both the lower and upper TI zones (right diagram). 
}
 \label{fig:hotspot_analysis}
\end{figure*}
%%%%%%%Fig.~\ref{fig:hotspot_analysis}

%%%%%%%%%%%%%%%%%%%%%%%%%%%%%%%%%%%%%%%%%%%%%%%%%%%%%%%%%%%%%%%%
\begin{figure*}[htb!]
 \centering
 \includegraphics[width=2.0\columnwidth]{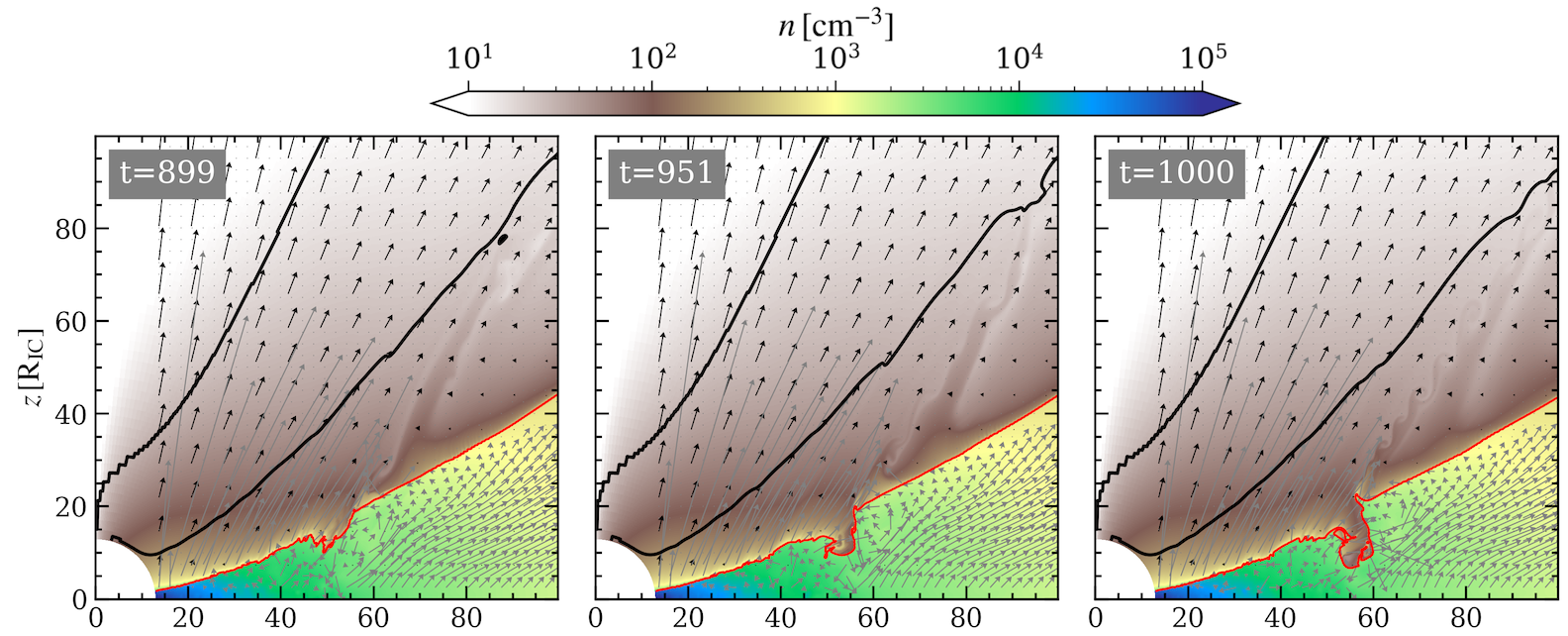}
 \includegraphics[width=2.0\columnwidth]{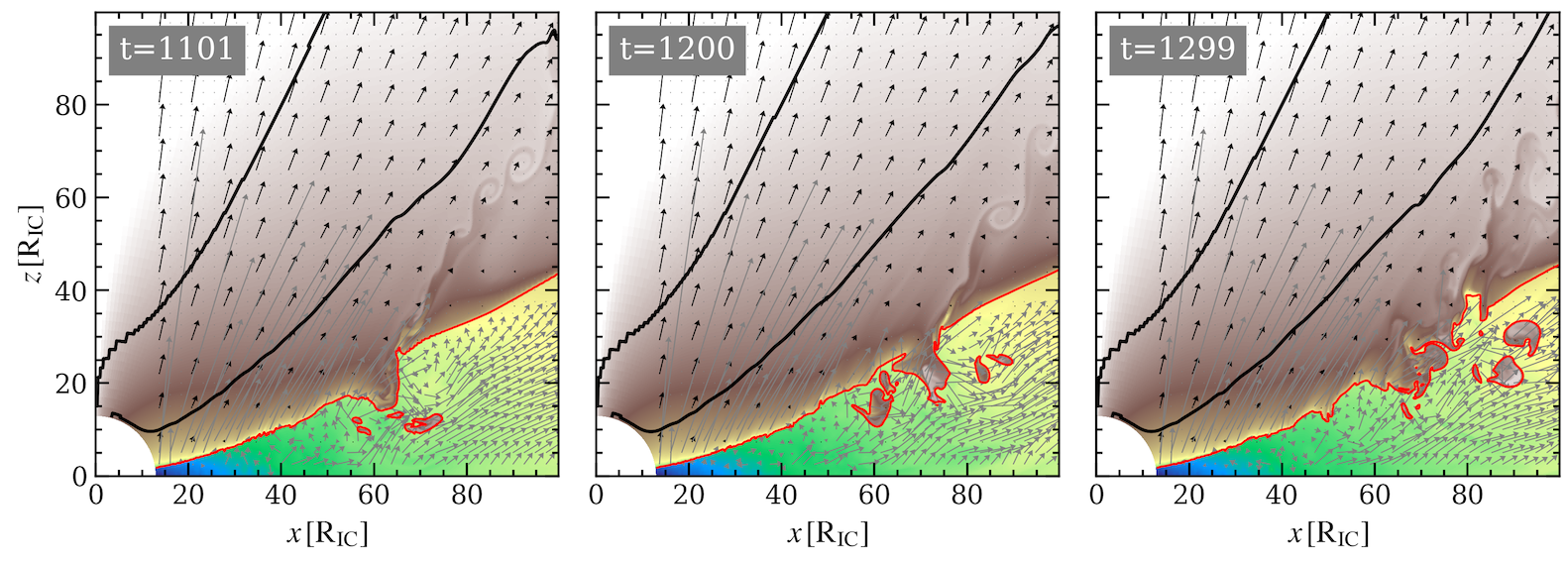} %2.025
 \caption{As in Fig.~\ref{fig:IAF_midres} but for the hi-res run.  
 Rather than through the growth of an embedded hot spot on the bottom edge of a vortex, the IAF here initially forms by hot gas from the wind region flowing into the atmosphere along the top edge of a vortex.  As the bottom panels show, embedded hot spots also form, indicating that both IAF formation channels operate simultaneously.  Additionally, in this hi-res run a turbulent wake forms in the wind downstream of the IAF and a vortex shedding process can be seen taking place.
 An animation of this is viewable at \url{https://trwaters.github.io/multiphaseAGNdiskwinds/}.
}
 \label{fig:IAF_hires}
\end{figure*}
%%%%%%%Fig.~\ref{fig:IAF_hires}

%%%%%%%%%%%%%%%%%%%%%%%%%%%%%%%%%%%%%%%%%%%%%%%%%%%%%%%%%%%%%%%%

%%%%%%%%%%%%%%%%%%%%%%%%%%%%%%%%%%%%%%%%%%%%%%%%%%%%%%%%%%%%%%%%
\begin{figure*}[htb!]
 \centering
 \includegraphics[width=2.\columnwidth]{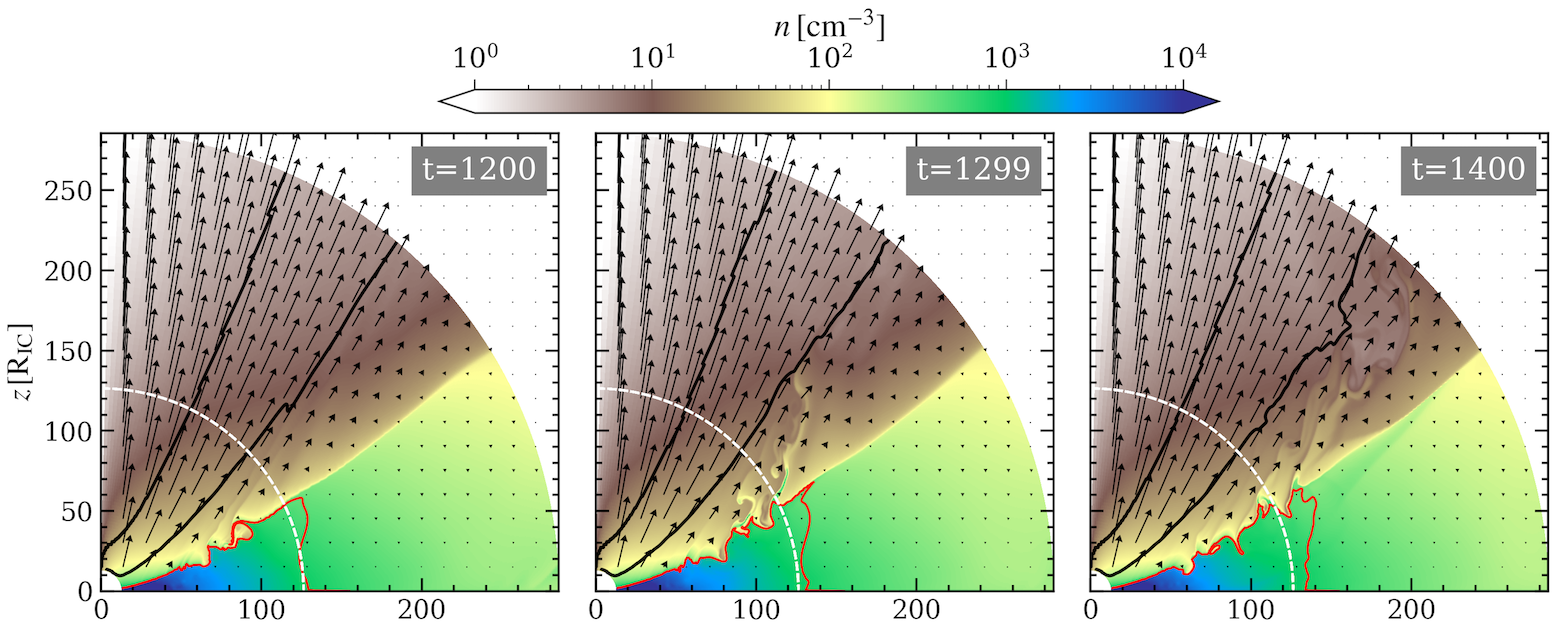}
 \includegraphics[width=2.\columnwidth]{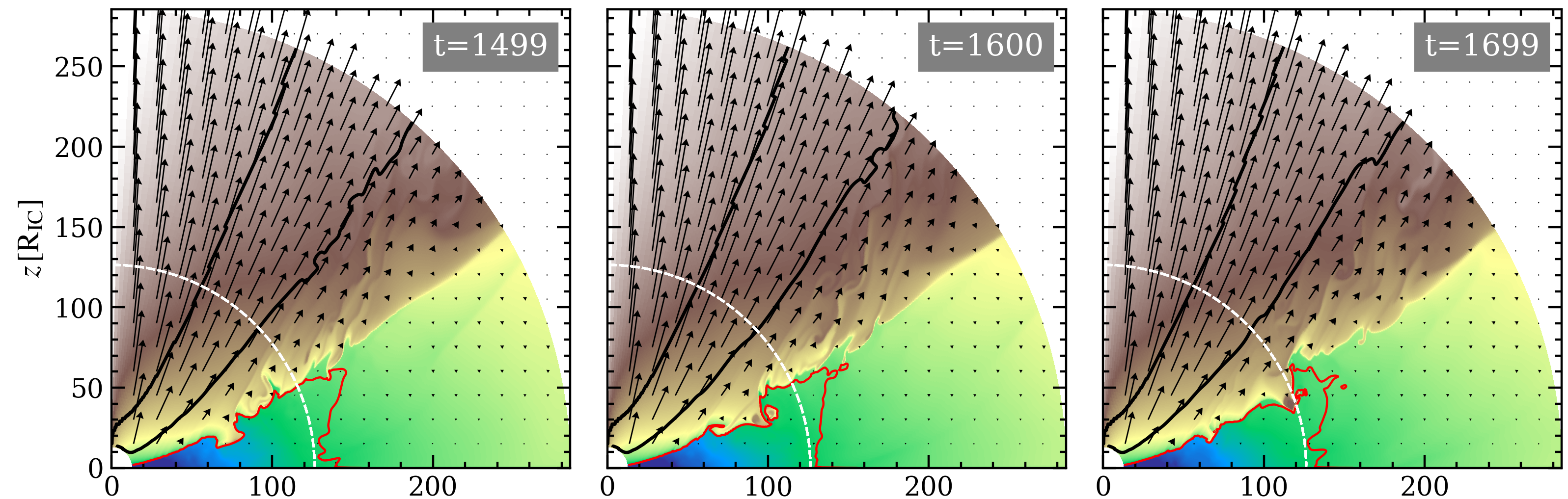}
 \includegraphics[width=2.\columnwidth]{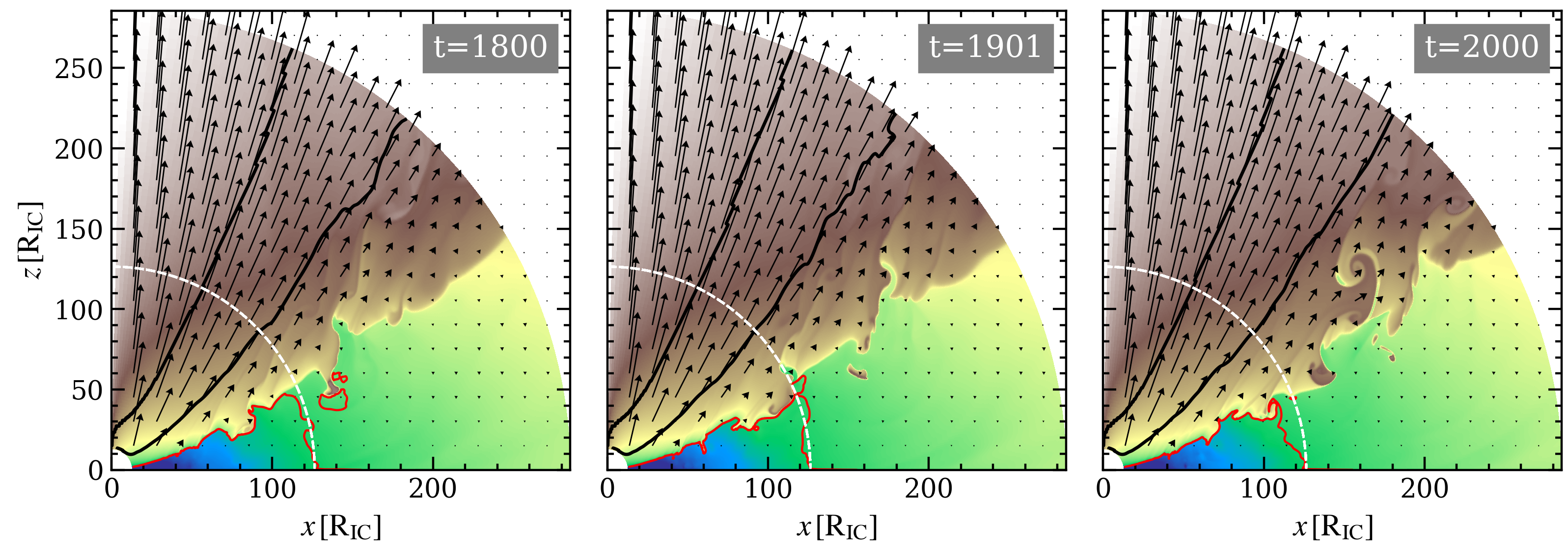}
 \caption{Density map snapshots as in Fig.~\ref{fig:IAF_midres} but showing the subsequent evolution of the flow.  The full domain is now plotted.  The white circular line marks the `unbound radius' $R_{\rm u}$ that approximates the location in the atmosphere where the contours $Be=0$ and $T=T_{\rm c,max}$ no longer coincide.  As shown by the red contour, $Be>0$ beyond $R_{\rm u}$; the flow at these distances becomes highly susceptible to TI.  The brown `spots' in the atmosphere at $t \geq 1600$ are thermally unstable hot bubbles.
}
\label{fig:IAFevolution}
\end{figure*}
%%%%%%%Fig.~\ref{fig:IAFevolution}

%%%%%%%%%%%%%%%%%%%%%%%%%%%%%%%%%%%%%%%%%%%%%%%%%%%%%%%%%%%%%%%%

\subsection{Numerical simulations of clumpy disk winds}
\label{sec:results}
Thus far we have examined solutions for a few different $\Xi_0$, all with $\rm{HEP_0} = 225$ and $\Gamma = 0.1$, corresponding to $r_0 \approx 1/3\,{\rm pc}$ for $M_{\rm bh} = 5.2\times 10^7\,M_\odot$, the mass of NGC~5548 \cite[e.g.,][]{Kriss19}.  
These domains have $r_{\rm in} = r_0/2 = 1.0\, R_{\rm IC}$ and $r_{\rm out}/r_{\rm in}  = 40$.  
Given the theoretical expectations from \S{\ref{sec:3bullets}}-\S{\ref{sec:Be}} for when TI can and cannot be triggered, the inner boundary 
of the computational domain need not begin at $R_{\rm IC}$ but should at least include the transition 
from bound to outflowing streamlines in the atmosphere, which occurs around $20\,R_{\rm IC}$ for $\rm{HEP_0} = 225$.
The outer boundary should extend beyond $R_{\rm u} = 140\,R_{\rm IC}\,(1-\Gamma)$, where clumpy wind dynamics is expected to occur. 
To have maximal numerical resolution at minimal cost, it is useful to minimize the dynamic range.
Based on these considerations, here we present solutions for $\rm{HEP_0} = 36$, which gives 
$r_0 = 13.0\, R_{\rm IC}$ for $\Gamma = 0.1$.  Our radial domain is then assigned as $r_{\rm in} = r_0$ 
with a dynamic range $r_{\rm out}/r_{\rm in}  = 22$, giving $r_{\rm out} = 1.55\,R_{\rm u}$.  

This fiducial setup was arrived at after analyzing more than a dozen simulations with different domain sizes and governing dimensionless parameters (${\rm HEP}_0$, $\Xi_0$, and $\Gamma$).
The processes at play are robust and hence captured by a single simulation.
Our numerical methods are detailed in Appendix~A; in summary, 
we solve the equations of non-adiabatic gas dynamics using the \athena code.
Our simulations are done in spherical coordinates assuming axisymmetry, and
we employ static mesh refinement (SMR).  We present results at
two resolutions: our \textit{mid-res} (\textit{hi-res}) run has three (four) levels of SMR.
We emphasize that these solutions are obtained without adding any perturbations to the flow.

To better enable a direct comparison with prior studies,
in our figures we quote times in units of the Keplerian orbital period at $R_{\rm IC}$
($2\pi R_{\rm IC}/\sqrt{G M_{\rm bh}/R_{\rm IC}}$):
\beq
t_{\rm kep}(R_{\rm IC}) = 3.53\times 10^2\,\f{\mu^{3/2}M_7}{(T_C/10^8\,{\rm K})}\, {\rm yrs}.
\label{eq:t_dyn}
\seq
With $\mu = 0.6$, $t_{\rm kep}(R_{\rm IC}) = 844\,{\rm yrs}$ for $M_{\rm bh} = 5.2\times 10^7\,M_\odot$, 
while $t_{\rm kep}(R_{\rm IC}) = 16.2\,{\rm yrs}$ for
$M_{\rm bh} = 10^6\,M_\odot$ (the black hole mass assumed in \citetalias{Dannen20}).
We continue to express distances in units of $R_{\rm IC}$, while 
number densities are computed in cgs units for $M_{\rm bh} = 5.2\times 10^7\,M_\odot$.

Because the dynamical time scales as $(r/R_{\rm IC})^{3/2}$, we see from Fig.~\ref{fig:IAF_midres} that a characteristic time span for the flow to visibly evolve is $\Delta t \approx 50$, consistent with the inner boundary being at $r_0 = 13\,R_{\rm IC}$ (i.e. $13^{3/2} \approx 47$).  
The other relevant timescale is the cooling time evaluated for temperatures near $T_{\rm c,max}$, the lowest temperature where gas can become thermally unstable.  As will be seen below, this typically occurs for gas near the atmosphere/wind interface
where number densities are $\sim 10^3\,{\rm cm^{-3}}$.  Re-evaluating \eqref{eq:tcool} gives
\beq
t_{\rm cool} = 1.42\times 10^2 \f{T/T_{\rm c,max}}{n_3\,C_{23}}\:{\rm yrs}.
\label{eq:tcool2}
\seq
This is comparable to the timescale for TI to operate within a single computational zone. 

A comparison of Fig.~\ref{fig:vortex} and Fig.~\ref{fig:IAF_midres} reveals that reducing $\rm{HEP_0}$ from 225 to 36 mostly preserves the dynamics in the disk wind.  Namely,
$\rm{HEP_0} = 36$ solutions are nearly equivalent to $\rm{HEP_0} = 225$ solutions at locations 
in the wind beyond where $\rm{HEP} < 36$ along the midplane.  
The location of vortices in the atmosphere, meanwhile, 
is somewhat sensitive to $\rm{HEP_0}$ because the dynamics of gas in the cold phase
depends more closely (due to the higher densities) on the value of $t_{\rm cool}$.

\subsubsection{The formation of IAFs}
\label{sec:IAFs}
Referring to Fig.~\ref{fig:IAF_midres}, the first appearance of unsteady behavior begins after $10^3\, t_{\rm kep}(R_{\rm IC})$.
Shown are six stages capturing how the first IAF in this run forms.  
The red contour here is the surface where $Be = 0$, which coincides with the surface 
where $\Xi = \Xi_{\rm c,max}$.  This implies that streamlines enter the TI zone with $Be < 0$, 
and so TI cannot easily amplify entropy modes (see \S{\ref{sec:Be}}).  However, there are several vortices 
in the atmosphere, and the flow along streamlines traversing a smaller vortex at 
$(x,z) \approx (75,25)\,R_{\rm IC}$ in the 1-st panel gets compressed enough to reach the temperature of the TI zone.  
Because it is circulating, the flow passes through the TI zone multiple times, giving more time for TI to operate.  
It is apparent from the 2-nd panel that a `hot spot' is forming due to the $Be = 0$ contour appearing 
within the atmosphere (as a red `dot').  This spot becomes a hot bubble by the 3rd panel.  As the bubble expands 
further in the next 3 frames, a thin layer of the atmosphere peels off, hence our referring to this 
as an irradiated atmospheric fragment.  

To show that the dynamics just described is attributed to TI, in Fig.~\ref{fig:hotspot_analysis}, 
we analyze the flow behavior on the $(T,\Xi)$-plane as in Fig.~\ref{fig:Scurve}.  In the top panels, 
three streamlines are plotted on zoomed-in versions of the 2-nd and 3-rd frames shown in Fig.~\ref{fig:IAF_midres}.  
All three pass through the hot spot, which is located near the bottom of the vortex 
where the streamlines bunch together, indicative of compressional heating.  The bottom left panel 
of Fig.~\ref{fig:hotspot_analysis} shows that these streamlines take several passes through the lower TI zone. 
The bottom right panel of Fig.~\ref{fig:hotspot_analysis} shows that the transition from hot spot to hot bubble 
coincides with the flow also passing through the upper TI zone.  The right-most tracks on the $(T,\Xi)$-plane 
that avoid the upper TI zone correspond to the streamlines exiting the vortex and entering the disk wind.  

A second formation channel for an IAF occurs in Fig.~\ref{fig:IAF_midres} also: rather than an embedded hot spot forming, 
the final two panels shows hot gas from the disk wind carving its way into the atmosphere.  
This channel is the first to occur in the hi-res run, as shown in Fig.~\ref{fig:IAF_hires}.  
Notice the IAF begins forming at an earlier time, which is not unexpected because higher resolution can permit faster 
growing entropy modes.  The depression at $r\approx 50\,R_{\rm IC}$ starts out as just a ripple on the surface 
of the atmosphere.  The ripples, in turn, are due to the small level of amplification that entropy modes undergo 
when gas with $Be < 0$ pass through the TI zone upon reaching this interface.  Consistent with our explanation in \S{\ref{sec:Be}}, the process that deepens the depression is not the direct growth of entropy modes along streamlines passing through the TI zone within the ripples, as there are ripples elsewhere that do not form depressions.  Rather, it is again dynamics associated with the vortex: instead of an embedded spot forming on a path through the bottom side of a vortex, pre-existing hot gas flows through the upper entry into the vortex.  
Thus, it is really the same physical mechanism at play --- flow through a vortex facilitating the growth of TI modes ---  and the bottom panels of Fig.~\ref{fig:IAF_hires} show that both of these IAF formation channels will in general occur simultaneously.    

\subsubsection{The evolution of IAFs}
Notice from both Fig.~\ref{fig:IAF_midres} and Fig.~\ref{fig:IAF_hires} that once an IAF is exposed to the disk wind, it begins to disintegrate, sending multiphase gas into the wind.  The classic turbulent flow patterns associated with vortex shedding accompany this process in the case of the second formation channel, when a small filament appears at the top of the more distant surface of the depression in the atmosphere.
This budding IAF acts as an obstruction to the wind, resulting in what clearly looks to be a K\'{a}rm\'{a}n vortex street in the downstream flow.  This is visible in the 3-rd, 4-th, and 5-th panels of Fig.~\ref{fig:IAF_hires}.  

The disintegration of the filaments/crests of IAFs combined with the downstream mixing and gradual heating of the shedded multiphase gas within their wakes will together be referred to as `evaporation'.  
We use this term to cast equal emphasis on the wake dynamics, which is what truly distinguishes smooth and clumpy evolution in terms of global wind diagnostics (see \S{\ref{sec:los_analysis}}).  Note that since we ignore the effects of thermal conduction, classical evaporation \citep{MB90} does not take place in these simulations.  The `evaporative' wakes that form downstream of IAFs are reminiscent of both the cometary-structured clumps that result from the photoevaporation of neutral gas clouds illuminated by massive stars \cite[e.g.,][]{NY19} and of the cloud destruction dynamics of wind-cloud interactions \cite[e.g.,][]{BB19}.  

Fig.~\ref{fig:IAFevolution} depicts the evolution of the flow on longer timescales, starting at $t=1200$, the final snapshot from Fig.~\ref{fig:IAF_midres}.  The IAF formation dynamics described above happens episodically, and the small scale IAFs shown up close in Fig.~\ref{fig:IAF_midres} gradually disrupt the flow at larger radii.
As seen in the $t=1600$ panel, an IAF can take on a tsunami-like appearance just within $R_{\rm u}$.  As shown in the final three panels, this IAF becomes increasingly filamentary and separates from the atmosphere, resulting in a smaller scale clump becoming entrained in the wind.  
This clump evaporates on a timescale of about $\Delta t = 100$, corresponding to a few thousand years for $M_{\rm bh} = 10^6\,M_\odot$ and $\sim 10^5\,{\rm yrs}$ for $M_{\rm bh} = 5.2\times 10^7\,M_\odot$ (see \eqref{eq:t_dyn}).  

This `early' evolution, while useful for examining the formation/disintegration dynamics of IAFs, can be considered `transient' behavior that eventually leads to a fully developed clumpy disk wind solution starting from smooth initial conditions.  As shown below, the fully developed solution features long lived hot bubbles evolving within the disk atmosphere at the base of even larger scale, tsunami-like IAFs.  In Fig.~\ref{fig:IAFevolution}, these bubbles are noticeable in the final five snapshots (see the brown gas to the lower left of the IAFs between $100\,R_{\rm IC}$ and $200\,R_{\rm IC}$).  

\begin{figure*}[htb!]
 \centering
 \includegraphics[width=1.85\columnwidth]{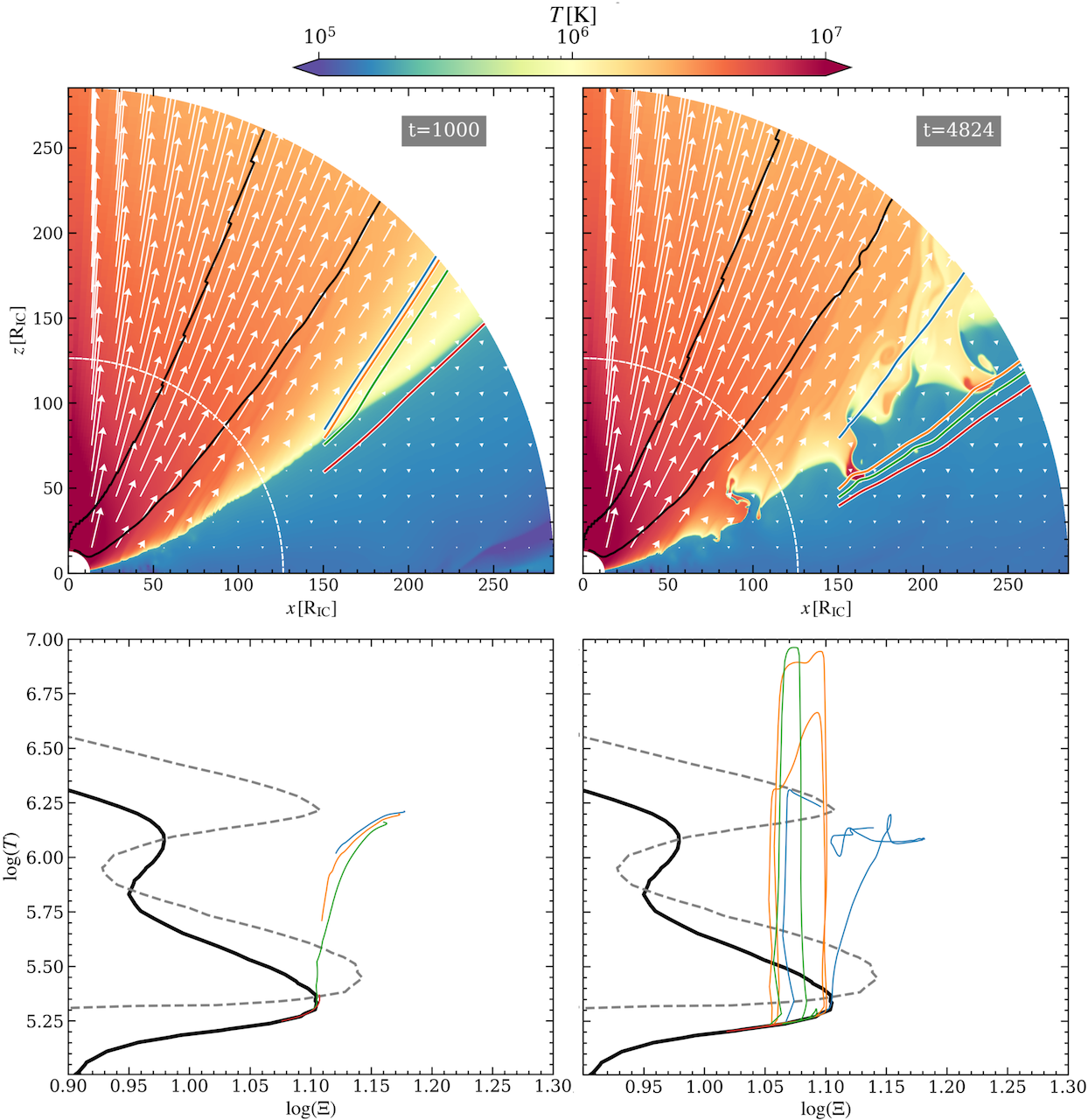}
 \caption{Comparison of the temperature structure and TI dynamics of smooth and clumpy wind stages.  
 {\it Top panels}: Colormaps of the temperature at $10^3$ orbital periods at $R_{\rm IC}$ when the solution is still smooth (left) and after evolving the solution for 1 orbital period at $r_{\rm out}$ when we end the simulation (right). 
 For each snapshot, we plot four streamlines anchored at various heights at $x = 150\,R_{\rm IC}$.  
 {\it Bottom panels}:
 Phase diagrams showing the tracks for the streamlines in the panels above.
 In both cases, the red streamline only occupies
 the cold branch of the S-curve.  In the clumpy wind snapshot, the green streamline passes through the inner hot bubble, 
 while the orange streamline passes through both the inner and outer bubbles, visible as two separate tracks through 
 the upper TI zone.  Both bubbles undergo runaway heating associated with TI.
}
\label{fig:smooth_vs_clumpy}
\end{figure*}
%%%%%%%Fig.~\ref{fig:smooth_vs_clumpy}

%%%%%%%%%%%%%%%%%%%%%%%%%%%%%%%%%%%%%%%%%%%%%%%%%%%%%%%%%%%%%%%%
\begin{figure*}[htb!]
 \centering
 \includegraphics[width=1.9\columnwidth]{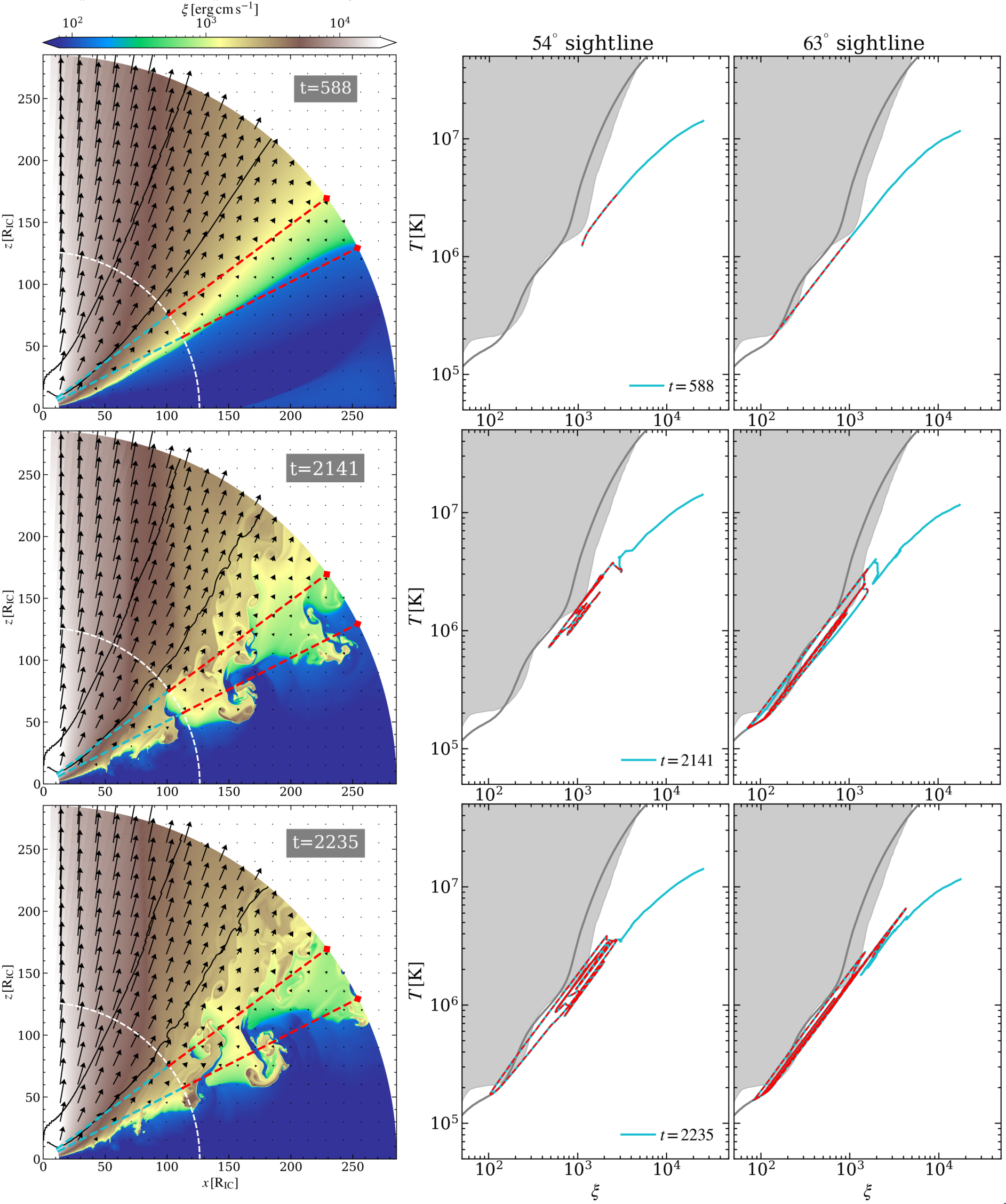}
 \caption{Comparison of the ionization structure and TI dynamics of smooth and clumpy wind stages along two different sightlines for the hi-res run.  
 {\it Left panels}: colormaps of the ionization parameter $\xi$ for an early smooth phase (top panel) and at two later stages when IAFs become fully developed.  Velocity arrows and contours are again overlaid.  The radius $R_{\rm u}$ is shown in white. Two sightlines at $54^{\degree}$ and $63^{\degree}$ from the polar axis are drawn; the red portions of these lines trace the radial range $r > R_{\rm u}$ used in subsequent figures.  
 {\it Right panels}: Corresponding phase diagrams of gas along the $54^{\degree}$ (left plot) and $63^{\degree}$ sightline (right plot).  The solid gray line is the S-curve.  The shaded gray area shows thermally unstable regions according to Balbus' criterion for TI; the boundary of this region defines the Balbus contour shown in previous phase diagrams.  The upper right panel shows that sightlines passing through both the disk atmosphere and the base of the disk wind span a large range of ionization states and only skim the instability region.  The smooth flow results in a continuous track on the phase diagram but would still be considered `multiphase' if observed along the $63^{\degree}$ sightline.
The subsequent clumpy flow regime shows an even larger range of ionization states.  Depending on the sightline and evolutionary stage, passing through an IAF can entail traversing brownish, yellowish, greenish, and bluish gas both upon entry and exit (corresponding to multiple passes through the instability region).
}
 \label{fig:xi_maps}
\end{figure*}
%%%%%%%Fig.~\ref{fig:xi_maps}

%%%%%%%%%%%%%%%%%%%%%%%%%%%%%%%%%%%%%%%%%%%%%%%%%%%%%%%%%%%%%%%%
\begin{figure}[htb!]
 \centering
 \includegraphics[width=0.95\columnwidth]{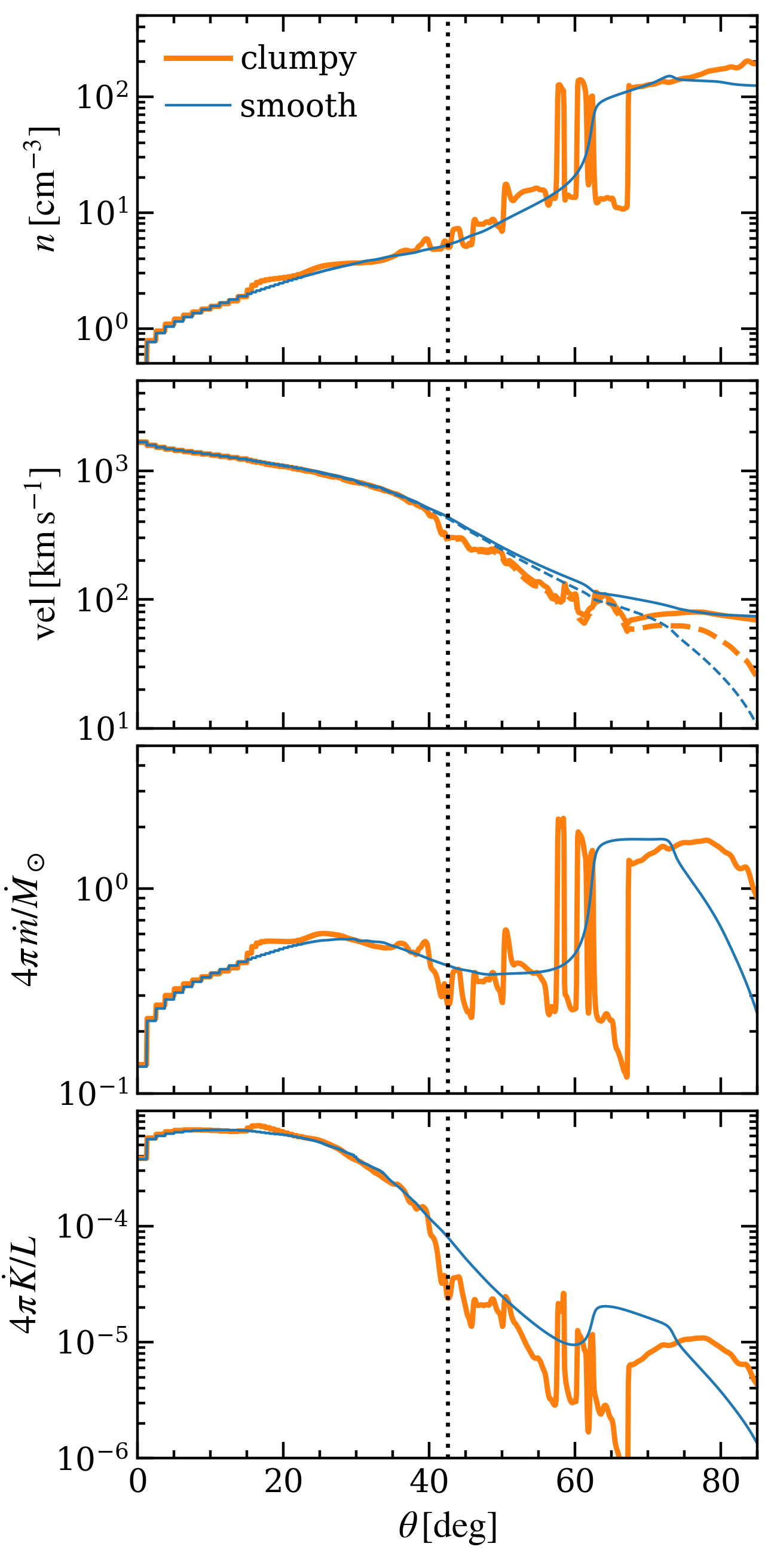}
 \caption{Comparison of wind properties along $\theta$ at $r_{\rm out}$ for smooth and clumpy stages of evolution.  These two states correspond to the $t=588$ and $t=2235$ snapshots shown in Fig.~\ref{fig:xi_maps}.  The solid and dashed lines in the second panel correspond to the total velocity $v$ and its radial component $v_r$, showing that the velocity field of the disk wind region is almost purely radial. 
 The next two panels show the mass loss rate and kinetic power. 
 Their integrated values are compared in Table~1. 
 Consistent with the table, we exclude the region $85^{\degree} < \theta < 90^{\degree}$ to allow for an underlying accretion disk.  The dotted vertical line divides the remaining domain equally.  
}
\label{fig:theta_profiles}
\end{figure}
%%%%%%%Fig.~\ref{fig:theta_profiles}

%%%%%%%%%%%%%%%%%%%%%%%%%%%%%%%%%%%%%%%%%%%%%%%%%%%%%%%%%%%%%%%%

\subsection{Smooth versus clumpy solutions}
\label{sec:los_analysis}
We now quantify the difference between smooth and clumpy solutions by comparing 
our runs at late times with an earlier phase when the flow resembles a steady 
state solution.  In Fig.~\ref{fig:smooth_vs_clumpy}, we present a streamline analysis similar to that of Fig.~\ref{fig:hotspot_analysis} but now focusing on the flow at large radii. 
The red streamline in both temperature maps traverses the atmosphere, and as seen 
on the accompanying phase diagrams, traces only points on the cold stable branch.  
For the smooth flow, neither the orange nor blue streamline enter a TI zone.  Only the green 
streamline that enters the wind through the atmosphere traverses the lower TI zone, and it does so 
with $\Xi$ increasing.  This is indicative of the flow still being marginally stable to TI 
at this stage despite this region being beyond $R_{\rm u}$.  

For the clumpy flow, the question arises, \textit{does TI operate only at the formation sites of 
IAFs or also within the IAFs as they are advected downstream?} 
In Fig.~\ref{fig:smooth_vs_clumpy}, the green streamline in the late time snapshot shows that 
TI is indeed operational in the distant flow.  This streamline passes through a hot bubble 
and is seen to occupy both the upper and lower TI zones.  Similarly, there are two separate tracks 
of the orange streamline in the upper TI zone, one reaching a lower temperature than the other.  
This is due to TI also taking place in the less hot bubble at $r\approx 250\,R_{\rm IC}$.  
Finally, the blue streamline passing through the crest of a fully developed IAF actually starts off 
in the upper TI zone, likely due to its footpoint being in hot gas from the nearby bubble.  
It passes isobarically to the cold phase as the IAF is traversed, and then pressure decreases 
downstream from there as it enters the yellow gas of the lower disk wind.   
Thus, while Fig.~\ref{fig:IAF_midres} shows that the dynamics of hot bubbles and IAFs are coupled, 
here we see that the bubbles themselves stay heated due to their being thermally unstable.  
This continued heating of the hot gas further expands the bubble, forcing more of the cool gas enveloping it
to enter the TI zone.

One can ask a related but more observationally relevant question: 
\textit{does a line of sight analysis reveal the same behavior, namely a paucity of gas at $\xi$ ranges 
corresponding to TI zones in smooth flows compared to clumpy ones?}  To answer this question, we switch 
to analyzing the high-resolution run because it yields the most accurate calculation of the absorption measure 
distribution (AMD).  Fig.~\ref{fig:xi_maps} shows maps of $\xi$, again for a smooth stage of evolution 
and also at two later, clumpy stages that capture how the crest of an IAF can break off, leaving an isolated, 
outflowing clump.  The upper sightline in the bottom panel passes through this clump, whereas the same sightline at $t=2141$ passes just above the crest.  Thus, there is a much broader range of $\xi$ at $t=2235$ 
compared to $t=2141$, as is also apparent from the corresponding phase diagrams, which are now plotted using 
the $(T,\xi)$-plane. There are again multiple passes through both TI zones for sightlines that intersect 
cold phase gas, implying higher column densities of gas in the temperature range of the lower TI zone, 
$T \approx 2-5\times10^6\,{\rm K}$.  For sightlines through the smooth solution, however, gas only occupies 
this temperature range if the upper atmosphere is skimmed. This contrast in what a distant observer will see 
indicates that smooth and clumpy wind models are very different, and both happen to defy the standard 
narrative that accompanies discussions of the AMD (see \S{\ref{sec:AMD}}).  

%%%%%%%%%%%%%%%%%%%%%%%
\begin{table} %[htb!]
\centering
\begin{tabular}{c||cc|cc}
\multicolumn{1}{c||}{\multirow{2}{*}{Quantity}} & \multicolumn{2}{c|}{Smooth} & \multicolumn{2}{c}{Clumpy}\\
& $<42.5^{\degree}$ & $>42.5^{\degree}$ & $<42.5^{\degree}$ & $>42.5^{\degree}$  \\\hline
$\langle v \rangle \: {\rm [km\, s^{-1}]}$  & 857 & 123 & 835 & 105 \\ 
$4\pi\,\dot{m}/\dot{M}_\odot$  & 0.13 & 0.57 & 0.13 & 0.62  \\ 
$4\pi\,\dot{K}/L\: [10^{-4}]$ &  1.04 & 0.11 & 1.04 & 0.07 \\ 
\end{tabular}
\caption{Global wind diagnostics of smooth versus clumpy wind stages corresponding to the $t=588$ and $t=2235$ snapshots in Fig.~\ref{fig:xi_maps}.
The top entries are the mean outflow speeds at $r_{\rm out}$, $\langle v \rangle = \int v_r(r_{\rm out}) \sin\theta\,d\theta/\int \sin\theta\,d\theta$.
The next entries are the integrated profiles shown in the lower panels of Fig.~\ref{fig:theta_profiles}:
$4\pi\,\dot{m}/\dot{M}_\odot$ is the mass loss rate in units of $M_\odot\,{\rm yr^{-1}}$ and
$4\pi\,\dot{K}/L$ is the kinetic luminosity in units of $L = 6.32\times10^{44}\,{\rm erg\,s^{-1}}$ (the luminosity for $M_{\rm bh} = 5\times 10^7\,M_\odot$ assuming $\eta = 0.1$).
The $>42.5^{\degree}$ values exclude a $5^{\degree}$ region above the midplane to allow for an underlying disk.  The entries for $\dot{m}$ and $\dot{K}$ account for midplane symmetry, so their sum gives $\dot{M}$ and $P_{\rm K}$, the total values over a $4\pi$ solid angle.}
\end{table}
%%%%%%%%%%%%%%%%%%%%%%%%%%%%%%%%%%%%%%%%%%%%%%%%%%%%%%%%%%%%%%%%%

On the other hand, in Fig.~\ref{fig:theta_profiles} we present a comparison of quantities along the outer boundary, which shows that smooth and clumpy wind stages hardly differ outside of the multiphase wind region. (The particular flow states compared here are top and bottom snapshots in Fig.~\ref{fig:xi_maps}.)
Within this region, the spikes in density in the top panel result in a somewhat larger mass loss rate, but this enhancement in the mass flux is outweighed by the drop in velocity (due to drag forces), yielding an overall reduction in kinetic luminosity.  
In Table~1, we list precise values for these global wind properties, which are
computed as 
$\dot{M} = 4\pi \int_0^{\pi/2} \dot{m}\,\sin\theta\,d\theta$
and 
$P_K = 4\pi \int_0^{\pi/2} \dot{K}\,\sin\theta\,d\theta$,
where 
$\dot{m} = \rho v_r r_{\rm out}^2$
and
$\dot{K} = (1/2)\rho v^2 v_r r_{\rm out}^2$.
These calculations require $v_r$ in addition to $\rho$ and $v = \sqrt{v_r^2 + v_\theta^2 + v_\phi^2}$, which is shown
by the dashed curve in the 2-nd from top panel of Fig.~\ref{fig:theta_profiles}.  
Notice that the profiles of $v_r$ and $v$ overlap above the disk atmosphere,
indicating that the disk wind regions are radially outflowing.
Within the atmosphere, $v_\phi$ is dominant, explaining 
the higher velocity for $\theta > 65^{\degree}$. 
In Table~1, for both smooth and clumpy states, we compute values to the left and right of the vertical line in Fig.~\ref{fig:theta_profiles}, which approximately separates the subsonic, multiphase flow from the highly ionized, supersonic flow above.  

That the solutions are so similar in terms of bulk wind properties is merely a reflection of the underlying dynamics serving mainly to restructure parts of the flow without affecting the efficiency of thermal driving.  Specifically, buoyancy effects lead to a rearrangement of matter along the atmosphere/wind interface, resulting in enhanced vertical rather than radial motion within the domain.  
This restructuring does not impact the fast, smooth flow at high latitudes, meaning it remains nearly undetectable in absorption (since it consists of almost fully ionized plasma).  The clumpiness in the multiphase wind region, a result of the evaporation dynamics within the wakes of the IAFs, leads to a broader range of ionization parameters --- all characteristic of warm absorbers --- implying that these new solutions may be important for the interpretation of AGN spectra.   

\subsection{Absorption diagnostics}
\label{sec:AMD}
Ionized AGN outflows can show absorption from ions that are widely separated in ionization energy.  
From the distribution of ionic column densities, it is possible to reconstruct the AMD \citep{Holczer07}, 
which is the distribution of hydrogen column density $N_{\rm H}$ as a function of the density ionization 
parameter, $\xi$. Conversely, from a model of the gas distribution along the line of sight, 
the AMD can be computed as the slope of the column density $N_{\rm H}$ when plotted against $\log(\xi)$,
\beq
{\rm AMD} \equiv \d{N_{\rm H}}{\log(\xi)}.
\seq
Because $N_H$ is a monotonically increasing function by definition, the AMD is a positive quantity that 
should be overall large in high density regions characterized by smaller ionization parameters, 
such as inside an AGN cloud.  The generic shape of the AMD across regions with sharp density gradients, 
such as the surface of clouds (IAFs in these models), is unclear.
We therefore present an in depth analysis of the AMD for the two sightlines discussed already in \S{\ref{sec:los_analysis}}. 
It is hoped that our discussion here provides a useful perspective into the ongoing debate over 
what shapes the AMD --- is it individual clouds, a stratified outflow, or as suggested 
by \cite{Behar09} and as we find here, 
individual clouds (IAFs) evolving within a stratified outflow?

To further assess the extent to which smooth and clumpy disk wind solutions differ 
in terms of absorption signatures, we also present calculations 
of X-ray absorption line profiles for selected ions spanning a similar range 
of ionization parameters as the AMD.  
We evaluate the formal solution to the radiative transfer equation 
for an absorption line against a background intensity, $I_0$, that is uniform in space and frequency,
\begin{align}
    I_\nu = I_0\, e^{-\tau_\nu(\theta_0)} , 
\label{eq4}
\end{align}
where $\tau_\nu(\theta_0) = \int \kappa_\nu(r,\theta_0)\, \rho(r,\theta_0)\,dr$ 
is the optical depth at $r_{\rm out}$ for a fixed inclination $\theta_0$ 
(the viewing angle for a distant observer).  These calculations therefore 
neglect contributions due to the wind's intrinsic emission along the line of sight, 
as well as scattering into the line of sight.
Following the methods of \cite{Waters17}, we evaluate 
$\kappa_\nu(r) = \kappa_{0}(r) \phi(\nu)/\phi(\nu_D)$ by post-processing 
our simulation results to Doppler broaden the lines according to 
the radial velocity and temperature profiles; 
$\phi(\nu)$ is the (radially dependent) Gaussian profile with a
width $\Delta\nu_0=\nu_0\,v_{\rm th}/c$ set by the thermal velocity 
$v_{\rm th}(r) = \sqrt{k T(r)/m_{\rm ion}}$ and with the 
line center frequency Doppler-shifted to
$\nu_D(r) = \nu_0[1+v_r(r)/c]$.
We use lookup tables generated from \xstar~output to obtain 
the line-center opacity $\kappa_{0}(r) = \kappa_{0}[\xi(r),T(r)]$ 
for each ion.  These opacity tables 
are generated from the same grid of photoionization calculations 
used to tabulate our net cooling function $\mathcal{L}$ that defines 
our S-curve.  See \cite{Ganguly21} for a more thorough description of these calculations, as well as for a detailed analysis of various factors affecting line formation.

\subsubsection{Absorption measure distribution (AMD)}
\begin{figure*}[htb!]
 \centering
 \includegraphics[width=1.02\columnwidth]{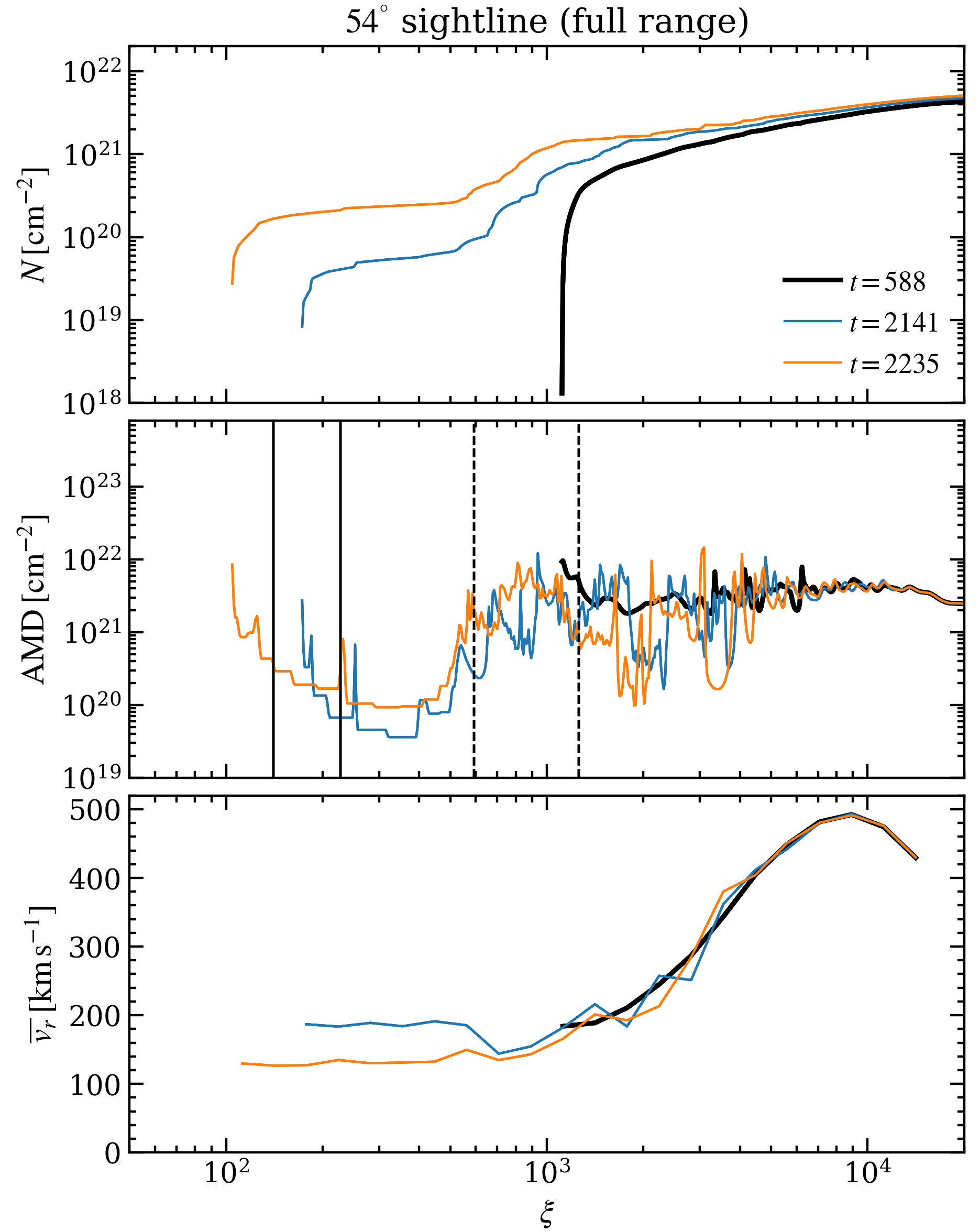}
 \includegraphics[width=0.95\columnwidth]{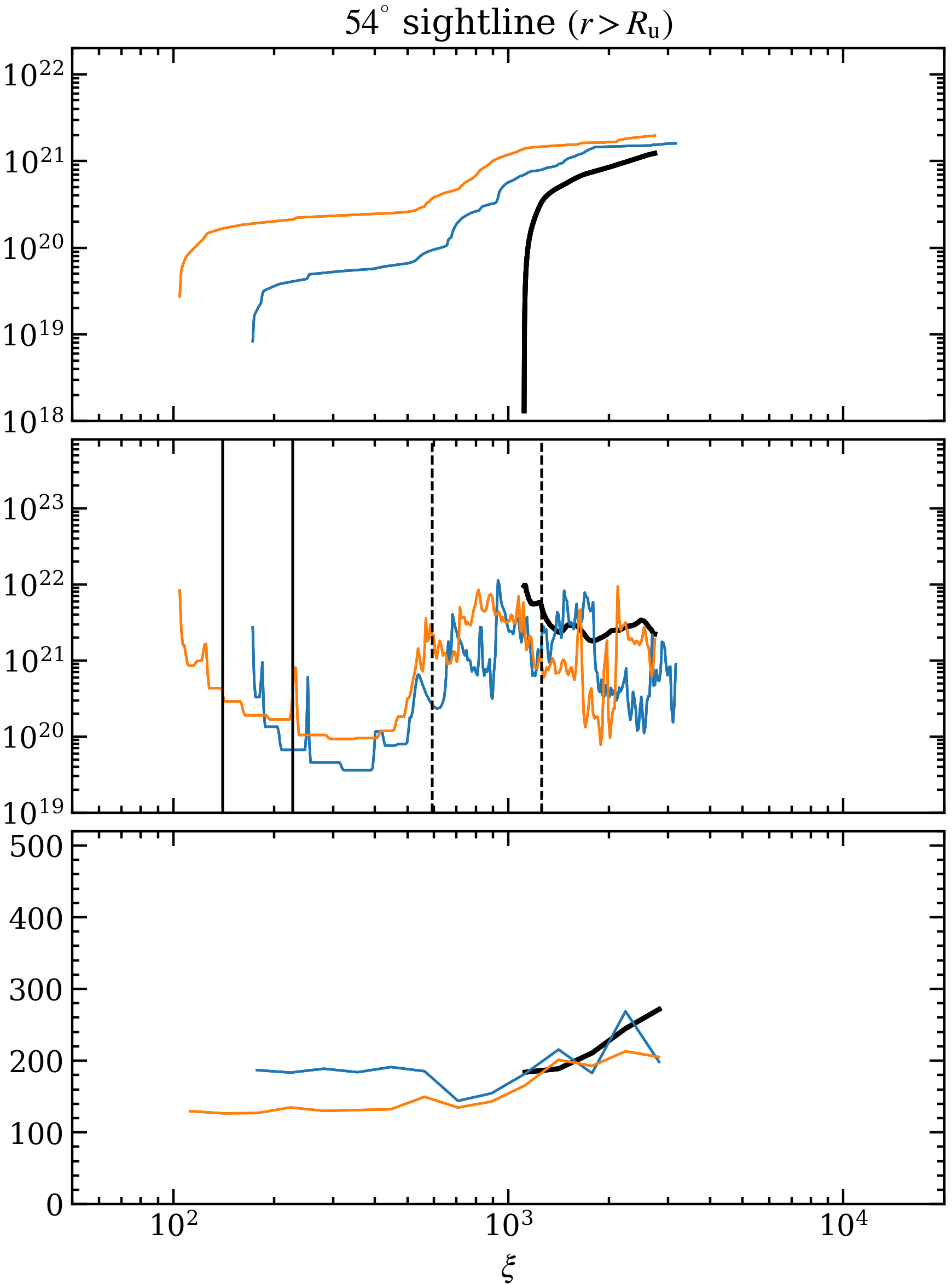}
 \includegraphics[width=1.02\columnwidth]{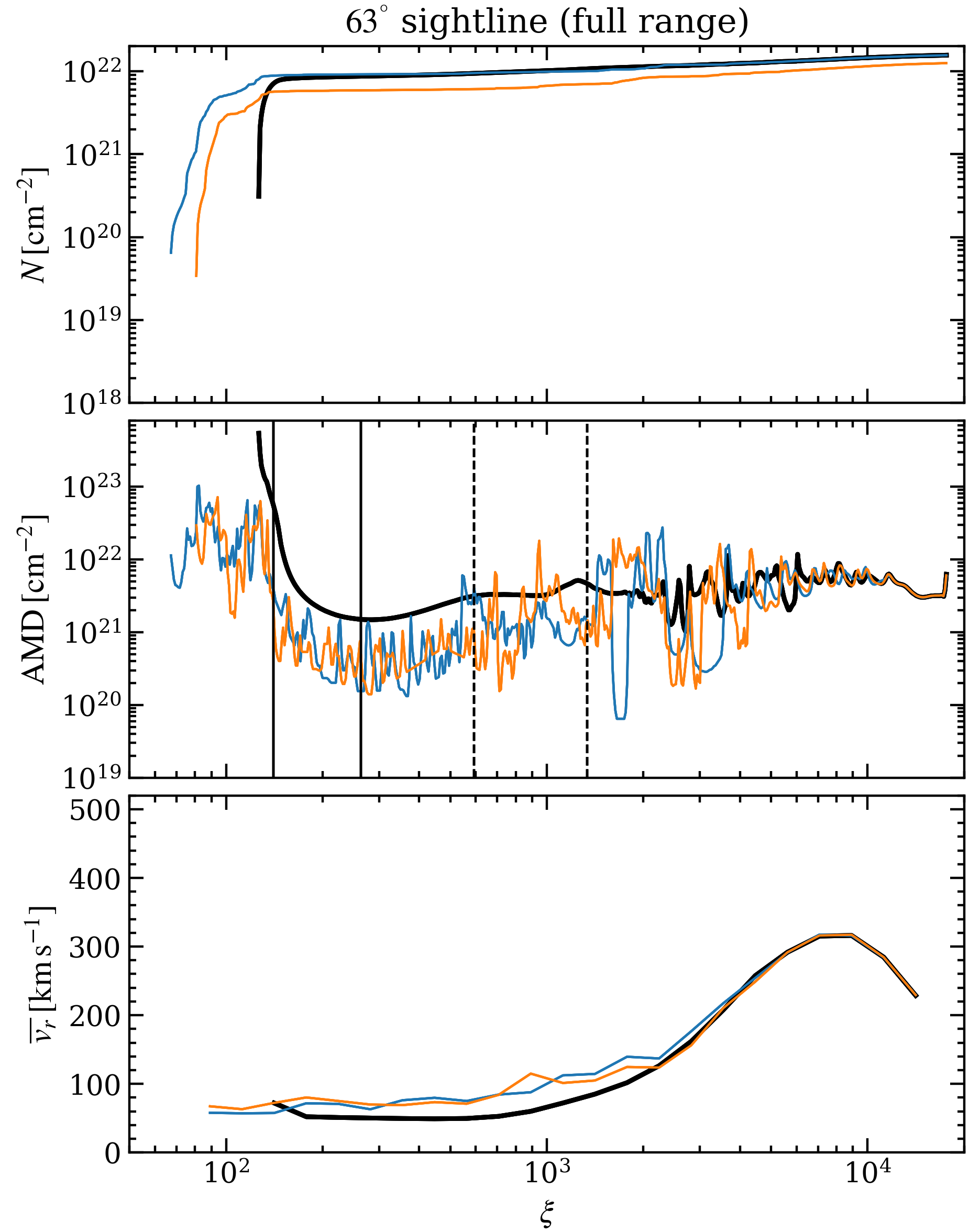}
 \includegraphics[width=0.95\columnwidth]{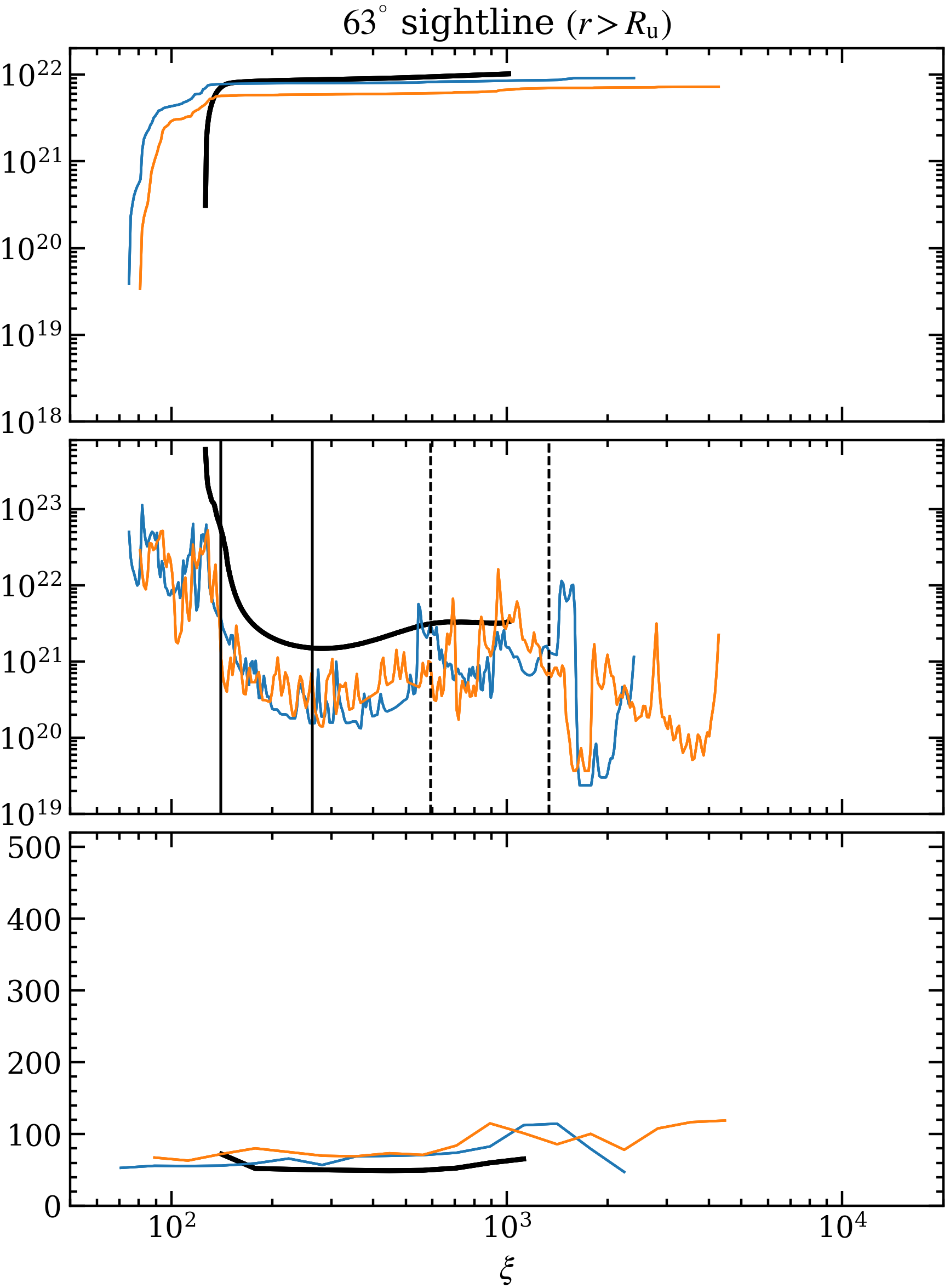}
 \caption{AMD analysis for the two sightlines shown in Fig.~\ref{fig:xi_maps}.  
 The top, middle, and bottom subpanels show the column density, the AMD, and the column density weighted 
 line of sight velocity (as defined in the text). Calculations for the full sightline are shown 
 in the left panels, while the right panels show a calculation only along the red portion 
 of the sightline shown in Fig.~\ref{fig:xi_maps}, covering the range $r > R_{\rm u}$.
 Black, blue, and orange curves correspond to the three snapshots shown in Fig.~\ref{fig:xi_maps}.
 The $\xi$-ranges of the upper and lower TI zones are marked by pairs of solid and dashed vertical lines 
 in the AMD plots.  
}
 \label{fig:AMDs}
\end{figure*}

We analyze the AMD for the particular states already examined in a dynamical context in \S{\ref{sec:los_analysis}}. 
The middle panels in the left column of Fig.~\ref{fig:AMDs} shows the AMD computed along the two sightlines plotted in Fig.~\ref{fig:xi_maps}.  
To aid our analysis, in the right middle panels we also show the AMD over just the outer portion of the sightline extending from $r > R_{\rm u}$ (the red portion in Fig.~\ref{fig:xi_maps}).  Additionally, in the upper panels we plot the cumulative column density to show that the shape of the AMD is consistent with the slope of these curves.  Finally, in the lower panels we plot the column density weighted radial velocity 
$\overline{v_r} \equiv \int_{\Delta} N\,v_r\,d\xi / \int_{\Delta} N\, d\xi$, 
where $\Delta = 0.1\,{\rm dex}$ denotes the bin size of $\log(\xi)$ over which the integrals were evaluated;  $\overline{v_r}$ is indicative of the expected blueshifts of absorption lines.  

Focusing first on the smooth state (upper panel in Fig.~\ref{fig:xi_maps}), notice that starting from the outer boundary, the two sightlines trace gas with increasing ionization as $r$ decreases.  These radial profiles are therefore in direct correspondence with the AMD.  For the $63^{\degree}$ sightline, the cold phase atmosphere gas (blue color) is all at the same low ionization and there is a large column of it compared to the less dense gas beyond (green through brown colors), explaining the more than two order of magnitude dip in the AMD.  The AMD of the $54^{\degree}$ sightline spans a much smaller range of $\xi$ and is overall small (a few $10^{21}\,{\rm cm^{-2}}$) as only low density gas is being probed. There is a dip upon \textit{exiting} the $\xi$-range of the upper TI zone (marked with vertical dashed lines) because the higher ionization gas beyond has a smaller rise in column density, as shown in the top panels of Fig.~\ref{fig:AMDs}.

Before discussing the AMD for the clumpy states, we assess how the analysis just given fits in with the most common discussion of the AMD in the literature (`the standard narrative' hereafter), in which reference is made to the theory of local TI.
Namely, dips in the AMD are interpreted as indicating thermally unstable $\xi$-ranges, the outcome of local TI being to replace these ranges by disparate regions of the S-curve that are connected isobarically.\footnote{Technically, in the theory of local TI, gas will not typically reach the upper stable branch of the S-curve due to the effects of thermal conduction \citep[see][]{PW15}.}
Below we will discuss in more detail local versus `dynamical' TI, but here we emphasize that the dip in the AMD is not due to dynamical TI because the flow is still smooth at this stage.
While the pronounced dip in the AMD for the $63^{\degree}$ sightline is merely a column density effect, it is not coincidental that it occurs near $\xi_{\rm c,max}$, the entry to the lower TI zone.
The flow structure along this sightline actually epitomizes the intrinsic connection between thermal driving and TI discussed in \S{2}.  To recap, the presence of a TI zone is responsible for the sharp transition from a bound atmosphere to a thermally driven wind at $\xi=\xi_{\rm c,max}$.  Gas with $T>T_{\rm c,max}$ is much more tenuous in a steady equilibrium state.  Hence, for sightlines through the atmosphere, the AMD will naturally exhibit a dramatic falloff once $\xi_{\rm c,max}$ is reached.  Compared to the standard narrative, therefore, pronounced dips in the AMD here are caused by the full structure (see \S{2}) of thermally driven wind solutions being probed.

The most basic result in Fig.~\ref{fig:AMDs} is that the AMD for clumpy states span a larger range of $\xi$.
The AMD always extends to lower $\xi$ for clumpy compared to smooth states because IAFs undergo compression.  It can extend to larger $\xi$ when gas within bubbles are probed.  We showed bubbles to be the primary sites were TI takes place, hence they are subjected to continuous runaway heating that increases $\xi$.  

%%%%%%%%%%%%%%%%%%%%%%%%%%%%%%%%%%%%%%%%%%%%%%%%%%%%%%%%%%%%%%%%
%%%%%%%Fig.~\ref{fig:AMDs}

Notice that beyond $\xi_{\rm c,max}$, the AMDs for both clumpy states feature dips or remain relatively flat between the pairs of vertical lines marking the `unstable' $\xi$-ranges that correspond to the upper and lower TI zones.  Dips can also occur beyond the upper TI zone.  For smooth states, the AMD rises slightly within the upper TI zone. 

These features are inconsistent with the standard narrative. 
Rather than featuring a dip within the unstable $\xi$-ranges, the AMD shape for smooth, thermally driven wind solutions is simpler still: gas continues to occupy the `unstable' $\xi$-range of the S-curve because this same range is actually stable just slightly off the S-curve.  As shown in the top panels of Fig.~\ref{fig:xi_maps}, gas does not actually pass through the upper TI zone; hence for $t=588$, the flow is formally stable in the range between the vertical dashed lines in Fig.~\ref{fig:AMDs}, and it lies below the equilibrium curve in a region of net radiative heating.  A steady state dynamical equilibrium is maintained by this heating being balanced by adiabatic cooling due to expansion.  For the $63^{\degree}$ sightline, gas also passes through the lower TI zone, but it has not yet become unstable due to the predominance of the stabilizing effects discussed in \S{\ref{sec:Be}}.  
In clumpy states, dips at large $\xi$ beyond the upper TI zone are simply due to the overheated gas within bubbles having low column densities.

To understand the AMD of the clumpy flow states in detail, 
it is helpful to contrast the regimes of local TI and dynamical TI.
The depletion of gas within TI zones --- the principle outcome of local TI --- still takes place within global outflow solutions once they cease to be smooth.
In local TI, there is finite supply of the initial unstable gas reservoir, while in global flows, the TI zone can be continually replenished with new gas.   
In papers that reference the theory of local TI when interpreting the AMD \citep[e.g.,][]{Behar09,Detmers11,Adhikari19}, the implicit assumption is therefore being made that thermal timescales are much shorter than dynamical ones to allow depletion to dominate replenishment.  This would entail, for example, that a hot bubble in pressure equilibrium with the cold atmosphere forms rapidly (compared to orbital timescales) at the onset of TI and then no longer evolves thermally, thereby keeping the TI zone devoid of gas.  Such a scenario is inconsistent with the force equation, however, because hot bubbles tend to buoyantly rise.  
The dynamics accompanying the bubbles appearing in our simulations indeed involves the effects of buoyancy, as this is what leads to the formation of IAFs (see \S{\ref{sec:IAFs}}).  Moreover, our simulations show that bubbles continue to expand but on timescales slow compared to orbital times, and we suspect that this is only possible when the rates for gas entering and exiting TI zones are closely balanced.

Based on this qualitative understanding of dynamical TI, we would have been surprised 
to find an association between dips in the AMD and the locations of TI zones.  Rather, 
the expected signature of dynamical TI is a dip occurring \textit{outside} of TI zones 
at even larger $\xi$.  
This dip probes the small column of gas within a hot bubble.  Referring 
to the lower ($63^{\degree}$) sightline in the bottom panel of Fig.~\ref{fig:xi_maps},
the corresponding partial AMD (orange curve in bottom right AMD panel of Fig.~\ref{fig:AMDs}) 
shows prominent dips at $\xi \approx 1.8\times 10^3$ and $\xi \approx 3.5\times 10^3$.  
The lower ionization dip is from the entrained bubble at $r\approx 200\,R_{\rm IC}$, 
while the higher ionization one is the darker brown gas between $R_{\rm u}$ 
and the first IAF encountered along this portion of the sightline.  This latter gas 
is seen to originate from another bubble located just below the sightline that is also subjected to TI. 
However, both of these dips become mostly filled when we include the full sightline 
because the gas at small radii covers the same ionization parameter range and is also denser.  
This analysis therefore casts serious doubt on the whole notion that the AMD can be used 
to draw any conclusions about TI.  In principle it can, but there is degeneracy 
with `normally' heated gas now that we have shown that TI should not lead to dips 
within the $\xi$-ranges of TI zones.

The AMD of the same sightline but for the $t=2141$ snapshot (blue $63^{\degree}$ line) 
does show a wide dip in the TI zone for the partial AMD calculation. 
This dip also gets filled in upon including the full sightline, 
but what is interesting is the red portion of the sightline traces only one region 
of yellowish/brownish gas --- that of the evaporating flow of the IAF just 
beyond $r = R_{\rm u}$.  This is the most highly ionized gas probed and is seen to be 
at the $\xi$ marking the location of the dip.  Furthermore, the evaporative layer 
of the other much thicker IAF along this sightline contains the less ionized 
(light green colored) gas with $\xi \approx 800$ that is within the same TI zone.  
The primary dip in the AMD near $\xi_{c,max}$ is unambiguously the dark green evaporation layer.
Thus, we have arrived at the expected AMD shape of a single IAF: an overall peak for 
$\xi < \xi_{c,max}$, a large dip near $\xi = \xi_{c,max}$ (that corresponds 
to `exiting the atmosphere' as in the AMD for the smooth phase)
followed by secondary dips at larger $\xi$ that often coincide with TI zones but are attributable to a sharp falloff in column density in the IAF's wake rather than to TI. 

The orange AMD curves in the top panels of Fig.~\ref{fig:AMDs} provide another example 
of this AMD shape.  This is the upper sightline in the bottom panel of Fig.~\ref{fig:xi_maps} 
that passes through an isolated clump.  Such clumps arise when the crest of an IAF becomes detached after being ablated by the wind.  
This is clearly the most low ionization gas along this sightline, showing that the primary dip 
in the AMD beginning near $\xi_{c,max}$ is the signature of this clump.  There would be a secondary dip 
in the upper TI zone but it is filled in by the larger column of brownish gas at smaller radii.
Notice that this evaporative structure is reminiscent of the cometary tails discussed in the context 
of occulation events \citep[e.g.,][]{Bianchi12}.  The evaporating wakes there are depicted as moving transverse to the line of sight rather than along it, but the ionization structure they inferred is nevertheless consistent 
with the dynamics taking place in our simulations.  

In summary, the AMD for clumpy multiphase disk wind models can be understood as a combination 
of that expected for a smooth outflow with that for one or more embedded IAFs (which are equivalent 
to discrete clouds when probed only along the line of sight).  
Specifically, the AMD shape has several variants depending on the sightline.
For sightlines intersecting cold phase gas, indicative of either passage through the atmosphere or an IAF, we find
\begin{itemize}
\setlength\itemsep{-0.5em}
\item a nearly flat distribution for $\xi < \xi_{\rm c,max}$ corresponding to compressed cold phase gas;
\item a pronounced dip just above $\xi_{\rm c,max}$, accompanied by somewhat larger blueshifts because IAFs are subjected to ram pressure by the disk wind.
\end{itemize}
For sightlines passing through hot bubbles, where TI is actively taking place,
\begin{itemize}
\item a dip should occur at $\xi$ larger than that of any TI zones (for $\log\xi > 3$ by our calculations). 
\end{itemize}
Finally, for sightlines passing above the dense IAFs but still probing their evaporative wakes,
\begin{itemize}
\item the low-$\xi$ range of the AMD does not extend to $\xi_{\rm c,max}$ and blueshifts are systematically higher. 
\end{itemize}
The last point was not discussed above, but it is evident from the blue curve 
in the upper panels of Fig.~\ref{fig:xi_maps} and is relevant to the synthetic line 
profile calculations that we now present.

%%%%%%%%%%%%%%%%%%%%%%%%%%%%%%%%%%%%%%%%%%%%%%%%%%%%%%%%%%%%%%%%
\begin{figure*}[htb!]
 \centering
 \includegraphics[width=0.601\columnwidth]{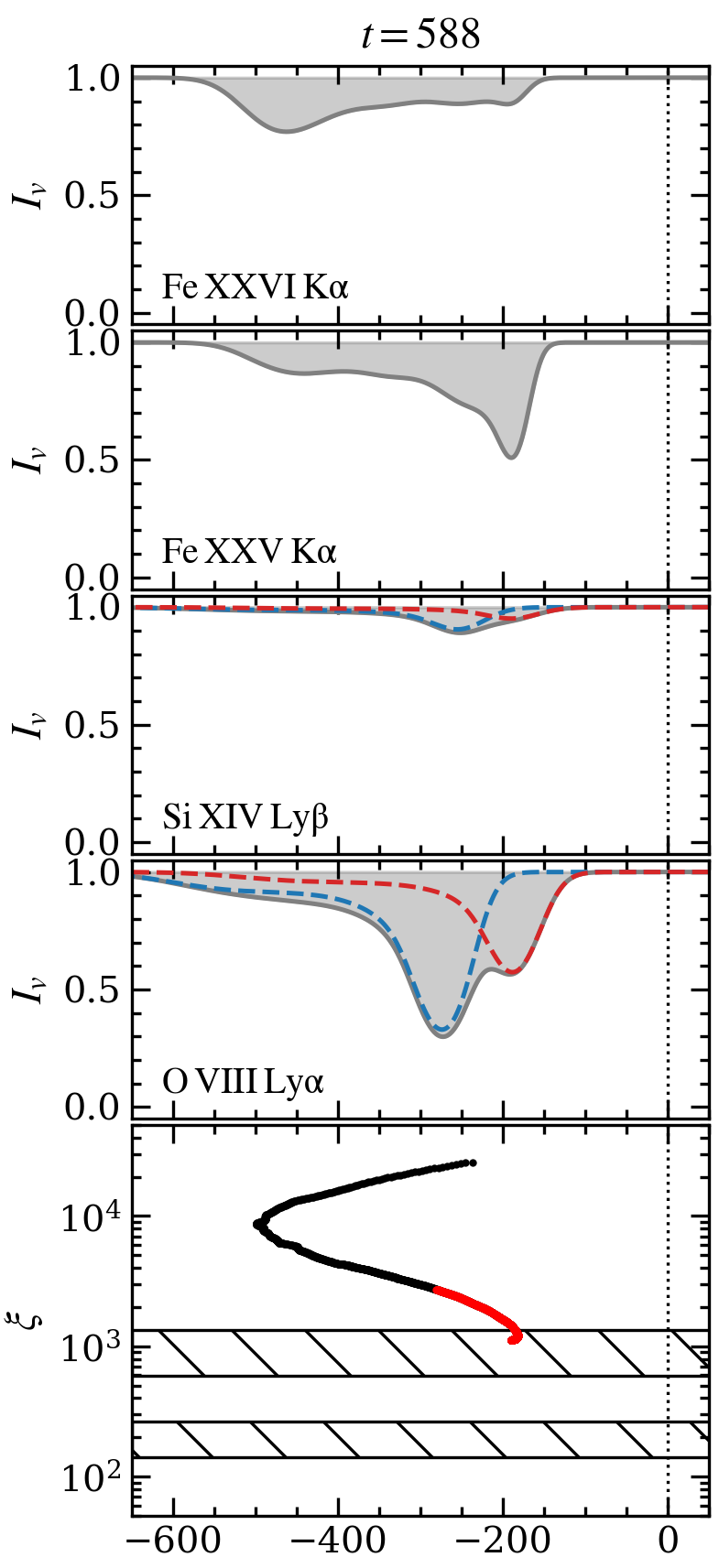}
 \includegraphics[width=0.5\columnwidth]{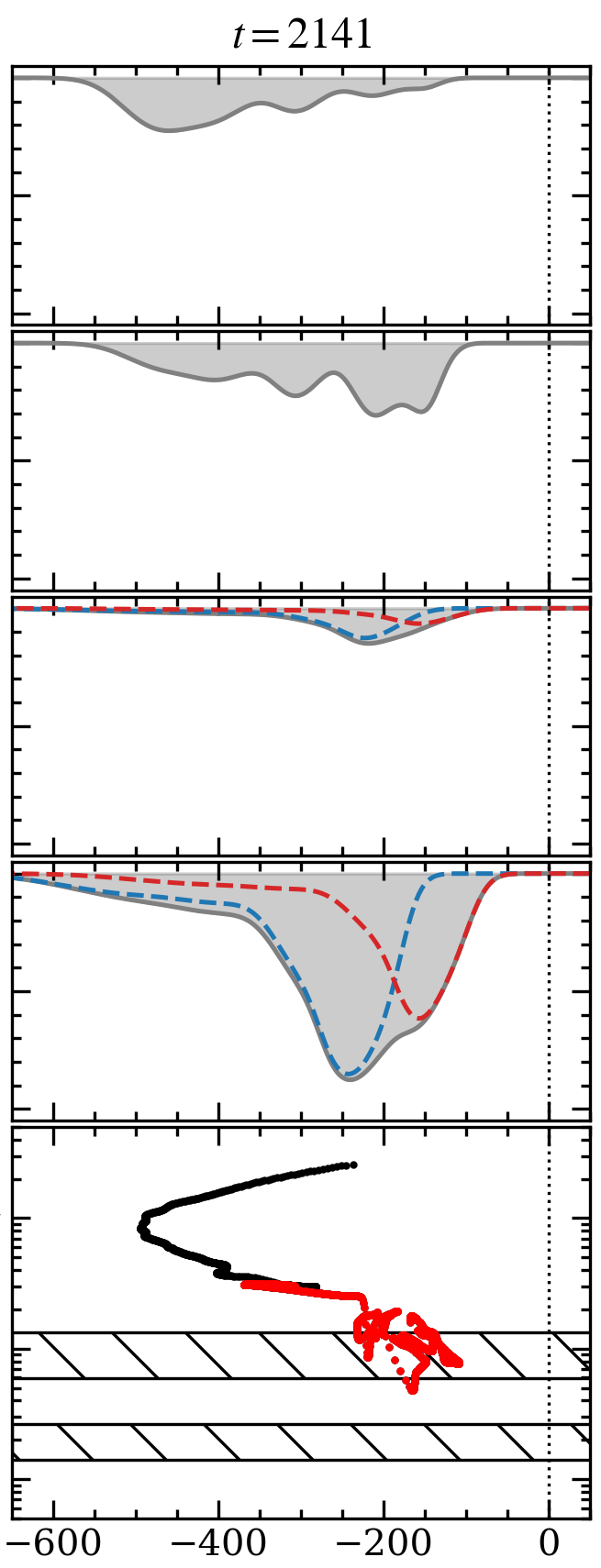}
 \includegraphics[width=0.5\columnwidth]{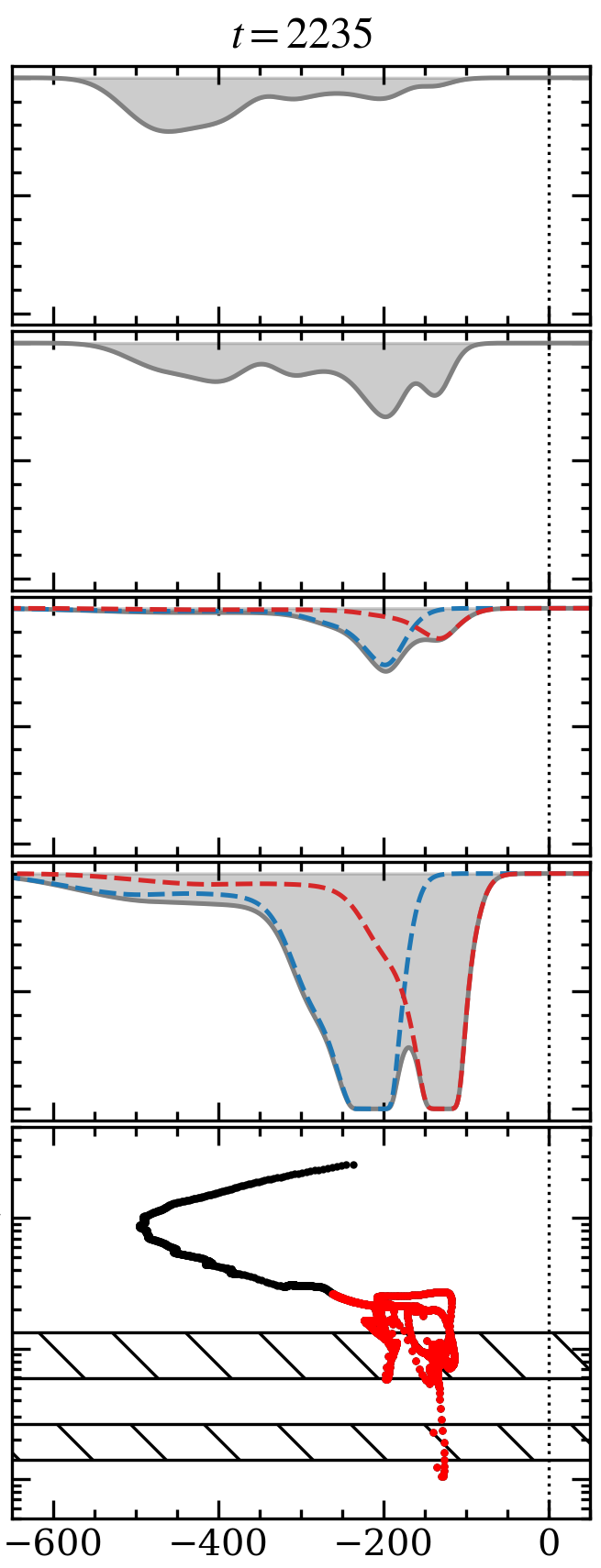}
 \includegraphics[width=0.601\columnwidth]{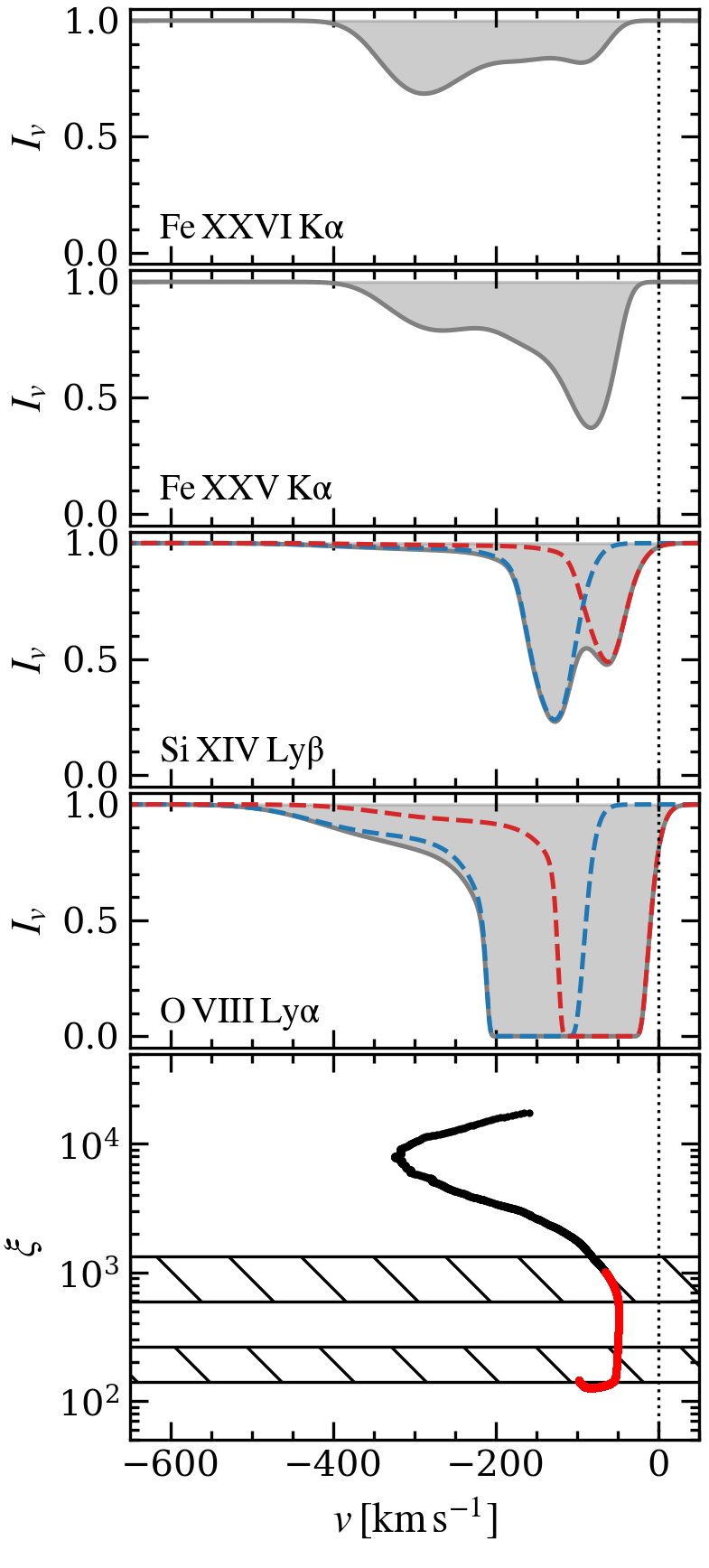}
 \includegraphics[width=0.5\columnwidth]{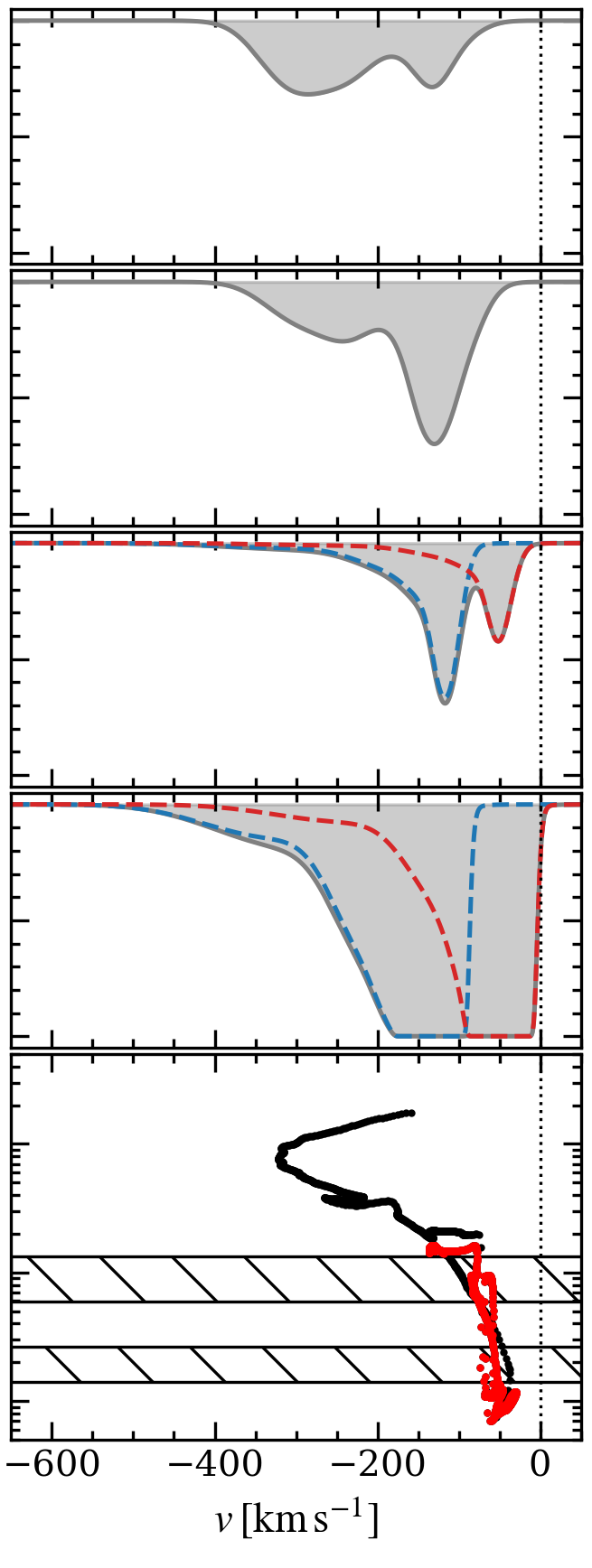}
 \includegraphics[width=0.5\columnwidth]{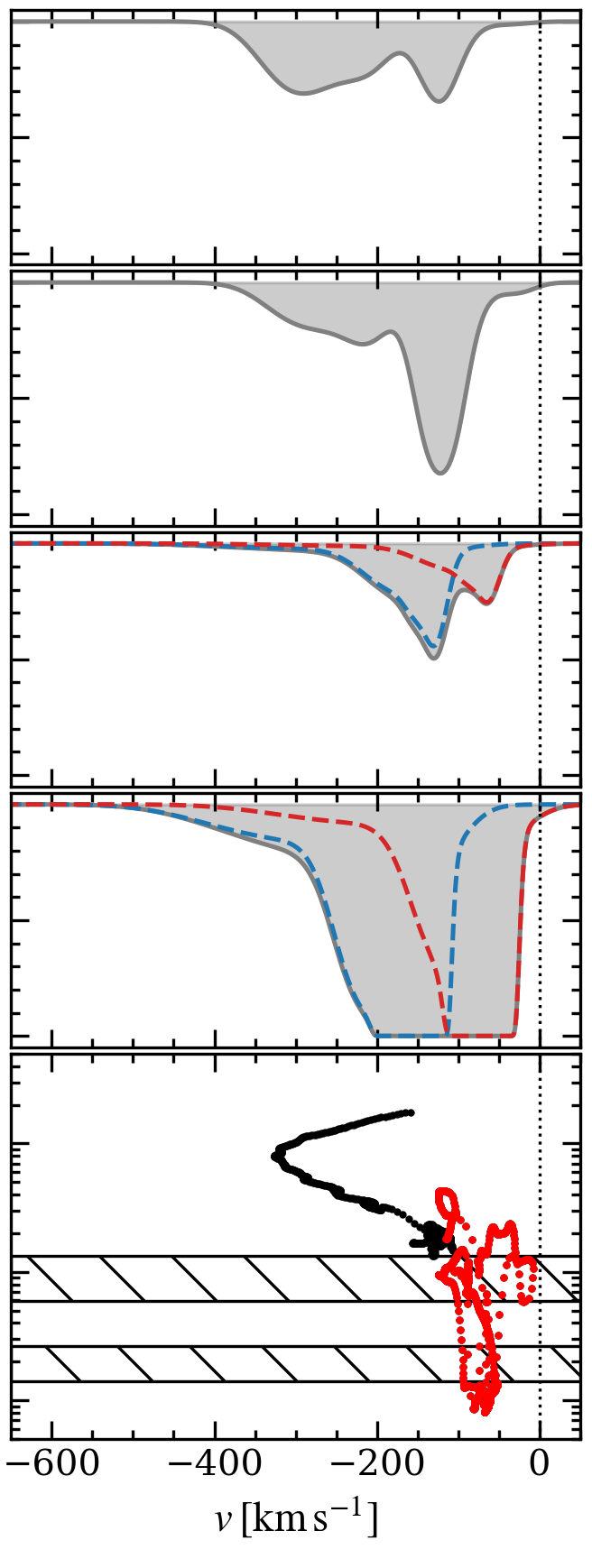}
 \caption{Synthetic X-ray absorption line profiles calculated for the $54^{\degree}$ (top panels) and $63^{\degree}$ (bottom panels) 
 sightline for each snapshot in Fig.~\ref{fig:xi_maps}.  
 Shown are profiles for the ${\rm K}\alpha$ resonance lines of \FeXXVI ~and \FeXXV ~followed by the doublets \SiXIV ~and \OVIII; the individual components of these blended doublets are shown as blue and red colors.  Bottom subpanels are scatter plots showing $\xi$ as a function of the line of sight velocity (the x-axis is $v_{\rm los} = -v_r$); red dots correspond to the red portions of the sightlines in Fig.~\ref{fig:xi_maps}.  Hashed areas mark the $\xi$-ranges of the upper/lower TI zones.
 }
 \label{fig:LPs}
\end{figure*}
%%%%%%%Fig.~\ref{fig:LPs}

\subsubsection{Synthetic absorption line profiles}
We frame our discussion of X-ray absorption line profiles around 
the four bullet points above.  These AMD properties 
correspond, respectively, to the following expectations:
\begin{itemize}
\setlength\itemsep{-0.25em}
\item Deeper absorption should occur for unsaturated lines in clumpy compared to smooth states 
for ions probing the cold phase gas.
\item There should be a spectral signature corresponding to the prominent dip near $\xi_{\rm c,max}$:
for ions with peak abundance at $\xi \gtrsim \xi_{\rm c,max}$, 
a gradual transition for core to wing as the line desaturates;  
for ions with peak abundance at $\xi < \xi_{\rm c,max}$, 
a sharp desaturation should occur right at $\xi_{\rm c,max}$.  
For the $54^{\degree}$ sightline here, this occurs at $v \approx 150-200\,{\rm km\,s^{-1}}$. 
\item Deeper absorption should occur at the lower velocities characterizing gas in the atmosphere
for ions directly probing gas in the hot bubbles (\FeXXVK\, and \FeXXVIK\, for our S-curve).  For ions probing the cold phase gas, less absorption should accompany bubbles due to the reduction in atmospheric column density caused by their presence.
\item Multiple absorption troughs should mimic the shapes of absorption line doublets for ions probing the evaporative wakes of the IAFs, due to the AMD showing dips at different velocities.   
\end{itemize}

In Fig.~\ref{fig:LPs}, line profile calculations are presented for ions with peak abundance 
at a few $10^5~{\rm K}$ where $\xi \approx \xi_{\rm c,max}$ (\OVIIIi),
at higher temperatures ($\sim 10^6~{\rm K}$) characteristic of the evaporative wakes of IAFs (\SiXIVi), 
and in the hot bubbles or the highly ionized gas at small radii (\FeXXV,\,\FeXXVI) where $T > 10^7~{\rm K}$.
The third and fourth bulleted expectations are readily seen to be realized for \FeXXVK\, and \FeXXVIK.  The abundance of \FeXXV\, peaks in what is essentially the brightest yellow gas in Fig.~\ref{fig:xi_maps}, while for \FeXXVI\, the peak is for the brown (higher ionization) gas, which has higher velocity.  This explains the overall line shapes for the $63^{\degree}$ sightline (bottom panels of Fig.~\ref{fig:LPs}).
Comparing the amounts of brown gas along the red portion of this sightline in Fig.~\ref{fig:xi_maps}, the $t=2235$ snapshot should show deeper absorption in \FeXXVIK\, at velocities corresponding to the bubbles ($v \approx 125\,{\rm km\,s^{-1}}$) compared to $t=588$.  This is clearly the case, and it occurs more prominently even for \FeXXVK, although for \FeXXVIK\, the effect serves to make it appear as though this singlet is blended with another line of comparable opacity.  

For the $54^{\degree}$ sightline, the shape of \FeXXVK\, at $t=2235$ is such that it masquerades as a blended 2:1 doublet (where the blue component has an oscillator strength twice that of the red component).  \SiXIV\, is an actual 2:1 doublet line with blended components.  The effects of clumpiness are mostly concealed for it because this line mainly traces the green gas in Fig.~\ref{fig:xi_maps}, which is confined to the outer portion of the sightline and hence rather localized in velocity space.  However, comparing the shapes of the \SiXIV\, profiles for the $63^{\degree}$ sightline at $t=588$ and $t=2235$, the effects of evaporation dynamics in IAFs can be inferred: a more gradual transition from core to wing implies increased ion opacity across a broader velocity range.  This reflects the underlying IAF dynamics, where the warmer gas in the wake is being accelerated to higher velocity.

This spectral signature for evaporation dynamics is even more prevalent in the line profiles for \OVIII, which is another prominent 2:1 X-ray doublet line. It is very optically thick in the cold phase gas (blue colors in Fig.~\ref{fig:xi_maps}), while it is marginally optically thin to evaporating gas (green and yellow colors).  For the $54^{\degree}$ sightline at $t=588$ and $t=2141$, Fig.~\ref{fig:LPs} therefore shows that this line is unsaturated because 
there is no cold phase gas present along the line of sight.  
For $t=2235$ in the upper set of plots, the line becomes saturated due to the dense clump in the wind (see Fig.~\ref{fig:xi_maps}).  The wing is no longer Guassian-like as in the profile at $t=588$.  This extended wing feature in clumpy compared to smooth snapshots is seen most clearly for the profiles of the $63^{\degree}$ sightline.  Note the difference with the extended wings of the \FeXXVIK\, profiles. \FeXXVI\, is not a direct tracer of evaporation dynamics, as its ${\rm K}\alpha$ line forms in the hottest gas far downstream in an IAF's wake.  

The \OVIII~line profiles for the $54^{\degree}$ sightline exhibit the expected behavior given in the first bullet above.  There is no cold gas along this sightline for the smooth state and progressively more in subsequent states.  We note that the \SiXIV~profiles are also consistent with our expectations despite showing opposite trends in moving from $t=588$ to $t=2235$ when comparing both sightlines.  Because this ion's abundance peaks in the evaporative wakes (green gas) rather than in the cold phase, it deepens in sightlines passing above rather than through the atmosphere.

As for the second bullet, here we have only considered ions with peak abundances near or above 
$\xi_{\rm c,max}$, so the core to wing transition is gradual rather than sharp.  In other words, a prominent dip in the AMD results in a sharp loss of opacity only if the ion abundance is not significantly increasing beyond the dip.  For high ionization UV lines, such as the doublet \CIVdbl, there is indeed a sharp core to wing transition \citep[see][where synthetic line profiles for our 1D radial wind solutions are presented]{Ganguly21}.

\section{Discussion}
\label{sec:discussion}
In this work, we found that the standard framework for modeling thermally driven disk winds allows steady
state solutions to be obtained only when the computational domain excludes the cold phase atmosphere and the accompanying 
large scale vortices that form at $r \gtrsim 15\,R_{\rm IC}$.  Clumpy disk wind solutions result otherwise, 
and we showed these to be analogs of the clumpy radial wind solutions obtained by \citetalias{Dannen20} as the clumpiness is again caused by TI.  Here we discuss this instance of dynamical TI in the larger context of AGN physics.

At scales well outside of the broad line region (BLR) as probed here, TI plays two related but distinct roles: 
(i) it first causes the formation of small hot spots in the upper atmosphere that subsequently expand, 
becoming larger hot bubbles; and 
(ii) it facilitates continuous heating of new gas entering the bubbles. 
As a result of (ii), the bubbles further expand and rise, causing a fragmented layer of the atmosphere 
to be lifted into the disk wind above.  
The resulting structure, which we refer to as an irradiated atmospheric fragment (IAF), forms crests and filaments that can break off, 
sending smaller clumps into the wind.  The evaporation of the IAF and any resulting clumps supplies 
the disk wind with a substantially higher column of lower ionization gas than is present in the smooth 
evolutionary stages at the same viewing angle. 
IAFs are large scale structures and the dynamics just described occurs on timescales spanning 
$10^3-10^4$ years.  According to these models, the system is frozen in a state like those shown 
in the bottom two panels of Fig.~\ref{fig:xi_maps} on timescales of years.

\subsection{Dynamical TI in the presence of other forces}
\label{sec:dynTI}
We previously showed that radiation forces alter the basic outcome of local TI in an anisotropic radiation field \citep{PW15}.
Line driving at the distances where IAFs form appears to be too weak (due to over-ionization and large optical depth in lines) to significantly affect these multiphase disk wind solutions \citep{Dannen19}.
It is useful nonetheless to draw a comparison with these previous cloud acceleration simulations to address the conflicting claims regarding the role of TI in both the BLR \cite[e.g.,][]{Beltrametti81,MathewsFerland87,Mathews90,Elvis17,Wang17,Matthews20} and intermediate to narrow line regions of AGNs \cite[e.g.,][]{Krolik84,Stern14,Goosmann16,Bianchi19,Borkar21}.
In \cite{PW15}, we were tracking a single cloud as it formed and accelerated (mainly by line driving), and we envisioned that globally, the environment of this cloud would be similar to that surrounding the isolated clump shown in the bottom panel of Fig.~\ref{fig:xi_maps}.
Rather than the result of condensation out of a warm unstable phase, this clump is an ejected piece of the crest of an IAF.  
`Cloud formation due to TI' has therefore taken on a new meaning in this paper, as clumps ejected from IAFs ultimately owe their existence to TI.  
It is thus necessary to distinguish between local TI, in which clouds form via condensation, and dynamical TI.
In this work, the latter is only directly responsible for forming hot bubbles, which then go on to fragment the cold phase gas in the upper layers of the disk atmosphere.  

At much smaller radii where line-driving becomes very important, it is first of all unclear if TI is even relevant.  
In the original discovery of clumpy disk winds by \cite{Proga98}, for example, the wind was taken to be isothermal, thereby showing that clumps can arise due to radiation forces and gravity alone. 
Additionally, the recent work of \cite{Matthews20} demonstrates that a combination of `microclumping' and density stratification shows promise for explaining the known properties of the BLR.  
However, it has been proposed that BLR clouds are the result of condensation out of the hot phase of line-driven winds \citep{Shlosman85}, i.e. local TI was envisioned. 

Let us suppose, therefore, that TI and line driving are interdependent in the BLR.
It then becomes unclear if the above distinction between dynamical TI and local TI still 
needs to be drawn. \cite{Waters19c} argue that the local TI approximation is valid in the BLR 
and furthermore show that if the flow dynamics can be modeled using multiphase turbulence simulations, 
then the resulting ionization properties are consistent with those of local optimally emitting cloud \citep[LOC; see e.g.,][]{Leighly07} models.
The properties of multiphase turbulence can likely resolve most of the shortcomings of the original two phase model of \cite{Krolik81} identified by \cite{MathewsFerland87} and \cite{Mathews90}, who concluded that BLR clouds are unlikely to be formed by TI.
The remaining issues depend on the validity of the local TI approximation, which will typically break down wherever gradients in the bulk flow become steep, e.g., within the transition layers expected between disks, atmospheres, coronae, and winds.  The line emitting gas is likely produced within such layers (or within the interfaces of embedded dense clumps), so these issues should be revisited after the much richer multiphase gas dynamics accompanying dynamical TI under strong radiation and/or magnetic forces is better understood.  

Our finding that the sign of the Bernoulli function at the entrance to a TI zone determines whether or not the flow is stable to TI is an important development that will hopefully lead to a more complete understanding of dynamical TI.  The need for a sign change in $Be$ may turn out to be unique to thermally driven flows at large distances where gravity is small.  
More fundamental is the accompanying interpretation that individual entropy modes propagating along streamlines get disrupted upon encountering a large \textit{gradient} in $Be$.  By considering Balbus' criterion for TI,
\beq
\left(\pd{\mathcal{L}/T}{T}\right)_p < 0,
\seq
and the fact that in a steady state, $\mathcal{L}/T = -T^{-1}\gv{v}\cdot \nabla Be$ by Bernoulli's theorem (see \eqref{eq:Be}),
we see directly that $\nabla Be$ along streamlines is relevant to the onset of TI.  A formal inquiry into this connection requires a Lagrangian perturbation analysis, however. 

Since a generalized Bernoulli function is central to the theory of MHD outflows \citep[e.g.,][]{Spruit96}, 
it will be interesting to determine if a jump in $Be$ caused by magnetic pressure alone can
inhibit TI.  
In extending these solutions to MHD, the disk atmospheres are expected to be unstable to the MRI.
It remains to be seen if the resulting MHD turbulence can upend the important role of the atmospheric vortices in this work, which result in the episodic production of IAFs. 
This is a pertinent issue to address, as these vortices can become much weaker upon including the pressure gradient necessary for a radial force balance in the midplane BC (see Appendix~A).  While this term is only a small correction for a cold disk and is therefore typically omitted (e.g., Tomaru et al. 2018; Higginbottom et al. 2018), our test runs showed that IAFs develop on significantly longer timescales upon including it.  
An MRI disk setup will not fully alleviate the ambiguity regarding this term.\footnote{Ambiguity arises because the midplane BC serves to approximate the angular momentum supply of the underlying disk but not its temperature, which is significantly colder than the actual temperature assigned to $\theta = \pi/2$.  Optical depth effects in the transition from disk to atmosphere will dictate the value of this pressure gradient, so radiation MHD is ultimately needed.}

We point out that by extending the current modeling framework to include an MRI disk setup, 
it becomes possible to investigate one of the most plausible theories envisioned 
for the origin of BLR clouds --- dense clumps being 
ejected from disk atmospheres by large scale magnetic fields \citep{EBS92,dKB95}. 
IAFs, if they can somehow still form at these small radii, can no longer be advected outward as gravity is too strong (i.e. the bound in \eqref{eq:HEPbound} is violated).  It is entirely conceivable, nevertheless, that they could facilitate the mass loading of cold phase atmospheric gas at the base of a magnetic flux tube, perhaps in concert with a combination of line driving and magnetic pressure forces.
However, solving the non-adiabatic MHD equations in the BLR is a very computationally 
demanding problem, even in 2D.  
The ${\rm HEP}_0$ of the BLR is found by rearranging \eqref{eq:r0},
\beq 
{\rm HEP}_0 = 3.73\times 10^4 \f{\mu(1-\Gamma)}{(T_0/10^5\,{\rm K})} \,(r_0/1\,{\rm ld})^{-1}.
\seq
For $M_{\rm bh} = 5.2\times 10^7\,M_\odot$ and a typical BLR radius of $10\,{\rm ld}$ \citep[characteristic of NGC 5548; see e.g.,][]{Kriss19}, ${\rm HEP}_0 \approx 10^4$.
Noting from \S{\ref{sec:background}} that $t_{\rm cool}(r_0)/t_{\rm kep}\propto {\rm HEP}_0^{-1/2}$,
applying the same methods used here would incur a timestep at least $10\times$ smaller than our current one, and our current one is already an order of magnitude smaller than the timestep necessary to perform global MRI disk simulations using ideal MHD.
Nevertheless, we aim to investigate such a role for dynamical TI in the near future.

\subsection{Limitations of the present models}
\label{sec:Limitations}
Our current solutions account for the gravity of the black hole, the radiation force due to electron scattering, and the effects of X-ray irradiation for optically thin flows. 
Several other processes need to be included to establish IAFs as a robust phenomenon in the environment of AGNs.
Perhaps the most important ones are geometrical: upon relaxing axisymmetry by running fully 3-D simulations, there is an additional degree of freedom afforded to TI that may allow IAFs to form at smaller radii.  Specifically, while the poloidal velocity field prevents the growth of entropy modes within $\sim15\,R_{\rm IC}$ in 2-D, entropy modes undergoing condensation in the $\phi-$direction will be subject to weaker stretching.  Overall this should increase the efficiency of thermal driving due to the enhanced heating accompanying clump formation.

Other neglected effects include those of self-gravity, radiative transfer, and dust opacity.
Of these, only dust opacity is expected to also strengthen the outflow.  Note that because dust can only survive inside the clumps, the contribution of dust opacity to the driving force can only be accurately assessed after determining the cold gas distribution from 3-D simulations.  Both radiative transfer and the disk self-gravity are expected to slow the outflow by leading to an effective increase in the local HEP value (the ratio of gravitational to thermal energy).  The former will likely reduce the net heating of cold gas and hence lower the thermal energy, while the latter should increase the local gravity.

\section{Summary and Conclusions}
\label{sec:summary}
In this paper, we have shown that the thermally unstable radial wind solutions obtained by \citetalias{Dannen20} have counterpart disk wind solutions.   This regime is found by modifying \citetalias{Dannen20}'s fixed-density spherical boundary condition to a `midplane' boundary condition where density varies as $r^{-2}$ and a near-Keplerian azimuthal velocity profile is assigned.
Because our understanding of TI derives mainly from studies performed in a local approximation (using periodic boundary conditions), we emphasize that these clumpy wind solutions are realizations of flows encountering `dynamical TI', where effects such as the stretching of perturbations due to the acceleration of the flow are more important than Field's criterion for determining stability.

Prior work on thermally driven disk winds has shown that within $\sim15\,R_{\rm IC}$, the presence of a TI zone on the S-curve simply aids 
heating to the Compton temperature by providing an efficient heating mechanism (see \S{\ref{sec:background}}).  The flow can enter the TI zone, 
meaning it becomes locally thermally unstable, but fundamentally the local theory of TI describes the evolution 
of individual entropy modes.  We showed in \S{\ref{sec:Be}} that by considering the evolution of such modes 
advected along streamlines, strong disruptive effects encountered as they enter the TI zone 
(such as the above mentioned stretching due to rapid acceleration) will prevent their growth.  Beyond the large scale vortices that are present 
in the atmospheres of our solutions within $80\,R_{\rm IC}$, streamlines in the atmosphere transition 
from being connected to the disk wind to being separately outflowing. 
On a phase diagram, this atmospheric outflow occupies the cold phase branch of the S-curve below the temperature of the TI zone.
The only actual pathway into the TI zone is along streamlines passing through the vortices themselves,
and the circulation permits multiple passes through the zone, giving TI more time to amplify entropy modes.
These streamlines are confined to a small range of radii relative to the domain size but nevertheless occupy a region spanning more than $10\,R_{\rm IC}$ (see Fig.~\ref{fig:hotspot_analysis}). 

The resulting nonlinear perturbations produce hot spots that then seed the formation of an IAF.  Namely, IAFs are what we call 
the large, filamentary and outward propagating vortical structures that arise as hot spots become hot bubbles and disrupt the transition layer between the atmosphere and disk wind.  
They are lifted into the disk wind by the expansion and buoyancy of the bubbles, which undergo continuous runaway heating as a consequence of remaining thermally unstable.  
Once exposed to the wind, the filamentary layers of the IAFs evaporate and supply the disk wind with multiphase gas over a large solid angle.  
Occasionally these layers break into smaller clumps that also evaporate before entering the supersonic flow.  Importantly, however, this evaporation dynamics occurs on timescales of thousands to hundreds of thousands of years in AGNs, so IAFs can provide a physical explanation for the fact that warm absorber type outflows are very common.

Our simulations show that the IAFs transition from mere bulges and depressions in the atmosphere 
to large scale filaments right around the `unbound radius',
\beq
R_{\rm u} \approx 3.03\times 10^2\,\f{T_C/10^8\,\rm{K}}{T_{\rm c,max}/10^5\,\rm{K}}\, R_{\rm IC}(1-\Gamma),
\seq
which was identified as the distance beyond which our radial wind solutions from \citetalias{Dannen20} can become clumpy.
We note that $R_{\rm IC}(1-\Gamma)$ is the generalized Compton radius upon accounting for radiation pressure \citep{Proga02} 
and that $R_{\rm u}$ (like $R_{\rm IC}$) is a property of the S-curve.  Therefore, the theory of multiphase thermally 
driven disk winds is predictive: this radius is sensitive to both the luminosity and the SED, 
making it possible to compare $R_{\rm u}$ with the inferred locations to warm absorbers across a population of AGNs 
with different values of $\Gamma$ and $T_{C}/T_{\rm c,max}$.  

In conclusion, here we have uncovered what are likely the simplest type of multiphase disk wind solutions.
It is also noteworthy that this year marks 25 years since the pioneering numerical work by \cite{Woods96}, 
whose hydrodynamical simulations of thermally driven winds not only confirmed and refined the theory developed 
by \citetalias{Begelman83}, but also established a framework for building self-consistent disk wind models.  
Here we have applied this framework to further extend the basic theory of thermally driven disk winds 
by clarifying the role of dynamical TI and its relation to local TI.  
We view this development as a building block toward a much broader theory 
of multiphase disk winds in AGNs.  A full theory should encompass multiple wind launching mechanisms (e.g., thermal-driving, line-driving, magnetic-driving, and shock-driving) as well as different ways that gas can become multiphase 
(e.g., clumps can alternatively arise from compression due to shocks or radiation pressure).  

\section{Acknowledgements}
We thank the anonymous referee for a constructive report that led us to improve the organization of \S{3} and to add \S{4.2}.
We thank Sergei Dyda for regular discussions over the course of this project.  TW thanks 
Hui Li and the Theoretical Division at Los Alamos National Laboratory (LANL) for allowing 
him to retain access to the institutional computing (IC) clusters.  These calculations were
performed under the LANL IC allocation award \verb+w19_rhdccasims+.
Support for this work was provided by the National Aeronautics Space Administration under ATP grant NNX14AK44G and through Chandra Award Number TM0-21003X issued by the Chandra X-ray 
Observatory Center, which is operated by the Smithsonian Astrophysical Observatory for 
and on behalf of the National Aeronautics Space Administration under contract NAS8-03060. 
\par
\bibliographystyle{aasjournal}
% \bibliography{progalab-shared}

\appendix 
\section{Numerical Methods} \label{sec:num-meth}
Using \athena \citep{Stone20}, we solve the equations of non-adiabatic gas dynamics, 
accounting for the forces of gravity and radiation pressure due to electron scattering opacity:
\beq
   \frac{D\rho}{Dt} = -\rho \nabla \cdot {\bf v},
\seq
\beq
   \rho \frac{D{\bf v}}{Dt} = - \nabla p - \f{G\,M_{\rm bh}\rho}{r^2}(1-\Gamma),
\seq
\beq
   \rho \frac{D\mathcal{E}}{Dt} = -p \nabla \cdot {\bf v} - \rho \mathcal{L}.
\seq
For an ideal gas, $p=(\gamma-1)\rho\mathcal{E}$, and we set $\gamma=5/3$.
Our calculations are performed in spherical polar coordinates,
$(r,\theta,\phi)$, 
assuming axial symmetry about a rotational axis at $\theta=0^{\degree}$. 
The S-curve of our net cooling function $\mathcal{L}$ is shown in Fig.~\ref{fig:Scurve} and was obtained by running a large grid of photoionization calculations using \xstar~\citep[see][]{Dyda17} for the SED of NGC~5548 obtained by \cite{Mehdipour15}.  The abundances are those of \cite{Lodders09}, but for simplicity, in this work we set $\mu_H = 1$, keeping only $\mu = 0.6$.  

The net cooling term can be expressed as $\rho \mathcal{L} = n^2(C - H)$, 
and we interpolate from tables of the total cooling ($C$) and total heating ($H$) rates given in units of ${\rm erg\,cm^3\,s^{-1}}$; these tables were made publicly available by \cite{Dannen19}.  To obtain the clumpy outflow solutions in \citetalias{Dannen20}, we used bilinear interpolation to access values from the tables 
\citep[for details, see][]{Dyda17}.  We suspected that this basic algorithm could be somewhat inaccurate near the S-curve where $\mathcal{L}$'s are close to zero. 
In the testing phase of this work, we verified this suspicion, 
confirming that interpolation errors introduce perturbations into the flow that are continually amplified by TI, allowing even our 1D solutions to remain clumpy indefinitely.
Specifically, the \%-difference in rates evaluated near the S-curve where $C \approx H$ could exceed 1000\% based on a comparison of looking up rates tabulated from an analytic function (using a similar sampling as our actual tables).

We therefore implemented a GSL bicubic interpolation routine that we found reduces the percent differences for rates near the S-curve to less than 1\% and returns rates that are on average about $10\times$ more accurate.  
As discussed in \S{\ref{sec:Be}, all of the 1D radial wind solutions in \citetalias{Dannen20} reach a steady state with this improvement.

\subsection{Initial conditions, boundary conditions, and computational domain}
Our initial conditions are $\rho(r,\theta_{\rm max}) = \rho_0(r_0/r)^2$ (with $\rho_0 = \bar{m}\,n_0$), $\gv{v}(r,\theta_{\rm max}) = v_\phi(r) \hat{\phi}$, and $p(r,\theta_{\rm max}) = n_0k\,T_0 (r_0/r)^2$, where $\theta_{\rm max}$ denotes zones with cell edges at $\pi/2$ ($\tt{je}$ zones in \athena notation).   
All other zones have $\rho(r,\theta) = \rho_0 (T_C/T_0)(r_0/r)^2$, $\gv{v}(r,\theta) =  (1.1\,v_0\sqrt{1. - r_{\rm in}/r} + 10^{-4} v_0)\hat{r}$, and $p(r) = p(r,\theta_{\rm max})$, where $v_0 = \sqrt{GM_{\rm bh}/r_0}$.  

Reflecting BCs are applied at $\theta = \pi/2$.
In our fiducial runs, the rotation profile along this boundary is enforced to be sub-Keplerian to account for the radiation force, $v_\phi(r) = \sqrt{GM_{\rm bh}(1-\Gamma)/r}$ at $\theta = \pi/2$.  We also ran tests accounting for the contribution of the pressure gradient in the radial force balance along the midplane, given by
\beq
\f{v_\phi^2}{r} = \f{GM_{\rm bh}}{r^2}(1-\Gamma) + \f{1}{\rho}\f{dp}{dr}.
\seq
For the isothermal midplane BC that we use, $dp/dr = (-2/r)p(r)$ at $\theta = \pi/2$.
This midplane BC is applied inside our heating and cooling routine (implemented using {\tt EnrollUserExplicitSourceFunction()}).  We reset the midplane density, velocity, and pressure on every call ($2\times$ per timestep for a 2nd order integrator) to the above values.  Only the density and velocity are required to be reset, as \athena will find the same solution without specifying the pressure also, but only if a timestep $2\times$ smaller is used.  This stability constraint arises because the midplane contains the densest gas and thus gas with the shortest cooling times in the domain --- see Fig.~\ref{fig:Xi0runs}.  

We apply a `constant gradient' BC at $r_{\rm in}$, in which all primitive variables are linearly extrapolated from the first active zone into the ghost zones.  At $r_{\rm out}$, we apply modified outflow BCs that prevent inflows from developing at the boundary.  Specifically, $v_r$ is set to 0 in the ghost zones if it is found to be less than zero (in \athena notation, we set $\tt{prim(IVX,k,j,ie+i}) = 0$ if $\tt{prim(IVX,k,j,ie+i}) < 0$).  Without this latter BC, a transient disturbance will enter from the outer boundary after hundreds of orbital times at $R_{\rm IC}$ (but before IAFs develop).  

Our calculations utilize static mesh refinement (SMR) to give an effective grid size, $N_r\times N_\theta$ ($\tt{nx1 \times nx2}$ in \athena notation), of
$1056\times 576$ for our mid-res run and $2112\times 1152$ for our hi-res run.
Our base grid is $132 \times 72$, and SMR levels are set at 
$15^{\degree} < \theta < 90^{\degree}$,
$30^{\degree} < \theta < 90^{\degree}$, and
$35^{\degree} < \theta < 90^{\degree}$ for our mid-res run.
For our hi-res run, a fourth SMR level is applied at $35^{\degree} < \theta < 90^{\degree}$.
Our radial domain size is $r_{\rm in} = 13.1\,R_{\rm IC}$ and $r_{\rm out} = 285\,R_{\rm IC}$.

\subsection{Timestep constraint}
\label{sec:custom_timestep}
In \cite{Waters18}, we used a semi-implicit solver to include the heating and cooling source term as described by \cite{Dyda17}.  
For this work, we had to abandon this routine in favor of a simpler explicit solver after finding 
that the semi-implicit solve causes a small fraction of the coldest zones within the atmosphere 
to `jump off' the S-curve at every timestep when these zones are visualized 
on the $(T,\Xi)$-plane.  This behavior indicates that numerically, multiple temperature 
solutions can exist for the same timestep, $\Delta t$.  This is a generic property 
of implicit routines \citep{Townsend09}, with the simple remedy being to resort 
to an explicit solve.  Note that Townsend's exact integration routine requires $\mathcal{L}$ 
to be a function of temperature alone, whereas here, $\mathcal{L} = \mathcal{L}(T,\xi)$.  

An explicit solve requires implementing a custom timestep (using {\tt EnrollUserTimeStepFunction}) 
because $t_{\rm cool}$ is generally smaller than the CFL constraint on $\Delta t$ for zones near the midplane.  
This can result in these solutions becoming computationally very costly, and indeed 
it is prohibitively expensive to obtain exact analog solutions to those in \citetalias{Dannen20}, 
which are for $\xi_0 = 5$.  Our $\xi_0 = 5$ solution shown in Fig.~\ref{fig:Xi0runs} used 
only 2 levels of SMR and was not run long enough for the regions beyond $15\,R_{\rm IC}$ 
to reach a steady state. However, the cost is greatly reduced for the $\rm{HEP_0} = 36$ 
solutions presented in \S{\ref{sec:results}} because $t_{\rm cool}(r_0) \propto {\rm HEP_0}^{-2}$ (see \eqref{eq:tcool_r0}).

\section{The `unbound radius'}
Dividing \eqref{eq:Be} by $c_s^2$ and writing $\phi$ in terms ${\rm HEP}_0$ gives 
$Be/c_s^2 = M^2/2 + 1/(\gamma - 1) - {\rm HEP}_0 (T_0/T)(r_0/r)$, where $M$ is the Mach number.   
As shown by the red contours in Fig.~\ref{fig:IAF_midres} through Fig.~\ref{fig:IAFevolution}, the locus of points at which $Be$ equals zero coincides with those where $T$ reaches $T_{\rm c,max} \equiv T(\Xi_{\rm c,max})$ up until some distance.  Setting $Be=0$ defines this contour implicitly as   
\beq
r = r_0\, {\rm HEP}_0 \f{T_0/T_{\rm c,max}}{M^2_{\rm c,max}/2 + 1/(\gamma-1)},
\seq
where $M_{\rm c,max}$ is the Mach number evaluated at $T = T_{\rm c,max}$.
The `unbound radius' we are after is the transition radius where surfaces of $Be=0$ and $T = T_{\rm c,max}$ no longer coincide in Fig.~\ref{fig:IAFevolution}.
It lies interior to the sonic surface, so a lower limit is found by setting $M_{\rm c,max} = 1$ in the above expression.  This defines a characteristic radius
\beq
R_{\rm u} \equiv 2\,r_0\, {\rm HEP}_0 \f{\gamma-1}{\gamma+1}\f{T_0}{T_{\rm c,max}}.
\seq
Eliminating $r_0\, {\rm HEP}_0$ in favor of $R_{\rm IC}$ using \eqref{eq:r0} gives the expression quoted in \eqref{eq:Ru}.
\end{document}